\documentclass[pra,superscriptaddress,preprintnumbers,showpacs,amsmath,amssymb]{revtex4}
\usepackage{graphicx}
\usepackage{subfigure}

\def\be{\begin{equation}}       \def\ee{\end{equation}}
\def\bea{\begin{eqnarray}}      \def\eea{\end{eqnarray}}

\begin{document}

\title{Dynamics of entanglement in a two-dimensional spin system}

\author{Qing Xu}
\affiliation{Department of Physics, Purdue University, West Lafayette, Indiana 47907, USA}
\author{Gehad Sadiek}
\affiliation{Department of Physics, King Saud University, Riyadh 11451, Saudi Arabia}
\affiliation{Department of Physics, Ain Shams University, Cairo 11566, Egypt}
\author{Sabre Kais\footnote{Corresponding author: kais@purdue.edu}}
\affiliation{Department of Physics, King Saud University, Riyadh 11451, Saudi Arabia}
\affiliation{Department of Chemistry and Birck Nanotechnology center,
Purdue University, West Lafayette, Indiana 47907, USA}

\date{\today}

\begin{abstract}
We consider the time evolution of entanglement in a finite two dimensional transverse Ising model. The model consists of a set of 7 localized spin-$\frac{1}{2}$ particles in a two dimensional triangular lattice coupled through nearest neighbor exchange interaction in presence of an external time dependent magnetic field. The magnetic field is applied in different function forms: step, exponential, hyperbolic and periodic. We found that the time evolution of the entanglement shows an ergodic behavior under the effect of the time dependent magnetic fields. Also we found that while the step magnetic field causes great disturbance to the system creating rabid oscillations, the system shows great controllability under the effect of the other magnetic fields where the entanglement profile follows closely the shape of the applied field even with the same frequency for periodic fields. This follow up trend breaks down as the strength of the field, the transition constant for exponential and hyperbolic, or frequency for periodic field increase leading to rapid oscillations. We observed that the entanglement is very sensitive to the initial value of the applied periodic field, the smaller the initial value the less distorted is the entanglement profile. Furthermore, the effect of thermal fluctuations is very devastating to the entanglement which decays very rapidly as the temperature increases. Interestingly, although large value of the magnetic field strength may yield small entanglement, it was found to be more persistent against thermal fluctuations than the small field strengths.
\end{abstract}


\maketitle

\section{Introduction}
Quantum entanglement lies in the heart of quantum theory and is of fundamental role in modern physics \cite{Peres1993}. Entanglement is a nonlocal correlation between two (or more) quantum systems such that the description of their states has to be done with reference to each other even if they are spatially well separated. Understanding and quantifying entanglement may provide an answer for many questions regarding the behavior of complex quantum systems \cite{Kais2007}. Particularly, entanglement is considered as the physical property responsible for the long-range quantum correlations accompanying a quantum phase transition in many-body systems at zero temperature \cite{Sondhi1997,Osborne2002,ZhangJF2009,Osenda2003,HuangZ2004}.
Particular fields where entanglement plays a crucial role are quantum teleportation, quantum cryptography and quantum computing, where it is considered as the physical basis for manipulating linear superpositions of the quantum states to implement the different proposed quantum computing algorithms \cite{Nielsen2000,Boumeester2000}. Different physical systems have been proposed as reliable candidates for the future technology of quantum computing and quantum information processing \cite{Barenco1995,Vandersypen2001,Chuang1998,Jones1998,Cirac1995,Turchette1995,Averin1997,WeiQ2010}. The main task in each one of these systems is to specify certain quantum degree of freedom as the qubit and find a controllable coupling mechanism to form an entanglement among these qubits to perform efficient quantum computing processes.

Multiparticle systems are of central interest in the field of quantum information, in particular, quantification of the entanglement contained in their quantum states. However, quantum states and entanglement are very fragile due to the induced decoherence caused by the inevitable coupling to the environment. Decoherence is considered as one of
the main obstacles toward realizing an effective quantum
computing system \cite{Zurek1991}. The main effect of decoherence is to
randomize the relative phases of the possible states of the considered system.
Quantum error correction \cite{Shor1995} and decoherence free subspace \cite{Bacon2000,Divincenzo2000} have been proposed to protect the quantum property during the computation process. Still offering a potentially ideal protection against environmentally induced decoherence is difficult. In NMR quantum computers, a series of magnetic pulses were applied to a selected nucleus of a molecule to implement quantum gates \cite{Doronin2002}. Moreover, a spin-pair entanglement is a reasonable measure for decoherence between the considered two-spin system and the environmental spins. The coupling between the system and its environment leads to decoherence in the system and vanishing of entanglement between the two spins. Evaluating the entanglement remaining in the considered system helps us to understand the behavior of the decoherence between the considered two spins and their environment \cite{Lages2005}.

In previous works, the evolution of entanglement in a one-dimensional spin system in presence of different forms of external magnetic fields, modeled by the XY Hamiltonian, was studied \cite{HuangZ2005,HuangZ2006}. It was found that the entanglement can be localized between nearest-neighbor qubits for certain values of the external time-dependent magnetic fields. Moreover, as known for magnetization of this model, the entanglement showed nonergodic behavior, i.e. it does not approach its equilibrium value at the infinite time limit. Also, the same system was investigated considering a time-dependent exchange coupling between neighboring spins \cite{Sadiek2010}. It was found that the asymptotic behavior of entanglement at the infinite time limit is very sensitive to the initial values of the coupling and magnetic field and for particular choices we may create finite asymptotic entanglement regardless of the final values of coupling and magnetic field. The quantum effects in the Ising model case showed persistence in the vicinity of both its critical phase transition point and zero temperature as it evolves in time.

The study of quantum entanglement in two-dimensional systems possesses a number of extra problems compared with systems of one dimension. The particular one is the lack of exact solutions. The existence of exact solutions has contributed enormously to the understanding of the entanglement for 1D systems \cite{Lieb1961,Sachdev2001,HuangZ2005,Sadiek2010}. Studies can be carried out on interesting but complicated properties, applied to infinitely large system, and so forth use finite scaling method to eliminate the size effects, etc. Some approximation methods, like Density matrix renormalization group (DMRG), are also only workable in one dimension \cite{Doronin2002,Xavier2010,Silva-Valencia2005,Capraro2002}. So when we carry out this two-dimensional study, no methods can be inherited from previous researches.  They heavily rely on numerical calculations, resulting in severe limitations on the system size and properties. For example, dynamics of the system is a computational-costing property. We have to think of a way to improve the effectiveness of computation in order to increase the size of research objects; mean while dig the physics in the observable systems. It may show the general physics, or tell us the direction of less resource-costing large scale calculations. In a previous work we have studied the entanglement in a 19-site two-dimensional transverse Ising model at zero temperature \cite{XuQ2010}. The spin-$\frac{1}{2}$ particles are coupled through an exchange interaction $J$ and subject to an external time-independent magnetic field $h$. We demonstrated that for such a class of systems the entanglement can be tuned by varying the parameter $\lambda=h/J$ and also by introducing impurities into the system. The system showed a quantum phase transition at a specific critical value of the parameter $\lambda_c$.

In this paper, we consider the dynamics of entanglement in a two-dimensional spin system, where spins are coupled through an exchange interaction and subject to an external time-dependent magnetic field. Four forms of time-dependent magnetic field are considered: step, exponential, hyperbolic and periodic. To tackle down the problem, we introduce two calculation methods: step by step time-evolution matrix transformation and step by step projection. We compare them side by side, in short, besides the exactly same results, step by step projection method turned out to be twenty times faster than the matrix transformation. One section of the paper will expound the scalability of this method, where it sheds light on studies of larger systems. The finite temperature effect is also considered to simulate more realistic systems. We show that the system entanglement behaves in an ergodic way in contrary to the one-dimensional Ising system. The system shows great controllability under all forms of external magnetic field except the step function one which creates rapidly oscillating entanglement. This controllability is shown to be breakable as the different magnetic field parameters increase. Also it will be shown that the mixing of even a few excited states by small thermal fluctuation is devastating to the entanglement of the ground state of the system. The critical temperature at which the concurrence vanishes depends significantly on the value of the magnetic field strength, smaller value yield smaller entanglement but higher critical temperature.

This paper is organized as follows. In the next section we present our model and discuss the two different approaches to evaluate the entanglement. In sec. III we present and discuss our results for the entanglement of the system under the effect of the different magnetic fields forms at zero temperature. The dynamics of thermal entanglement is considered in sec. IV. In sec. V we explain part of our key results in the light of Fermi's golden rule and adiabatic approximation. The extension of our work to larger size spin systems is discussed in sec. V. We
conclude in Sec. VI and discuss future directions.

%%%%%%%%%%%%%%%%%%%%%%%%%%%%%%%%%%%%%%%%%%%%%%%%%%%%%%%%%%%%%%%%%%%%%%%%%
\section{Solution of the time-dependent two-dimensional Ising model}
%%%%%%%%%%%%%%%%%%%%%%%%%%%%%%%%%%%%%%%%%%%%%%%%%%%%%%%%%%%%%%%%%%%%%%%%%

%%%%%%%%%%%%%%%%%%%%%%%%%%%%%%%%%%%%%%%%%%%%%%%%%%%%%%%%%%%%%%%%%%%%%%%%%
\subsection{Model}
%%%%%%%%%%%%%%%%%%%%%%%%%%%%%%%%%%%%%%%%%%%%%%%%%%%%%%%%%%%%%%%%%%%%%%%%%

We consider a set of 7 localized spin-$\frac{1}{2}$ particles in a two
dimensional triangular lattice coupled through exchange interaction $J$ and
subject to an external time-dependent magnetic field of strength $h(t)$. The
Hamiltonian for such a system is given by
\begin{equation}
H=-\sum_{<i,j>}J_{i,j}\sigma_{i}^x\sigma_{j}^x-h(t)\sum_{i}\sigma_{i}^z,
\end{equation}
where $<i,j>$ is a pair of nearest-neighbors sites on the lattice,
$J_{i,j}=J$ for all sites. For this model it is convenient to define a dimensionless coupling constant $\lambda=h/J$. We apply different forms of the magnetic field as a function of time: step function, exponential, hyperbolic and periodic.

%%%%%%%%%%%%%%%%%%%%%%%%%%%%%%%%%%%%%%%%%%%%%%%%%%%%%%%%%%%%%%%%%%%%%%%%%
\subsection{The evolution operator}
%%%%%%%%%%%%%%%%%%%%%%%%%%%%%%%%%%%%%%%%%%%%%%%%%%%%%%%%%%%%%%%%%%%%%%%%%

According to quantum mechanics the transformation of $|\psi_i(t_{0})\rangle$, the state vector at the initial instant $t_0$, into $|\psi_i(t)\rangle$, the state vector at an arbitrary instant, is linear \cite{Cohen-Tannoudji2005}. Therefore there exists a linear operator $U(t,\, t_{0})$ such that:
\be
\label{U_vector}
|\psi_i(t)\rangle=U(t,\, t_0)\,|\psi_i(t_0)\rangle.
\ee
This is, by definition, the evolution operator of the system. Substituting Eq. (\ref{U_vector}) into the Schor\"{o}dinger equation, we obtain:
\be
i\hbar\frac{\partial}{\partial t}U(t,t_0)|\psi(t_0)\rangle=H(t)U(t,t_0)|\psi(t_0)\rangle,
\ee
which means
\be
i\hbar\frac{\partial}{\partial t}U(t,t_0)=H(t)U(t,t_0).
\ee
Further taking the initial condition
\be\label{U_initial}
U(t_0,t_0)=\mathbb{I},
\ee
the evolution operator can be condensed into a single integral equation:
\be\label{U_integral}
U(t,t_0)=\mathbb{I}-\frac{i}{\hbar}\int_{t_0}^t H(t')U(t',t_0)dt'.
\ee

When the operator $H$ does not depend on time, Eq. (\ref{U_integral}) can easily be integrated and at finally gives out
\be\label{U_time_independent}
U(t,t_0)=e^{-iH(t-t_0)/\hbar}.
\ee

%%%%%%%%%%%%%%%%%%%%%%%%%%%%%%%%%%%%%%%%%%%%%%%%%%%%%%%%%%%%%%%%%%%%%%%%%
\subsection{Step by step time-evolution matrix transformation}
%%%%%%%%%%%%%%%%%%%%%%%%%%%%%%%%%%%%%%%%%%%%%%%%%%%%%%%%%%%%%%%%%%%%%%%%%

\begin{figure}[htbp]
\begin{minipage}[c]{\textwidth}
 \centering
   \subfigure[]{\label{Step_by_step1}\includegraphics[width=7 cm]{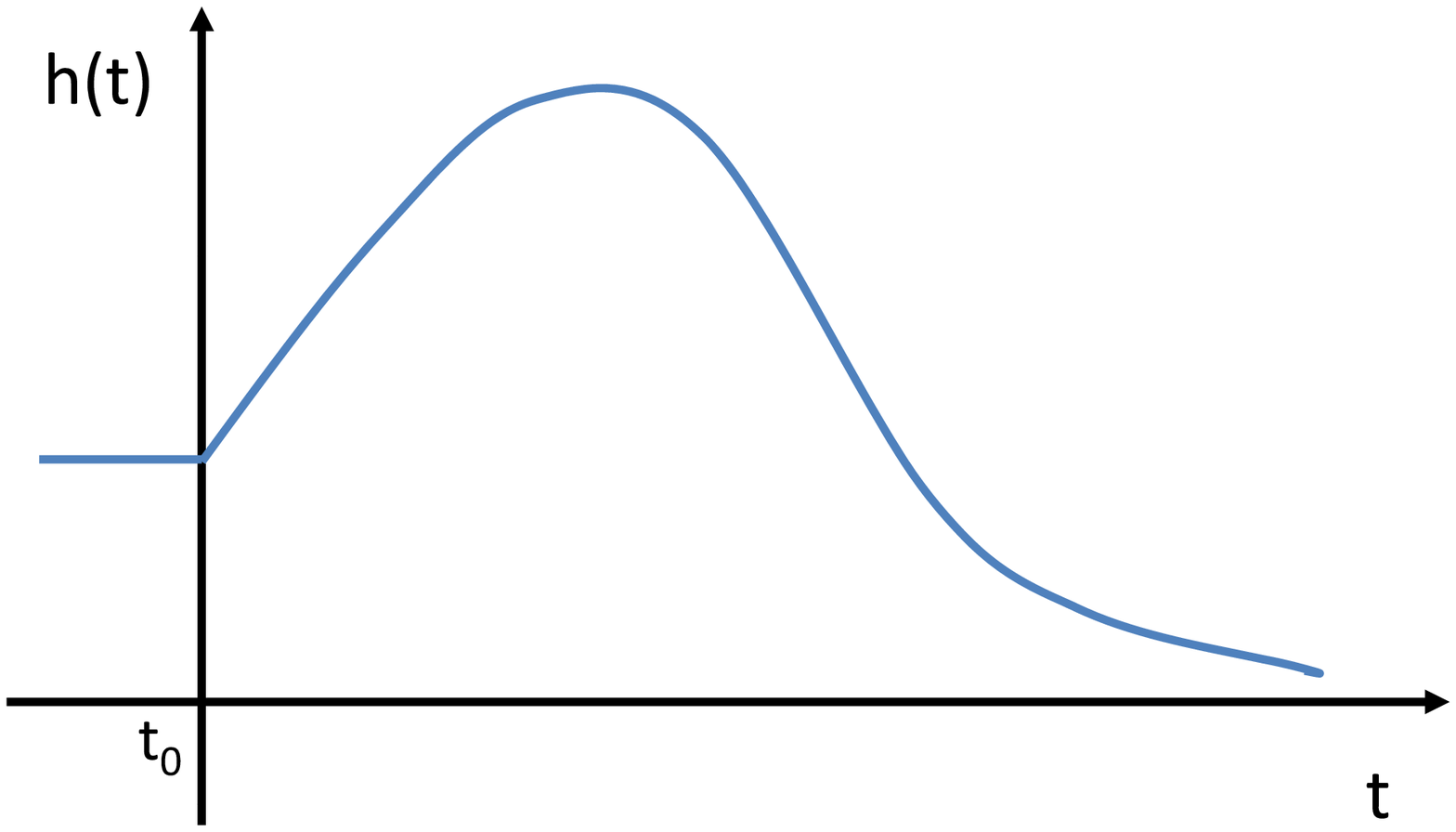}}\quad
   \subfigure[]{\label{Step_by_step2}\includegraphics[width=7 cm]{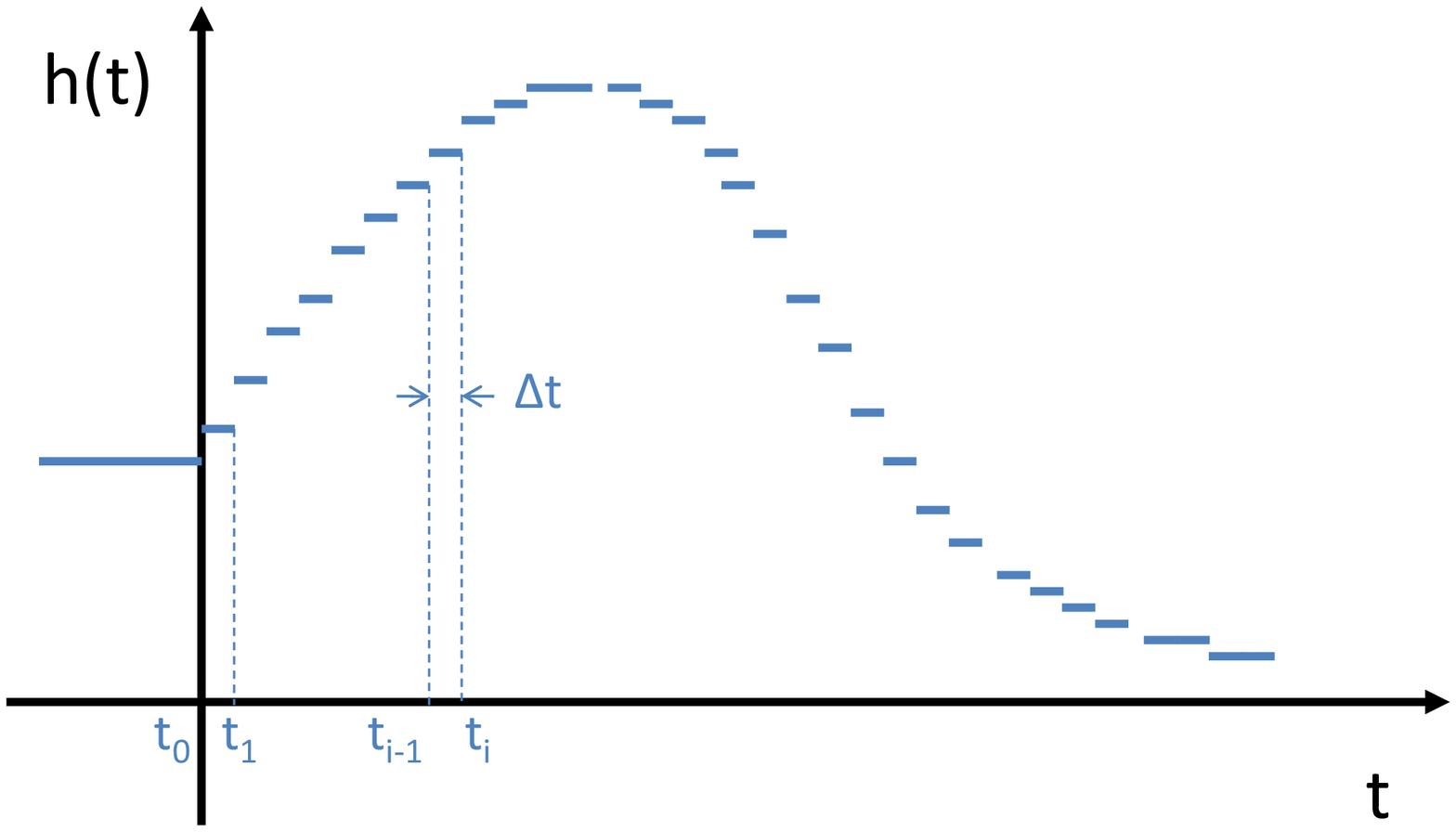}}
   \caption{{\protect\footnotesize (Color online) Divide an arbitrary magnetic field function into small time intervals. Every time step is $\Delta t$. Treat the field within the interval as a constant. In all, we turn a smooth function into a collection of step functions, which makes the calculation of dynamics possible.}}
 \label{Step_by_step}
 \end{minipage}
\end{figure}

To unveil the behavior of concurrence at time $t$, we need to find the density matrix of the system at that moment, which can be obtained from
\be
\rho(t)=U(t)\rho(0)U^{\dagger}(t).
\ee
%%%%%%%%%%%%%%%%%%%%%%%%%%%%%%%%%%%%%%%%%%%%%%%%%%%%%%%%%%%%%%%%%%%%%%%%%
\begin{figure}[htbp]
 \centering
   \includegraphics[width=7 cm]{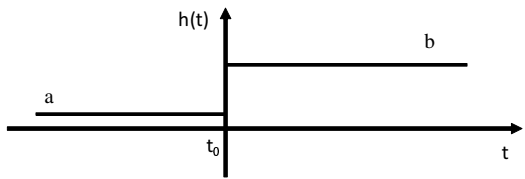}
   \caption{{\protect\footnotesize The external magnetic field in a step function form $h(t)=a\;@\;t\leq t_0$, $h(t)=b\;@\;t>t_0$.}}
 \label{Magnetic_fields_Step}
\end{figure}
%%%%%%%%%%%%%%%%%%%%%%%%%%%%%%%%%%%%%%%%%%%%%%%%%%%%%%%%%%%%%%%%%%%%%%%%%
Although Eq. (\ref{U_integral}) gives a beautiful expression for the evolution operator, in reality $U$ is hard to be obtained because of the integration involved. In order to overcome this obstacle, let us first consider the simplest time-dependent magnetic field: a step function of the form (Fig.~\ref{Magnetic_fields_Step})
\be
h(t)= a+(b-a)\theta(t-t_0) \; ,
\ee
where $\theta(t-t_0)$ is the usual mathematical step function defined by
\be
\theta(t-t_0)=\left\{
\begin{array}{lr}
0 & \qquad t\leq t_0 \\
1 & \qquad t>t_0
\end{array}.
\right.
\ee
At $t_0$ and before, the system is time-independent since $H_a \equiv H(t\leq t_0)=-\sum_{<i,j>}\sigma_{i}^{x}\sigma_{j}^{x}-a\Sigma_{i}\sigma_{i}^{z}$. Therefore we are capable of evaluating its ground state and density matrix at $t_0$ straightforwardly. For the interval $t_0$ to $t$, the Hamiltonian $H_b \equiv H(t>t_0)=-\sum_{<i,j>}\sigma_{i}^{x}\sigma_{j}^{x}-b\Sigma_{i}\sigma_{i}^{z}$ does not depend on time either, so Eq. (\ref{U_time_independent}) enables us to write out
\be
U(t,t_0)=e^{-iH(t>t_0)(t-t_0)/\hbar},
\ee
and therefore
\be
\rho(t)=U(t,t_0)\rho(t_0)U^{\dagger}(t,t_0).
\ee
Starting from here, it is not hard to think of breaking an arbitrary magnetic function into small time intervals, and treating every neighboring intervals as a step function. Comparing the two graphs in Fig.~\ref{Step_by_step}, the method has just turned a ski sliding into a mountain climbing.
%%%%%%%%%%%%%%%%%%%%%%%%%%%%%%%%%%%%%%%%%%%%%%%%%%%%%%%%%%%%%%%%%%%%%%%%%%%%%%%%%%%%%%
%%%%%%%%%%%%%%%%%%%%%%%%%%%%%%%%%%%%%%%%%%%%%%%%%%%%%%%%%%%%%%%%%%%%%%%%%%%%%%%%%%%%%%
Assuming each time interval is $\Delta t$, setting $\hbar=1$ then
\bea
U(t_i,t_0)|\psi_0\rangle&=&U(t_{i},t_{i-1})U(t_{i-1},t_{i-2})...U(t_1,t_0)|\psi_0\rangle,\\
U(t_i,t_0)&=&\prod_{k=1}^{i}exp[-iH(t_k)\Delta t],\\
U(t_i,t_0)&=&exp[-iH(t_i)\Delta t]U(t_{i-1}-t_0).
\eea
Here we avoided integration, instead we have chain multiplications which can be easily realized as loops in computational calculations. This is a common numerical technique; desired precisions can be achieved via proper time step length adjustment.
%%%%%%%%%%%%%%%%%%%%%%%%%%%%%%%%%%%%%%%%%%%%%%%%%%%%%%%%%%%%%%%%%%%%%%%%%
\subsection{Step by step projection}
%%%%%%%%%%%%%%%%%%%%%%%%%%%%%%%%%%%%%%%%%%%%%%%%%%%%%%%%%%%%%%%%%%%%%%%%%

Step by step matrix transformation method successfully breaks down the integration, but still involves matrix exponential, which is numerically resource costing. We propose a projection method to accelerate the calculations. Let us look at the step magnetic field again Fig. \ref{Magnetic_fields_Step}. For $H_a$, after enough long time, the system at zero temperature is in the ground state $|\phi\rangle$ with energy, say, $\varepsilon$. We want to ask how will this state evolves after the magnetic field is turned to the value b? Assuming the new Hamiltonian $H_b$ has $N$ eigenpairs $E_i$ and $|\psi_{i}\rangle$. The original state $|\phi\rangle$ can be expanded in the basis $\{|\psi_{i}\rangle\}$:
\be
|\phi\rangle=c_{1}|\psi_{1}\rangle+c_{2}|\psi_{2}\rangle+...+c_{N}|\psi_{N}\rangle,
\ee
where
\be
c_i=\langle\psi_i|\phi\rangle.
\ee
When $H$ is independent of time between $t$ and $t_0$ then we can write
\be
U(t,\, t_{0})\,|\psi_{i,t_0}\rangle=e^{-iH(t>t_0)(t-t_{0})/\hbar}|\psi_{i,t_0}\rangle=e^{-iE_{i}(t-t_{0})/\hbar}|\psi_{i,t_0}\rangle,
\ee
Now the exponent in the evolution operator is a number no longer a matrix. The ground state will evolve with time as
\bea\label{projection_sum}
|\phi(t)\rangle &=& c_{1}|\psi_{1}\rangle e^{-iE_{1}(t-t_0)}+c_{2}|\psi_{2}\rangle e^{-iE_{2}(t-t_0)}+...+c_{N}|\psi_{N}\rangle e^{-iE_{N}(t-t_0)}\nonumber\\
&=& \sum_{i=1}^{N}c_{i}|\psi_{i}\rangle e^{-iE_{i}(t-t_0)}.
\eea
and the pure state density matrix becomes
\be
\rho(t)=|\phi(t)\rangle\langle\phi(t)|.
\ee

Again any complicated function can be treated as a collection of step functions. When the state evolves to the next step just repeat the procedure to get the following results. Our test shows, for the same magnetic field both methods give the same results, but projection is much (about 20 times faster) than matrix transformation. This is a great advantage when the system size increases. But this is not the end of the problem. The summation is over all the eigenstates. Extending one layer out to $19$ sites, fully diagonalizing the $2^{19}$ by $2^{19}$ Hamiltonian and summing over all of them in every time step is breath taking. In a later section, we will show how to further improve the method to shrink the amount of calculation which will pave the way towards larger systems.

%%%%%%%%%%%%%%%%%%%%%%%%%%%%%%%%%%%%%%%%%%%%%%%%%%%%%%%%%%%%%%%%%%%%%%%%%
\subsection{Entanglement of formation}
%%%%%%%%%%%%%%%%%%%%%%%%%%%%%%%%%%%%%%%%%%%%%%%%%%%%%%%%%%%%%%%%%%%%%%%%%

We confine our interest to the entanglement of two spins, at any
position $i$ and $j$ \cite{Osterloh2002}. We adopt the entanglement of
formation,  a well known  measure of entanglement \cite{Wooters1998},
 to quantify our entanglement \cite{Kais2003}. All the information
needed in this case, at any moment t, is contained in the reduced density matrix
$\rho_{i, j}(t)$ which can be obtained form the entire system density matrix by integrating out all the spins states except $i$ and $j$. Wootters \cite{Wooters1998} has shown that, for a pair of binary qubits,
the concurrence $C$, which goes from $0$ to $1$, can be taken
as a measure of entanglement. The concurrence between sites $i$ and
$j$ is defined as
\begin{equation}
\label{concurrence}
C(\rho)=max\{0,\epsilon_1-\epsilon_2-\epsilon_3-\epsilon_4\},
\end{equation}
where the $\epsilon_i$'s are the eigenvalues of the Hermitian matrix
$R\equiv\sqrt{\sqrt{\rho}\tilde{\rho}\sqrt{\rho}}$ with
$\tilde{\rho}=(\sigma^y \otimes
\sigma^y)\rho^*(\sigma^y\otimes\sigma^y)$ and $\sigma^y$ is the
Pauli matrix of the spin in y direction.
For a pair of qubits the entanglement can be written as,
\begin{equation}
\label{entanglement}
E(\rho)=\epsilon(C(\rho)),
\end{equation}
where $\epsilon$ is a function of the ``concurrence'' $C$
\begin{equation}
\epsilon(C)=h\left(\frac{1-\sqrt{1-C^2}}{2}\right),
\end{equation}
where $h$ is the binary entropy function
\begin{equation}
h(x)=-x\log_{2}x-(1-x)log_{2}(1-x).
\end{equation}
In this case, the entanglement of formation is given in terms of another entanglement measure, the concurrence C. The matrix elements of the reduced density matrix needed for calculating the concurrence can be obtained numerically using one of the two methods developed above.

%%%%%%%%%%%%%%%%%%%%%%%%%%%%%%%%%%%%%%%%%%%%%%%%%%%%%%%%%%%%%%%%%%%%%%%%%
\section{Dynamics of the spin system in a time-dependent magnetic field}
%%%%%%%%%%%%%%%%%%%%%%%%%%%%%%%%%%%%%%%%%%%%%%%%%%%%%%%%%%%%%%%%%%%%%%%%%

%%%%%%%%%%%%%%%%%%%%%%%%%%%%%%%%%%%%%%%%%%%%%%%%%%%%%%%%%%%%%%%%%%%%%%%%%
\subsection{Step magnetic field}
%%%%%%%%%%%%%%%%%%%%%%%%%%%%%%%%%%%%%%%%%%%%%%%%%%%%%%%%%%%%%%%%%%%%%%%%%

At first, we study the dynamic response of the 7-site spin system to the external step magnetic field (Fig. \ref{Magnetic_fields_Step}), the simplest form of time-dependent function, which is given by
\be
h(t)= a+(b-a)\theta(t-t_0) \; ,
\ee
In a previous work we have showed that this two dimensional spin system has a critical point at $\lambda=1.64$ and a maximum reachable concurrence at $\lambda=2.61$ \cite{XuQ2010}. Accordingly, we design here all kind of steps: big, small, jump, drop, happening before, after or cross the ``critical point'' (and ``maximum point''), trying to identify what will affect the system behavior and how.
%%%%%%%%%%%%%%%%%%%%%%%%%%%%%%%%%%%%%%%%%%%%%%%%%%%%%%%%%%%%%%%%%%%%%%%%%
\begin{figure}[htbp]
\begin{minipage}[c]{\textwidth}
 \centering
   \includegraphics[width=15 cm]{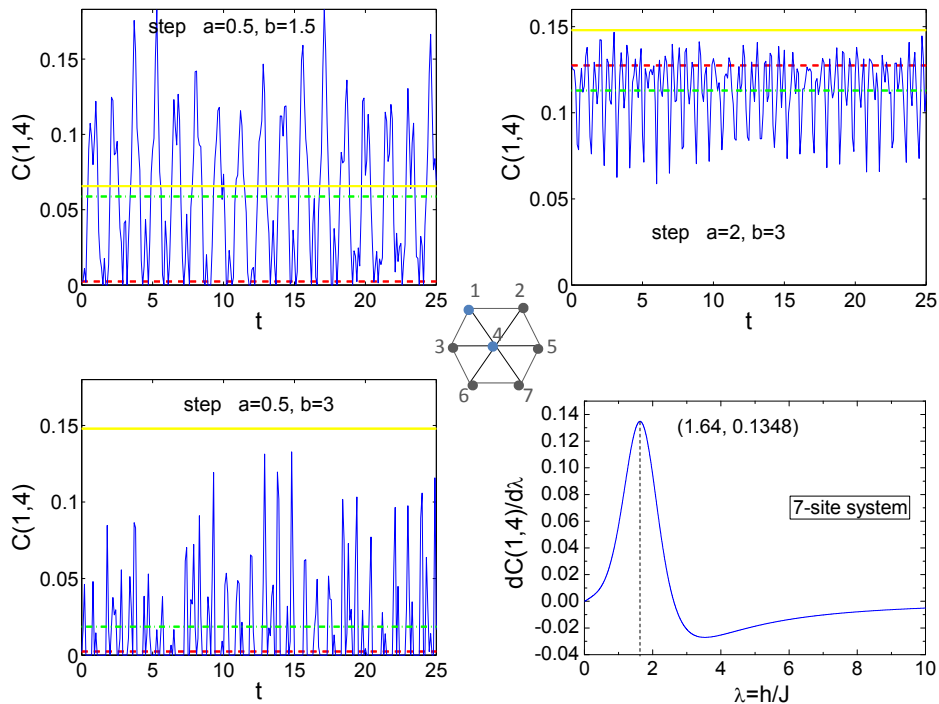}
   \caption{{\protect\footnotesize (Color online) Dynamics of C(1,4) in the 7-site system when the step magnetic field is changed from $a=0.5$ to $b=1.5$ (before the ``critical point'' $h=1.64$), from $a=2$ to $b=3$ (after) and from $a=0.5$ to $b=3$ (big step cross the ``critical point'').}}
 \label{before_after_cross_cp_14}
 \end{minipage}
\end{figure}
%%%%%%%%%%%%%%%%%%%%%%%%%%%%%%%%%%%%%%%%%%%%%%%%%%%%%%%%%%%%%%%%%%%%%%%%%
%%%%%%%%%%%%%%%%%%%%%%%%%%%%%%%%%%%%%%%%%%%%%%%%%%%%%%%%%%%%%%%%%%%%%%%%%
\begin{figure}[htbp]
\begin{minipage}[c]{\textwidth}
 \centering
   \includegraphics[width=15 cm]{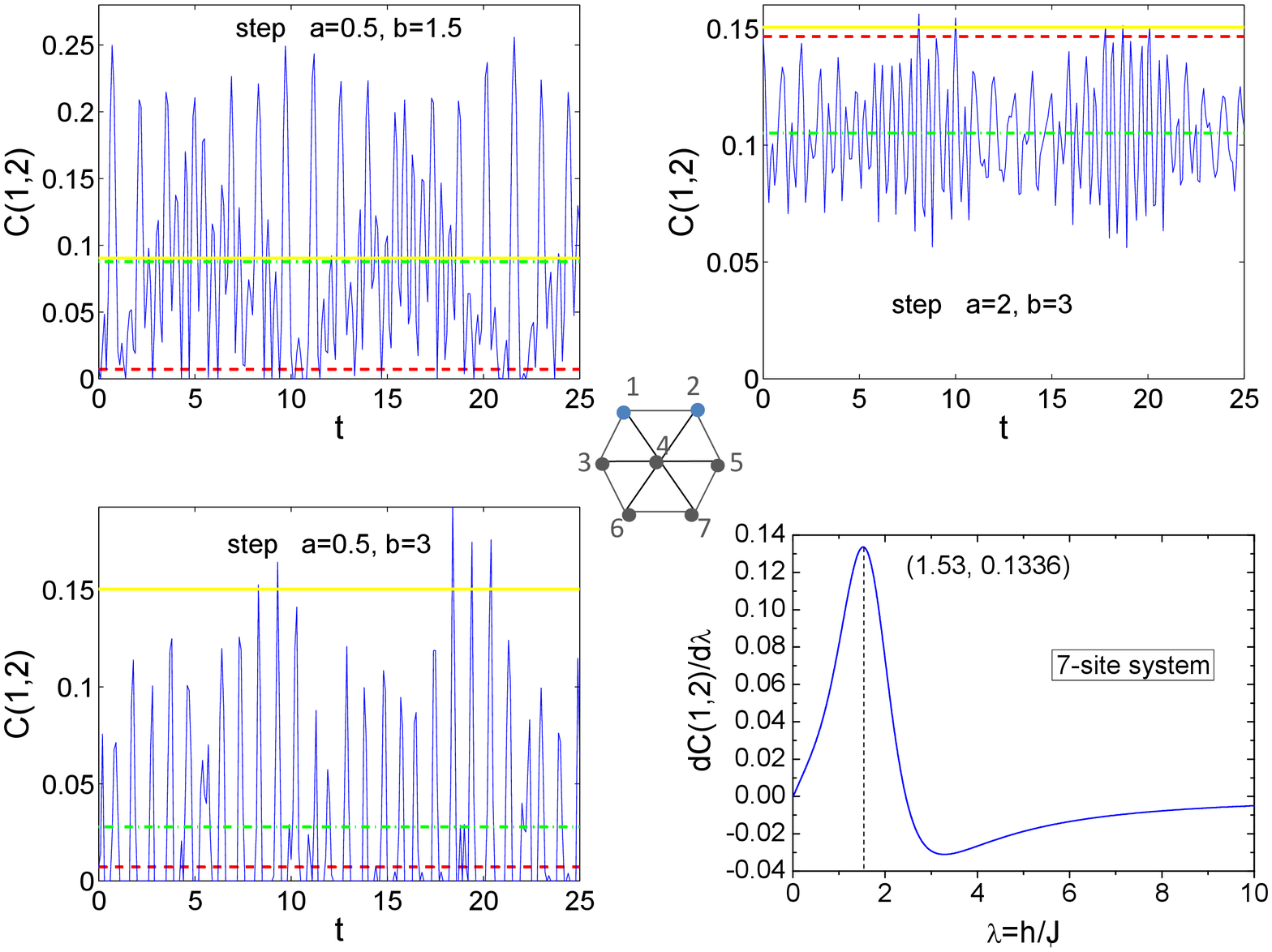}
   \caption{{\protect\footnotesize (Color online) Dynamics of C(1,2) in the 7-site system when the step magnetic field is changed from $a=0.5$ to $b=1.5$ (before the ``critical point'' $h=1.53$), from $a=2$ to $b=3$ (after) and from $a=0.5$ to $b=3$ (big step cross the ``critical point'').}}
 \label{before_after_cross_cp_12}
 \end{minipage}
\end{figure}
%%%%%%%%%%%%%%%%%%%%%%%%%%%%%%%%%%%%%%%%%%%%%%%%%%%%%%%%%%%%%%%%%%%%%%%%%

\begin{figure}[htbp]
\begin{minipage}[c]{\textwidth}
 \centering
   \includegraphics[width=15 cm]{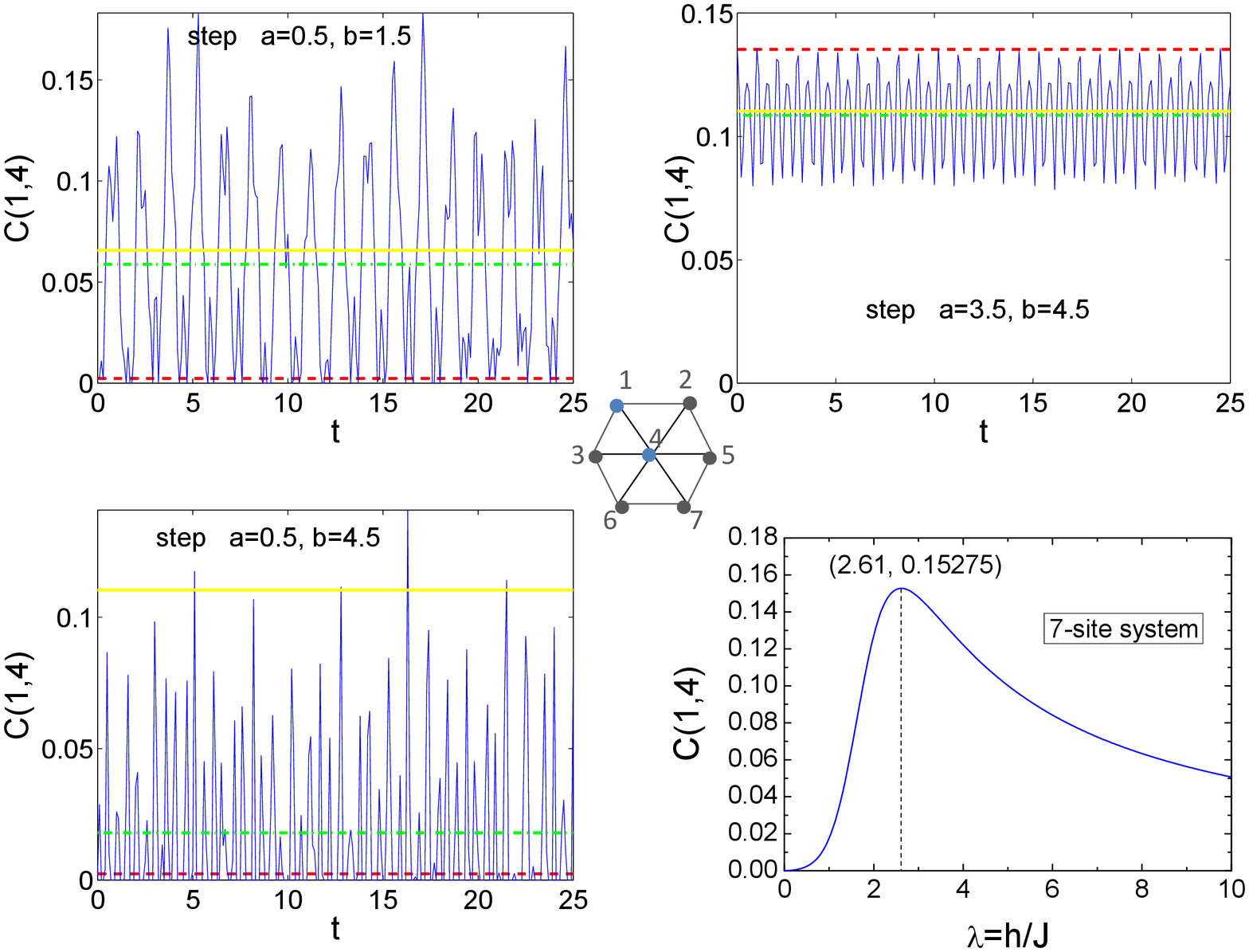}
   \caption{{\protect\footnotesize (Color online) Dynamics of C(1,4) in the 7-site system when the step magnetic field is changed from $a=0.5$ to $b=1.5$ (before the ``maximum point'' $h=2.61$), from $a=3.5$ to $b=4.5$ (after) and from $a=0.5$ to $b=4.5$ (big step cross the ``maximum point'').}}
 \label{before_after_cross_mp_14}
 \end{minipage}
\end{figure}

\begin{figure}[htbp]
\begin{minipage}[c]{\textwidth}
 \centering
   \includegraphics[width=15 cm]{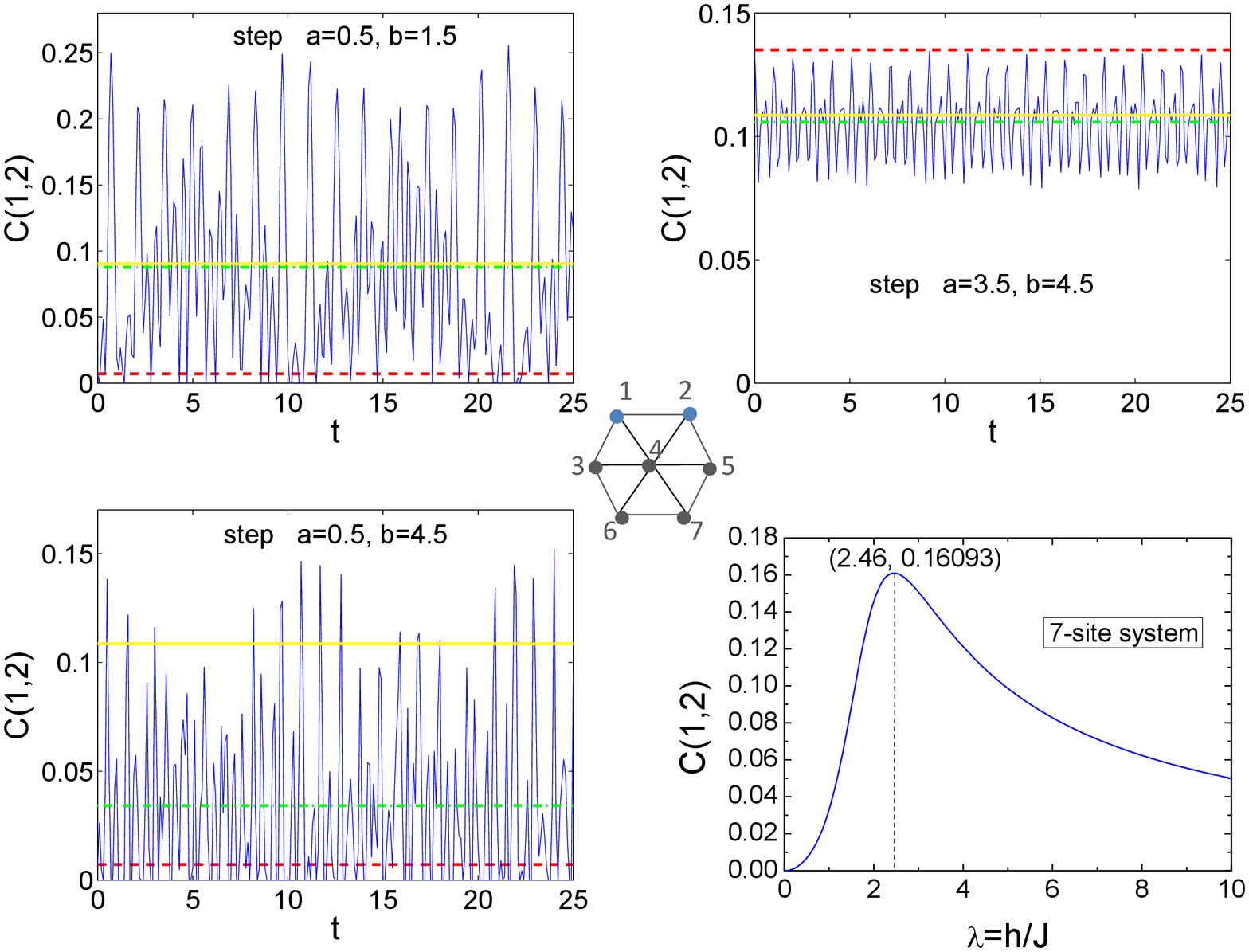}
   \caption{{\protect\footnotesize (Color online) Dynamics of C(1,2) in the 7-site system when the step magnetic field is changed from $a=0.5$ to $b=1.5$ (before the ``maximum point'' $h=2.46$), from $a=3.5$ to $b=4.5$ (after) and from $a=0.5$ to $b=4.5$ (big step cross the ``maximum point'').}}
 \label{before_after_cross_mp_12}
 \end{minipage}
\end{figure}

Figure \ref{before_after_cross_cp_14} displays the dynamics of the pairwise entanglement between sites 1 and 4 C(1,4)  when the magnetic field is changed from $a=0.5$ to $b=1.5$ (before the ``critical point'' $h=1.64$), from $a=2$ to $b=3$ (after) and from $a=0.5$ to $b=3$ (big step cross the ``critical point''). Thick and big oscillations appear in every graph. Dynamics of C(1,4) under similar design (Figure \ref{before_after_cross_mp_14}), but before, after and cross the ``maximum point'' $h=2.61$, shows resembling oscillations. In these graphs, we have plotted the concurrence corresponding to a constant magnetic field $h=a$ (red line), $h=b$ (yellow line) and the average value of the oscillating concurrence (green line). In the same manner, the dynamics of the concurrence C(1,2) is explored in Figs. \ref{before_after_cross_cp_12} and \ref{before_after_cross_mp_12} which shows a very similar behavior to the C(1,4) case. Examining the dynamics of the next nearest neighbor concurrences C(1,5) and C(1,7) in the same way as we did with C(1,2) and C(1,4), we observed a very similar behavior with much much smaller value of the concurrence as expected for next nearest neighbors.

\begin{figure}[htbp]
\begin{minipage}[c]{\textwidth}
 \centering
   \subfigure{\includegraphics[width=6 cm]{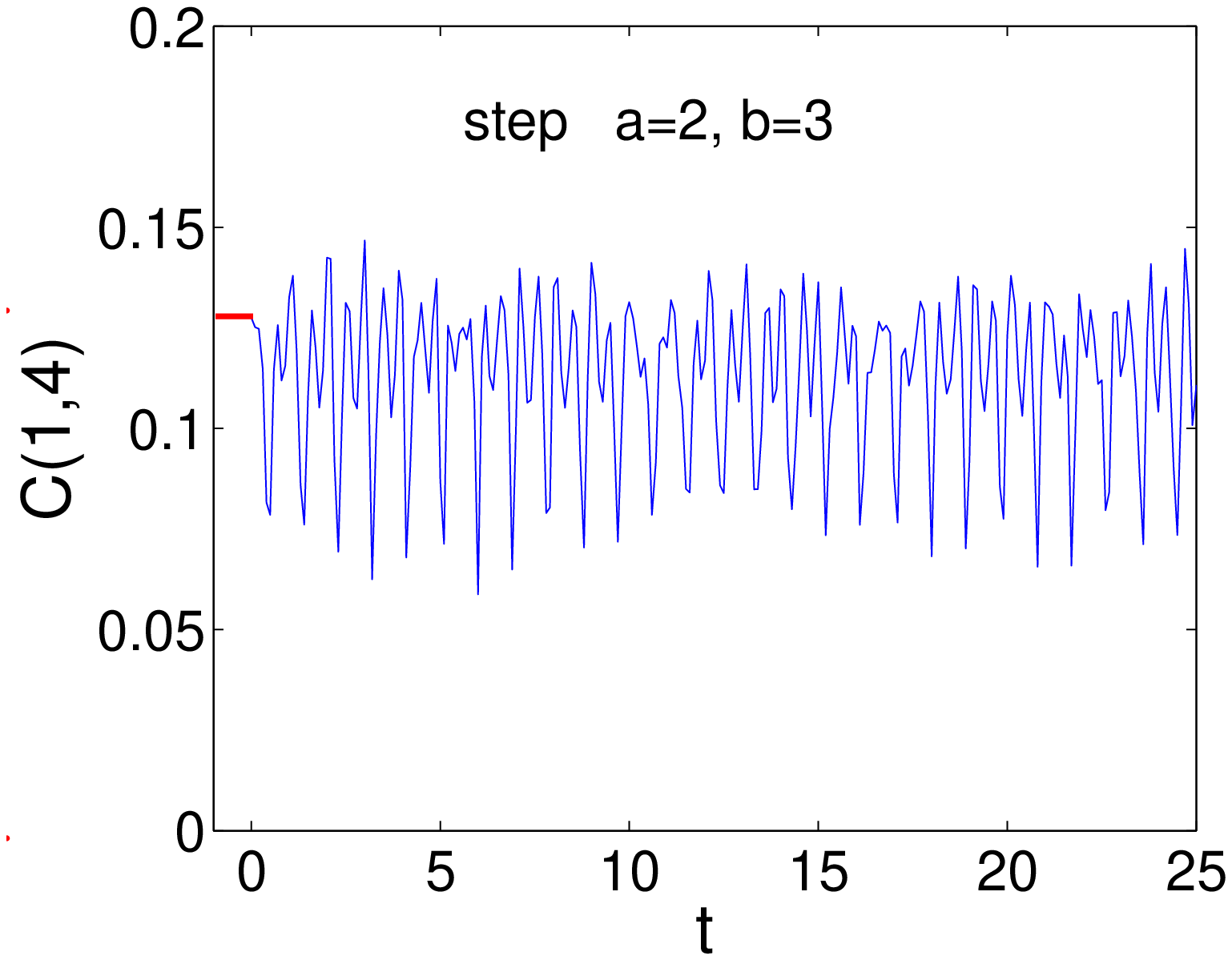}\quad
                \includegraphics[width=6 cm]{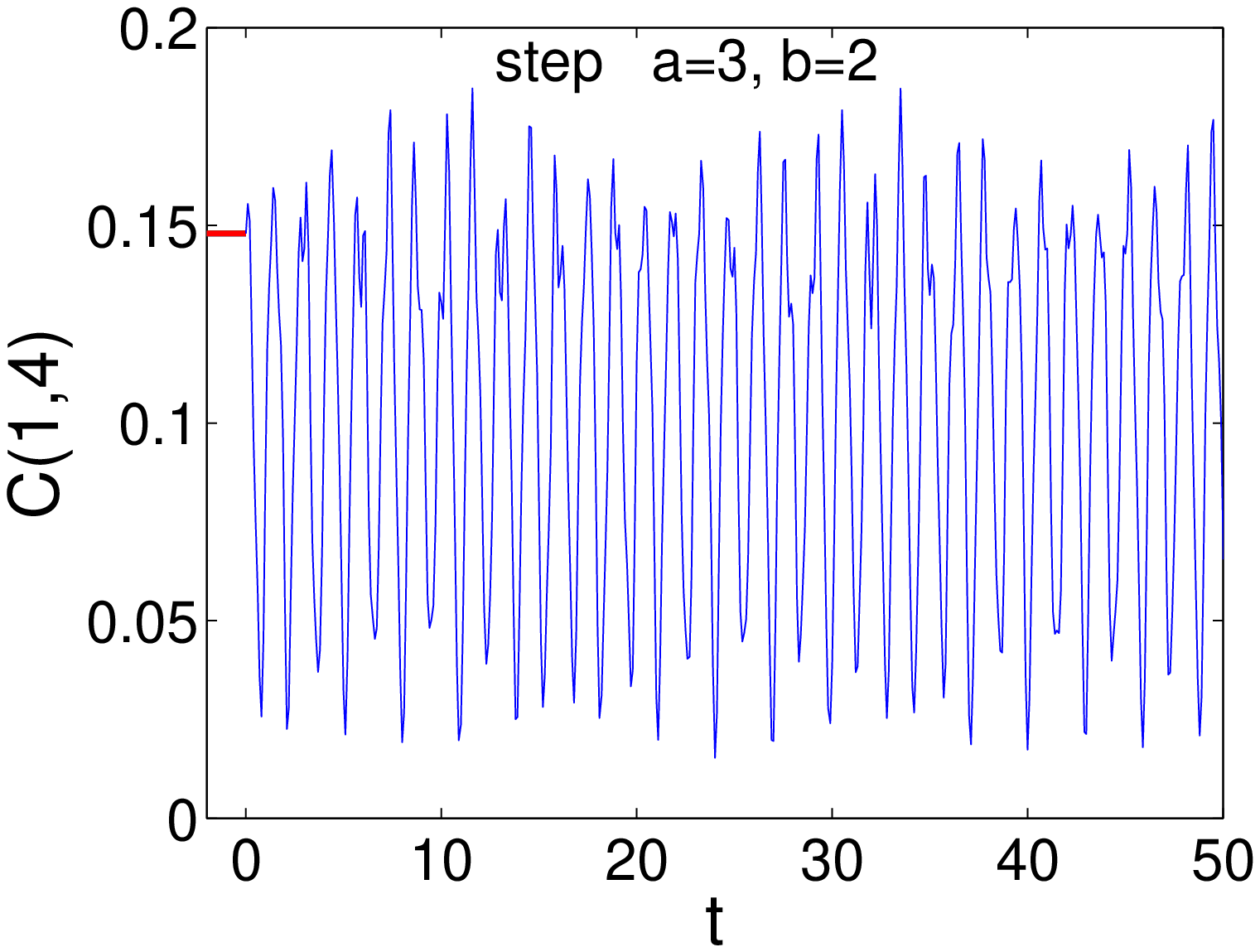}}\\
   \subfigure{\includegraphics[width=6 cm]{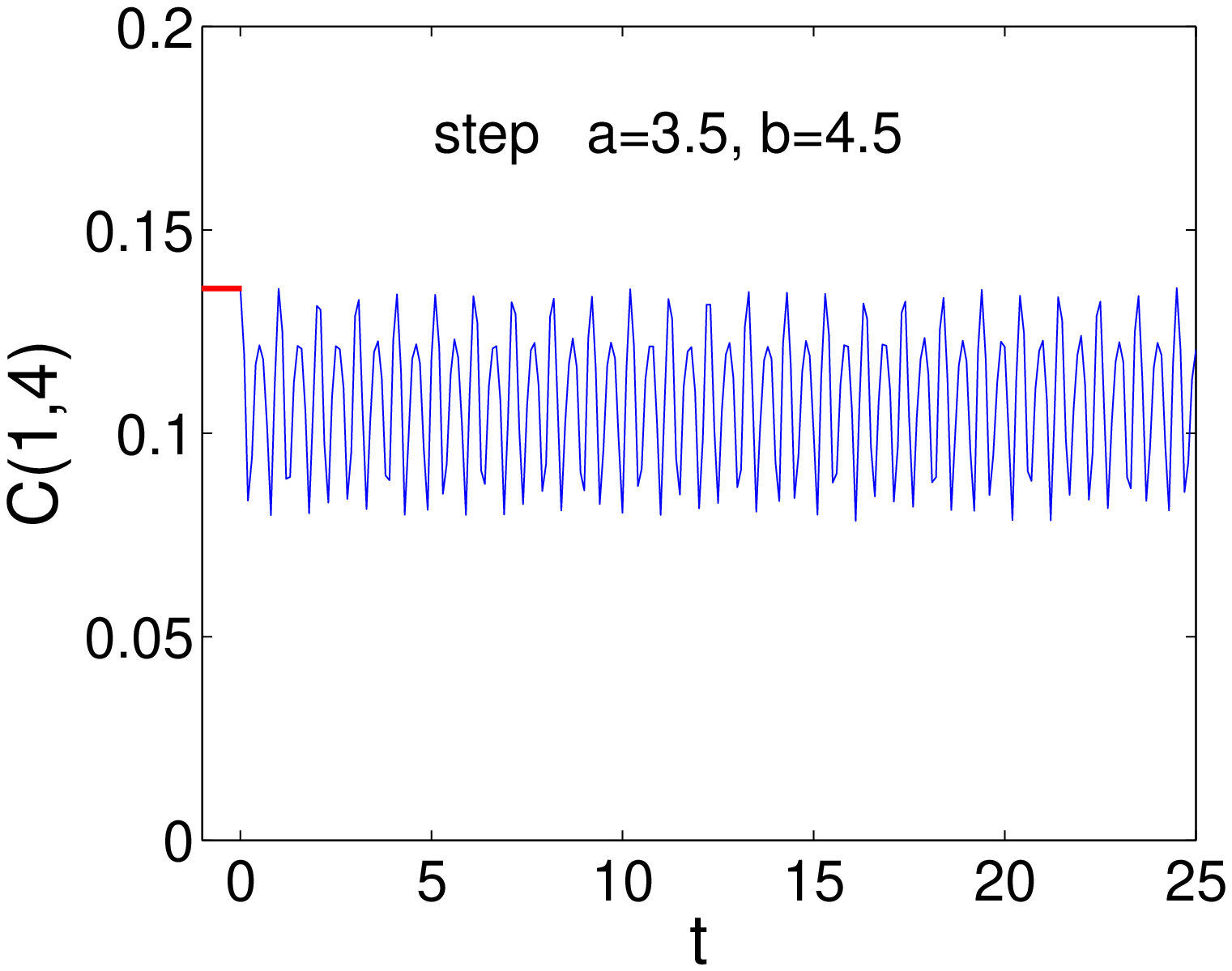}\quad
                \includegraphics[width=6 cm]{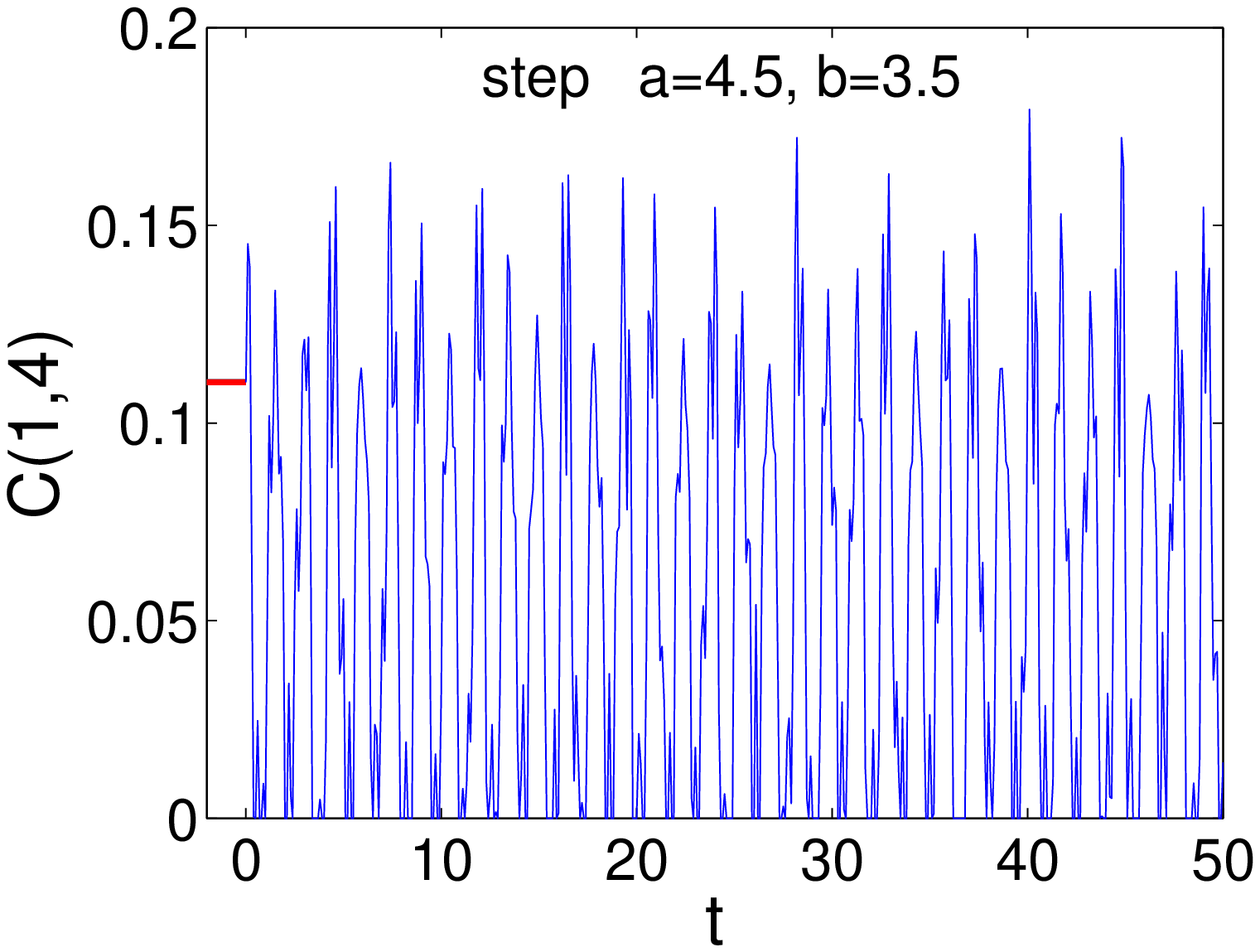}}
  \caption{{\protect\footnotesize (Color online) Dynamics of the system from one magnetic field value to a smaller one and the reverse process are different. But system still oscillates dramatically.}}
 \label{jump_drop}
 \end{minipage}
\end{figure}

Figure \ref{jump_drop} intends to check behaviors of the system going from one magnetic field value to a smaller one and the reverse process. Both of them show the same characteristic oscillations, although they are different. Because the projection method requires mapping the ground state in $h=a$ into all the eigenstates of the system in $h=b$; while we reverse the process, mapping becomes from the ground state
in $h=b$ to all the eigenstates of the system in $h=a$. The behavior of the system is not mirrored.
%%%%%%%%%%%%%%%%%%%%%%%%%%%%%%%%%%%%%%%%%%%%%%%%%%%%%%%%%%%%%%%%%%%%%%%%%
%%%%%%%%%%%%%%%%%%%%%%%%%%%%%%%%%%%%%%%%%%%%%%%%%%%%%%%%%%%%%%%%%%%%%%%%%
%%%%%%%%%%%%%%%%%%%%%%%%%%%%%%%%%%%%%%%%%%%%%%%%%%%%%%%%%%%%%%%%%%%%%%%%%
%%%%%%%%%%%%%%%%%%%%%%%%%%%%%%%%%%%%%%%%%%%%%%%%%%%%%%%%%%%%%%%%%%%%%%%%%%%%%%%%%%%%%%%%%%%%%%%%%%%%%%%%%%%%%%%%%%%%%%%%%%%%%%%%%%%%%%%%%%%%%%%%%%%%

No matter how we place the steps, all of them inevitably cause oscillation. This will be explained after we study all four kinds of external magnetic field. But now we can answer the question: what will happen to the entanglement of spins after the constant magnetic field is turned on, and can we benefit from that, say using the step magnetic field as an entanglement switch? The answer is negative, because the entanglements oscillate fast and relatively big among values. Simple step magnetic field does not provide a way to control or tuning the entanglement in this spin system.

%%%%%%%%%%%%%%%%%%%%%%%%%%%%%%%%%%%%%%%%%%%%%%%%%%%%%%%%%%%%%%%%%%%%%%%%%
\subsection{Exponential magnetic field}
%%%%%%%%%%%%%%%%%%%%%%%%%%%%%%%%%%%%%%%%%%%%%%%%%%%%%%%%%%%%%%%%%%%%%%%%%

\begin{figure}[htbp]
\begin{minipage}[c]{\textwidth}
 \centering
   \subfigure[]{\label{fig:he_a}\includegraphics[width=6 cm]{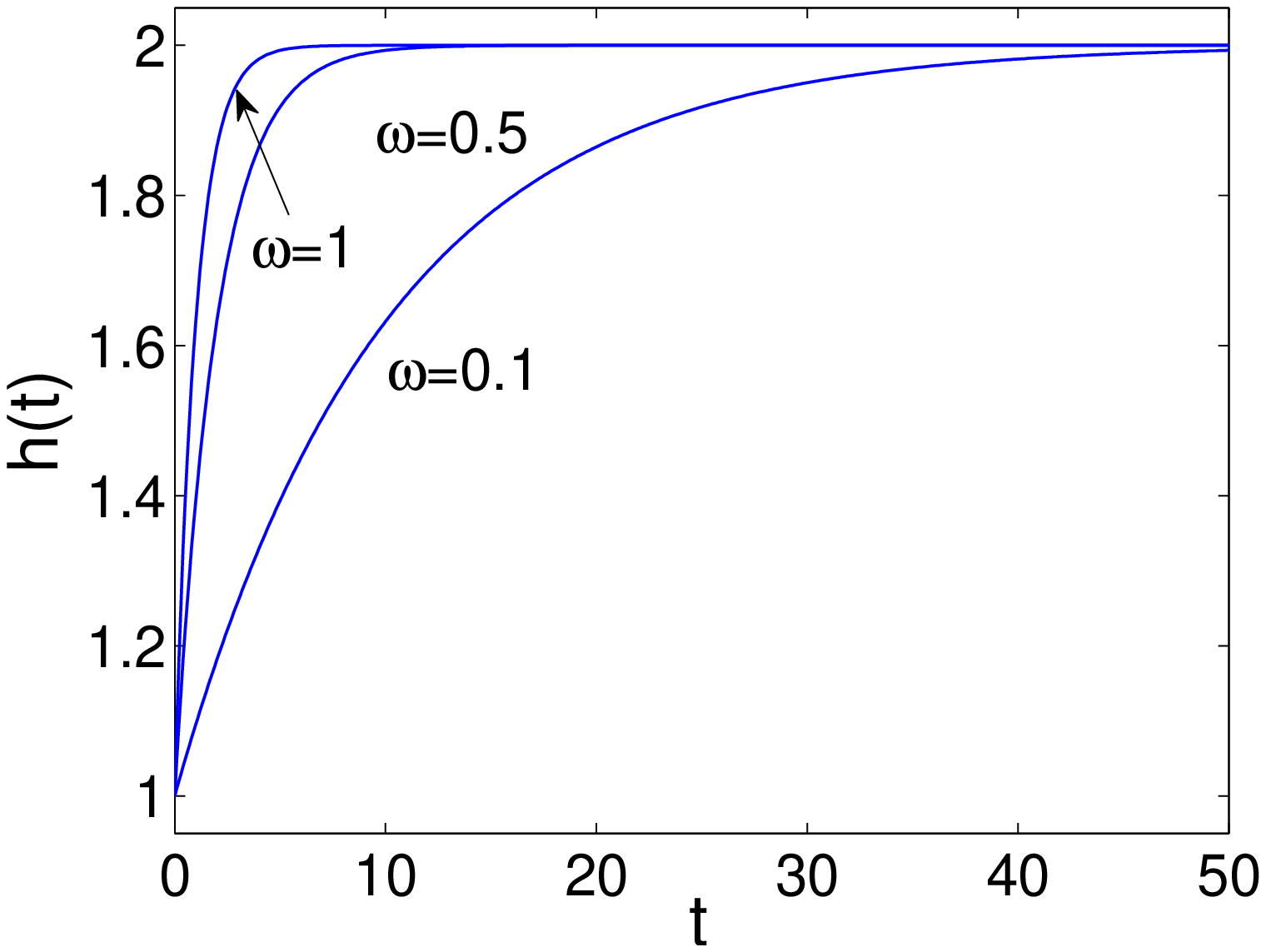}}\quad
   \subfigure[]{\label{fig:he_b}\includegraphics[width=6 cm]{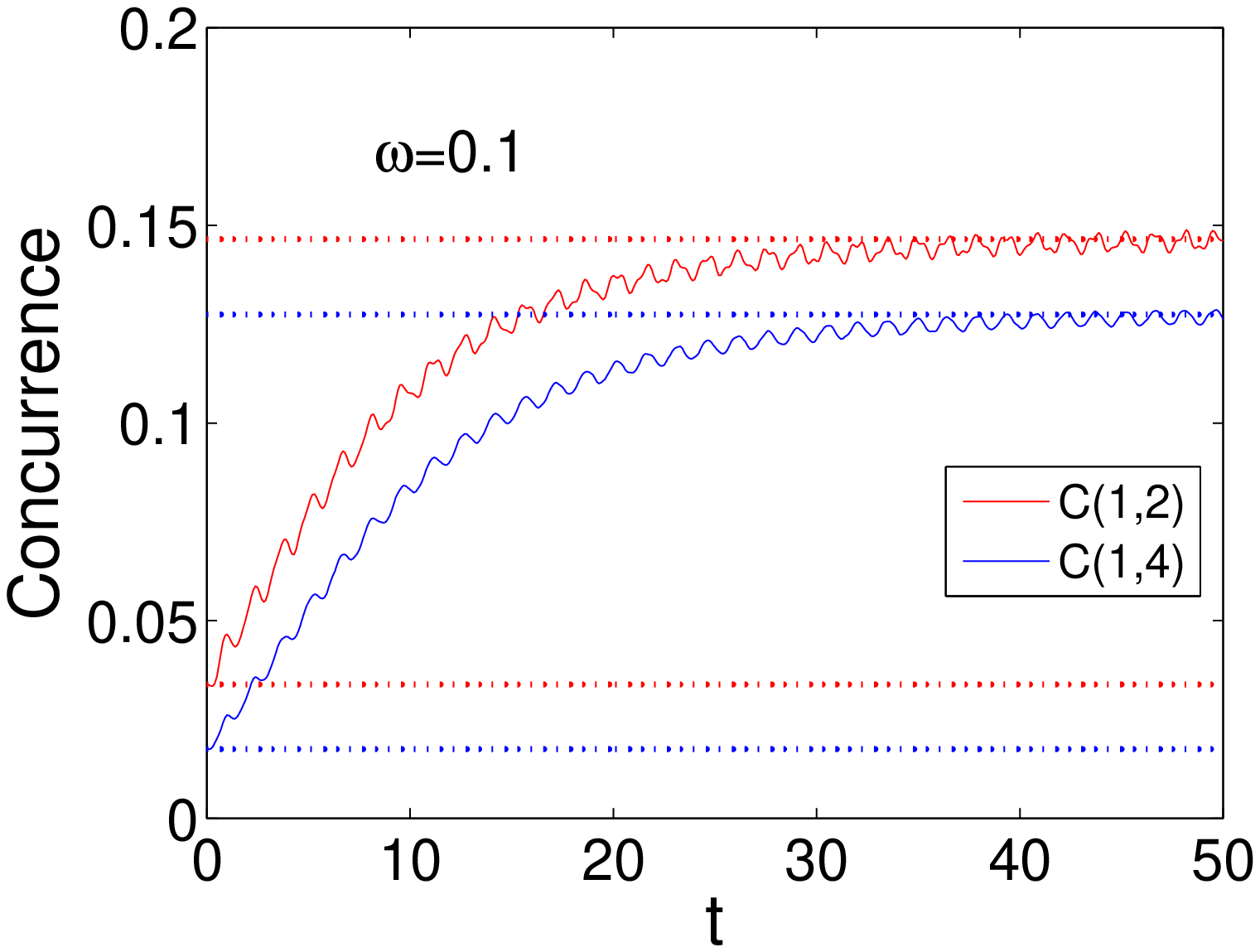}}\\
   \subfigure[]{\label{fig:he_c}\includegraphics[width=6 cm]{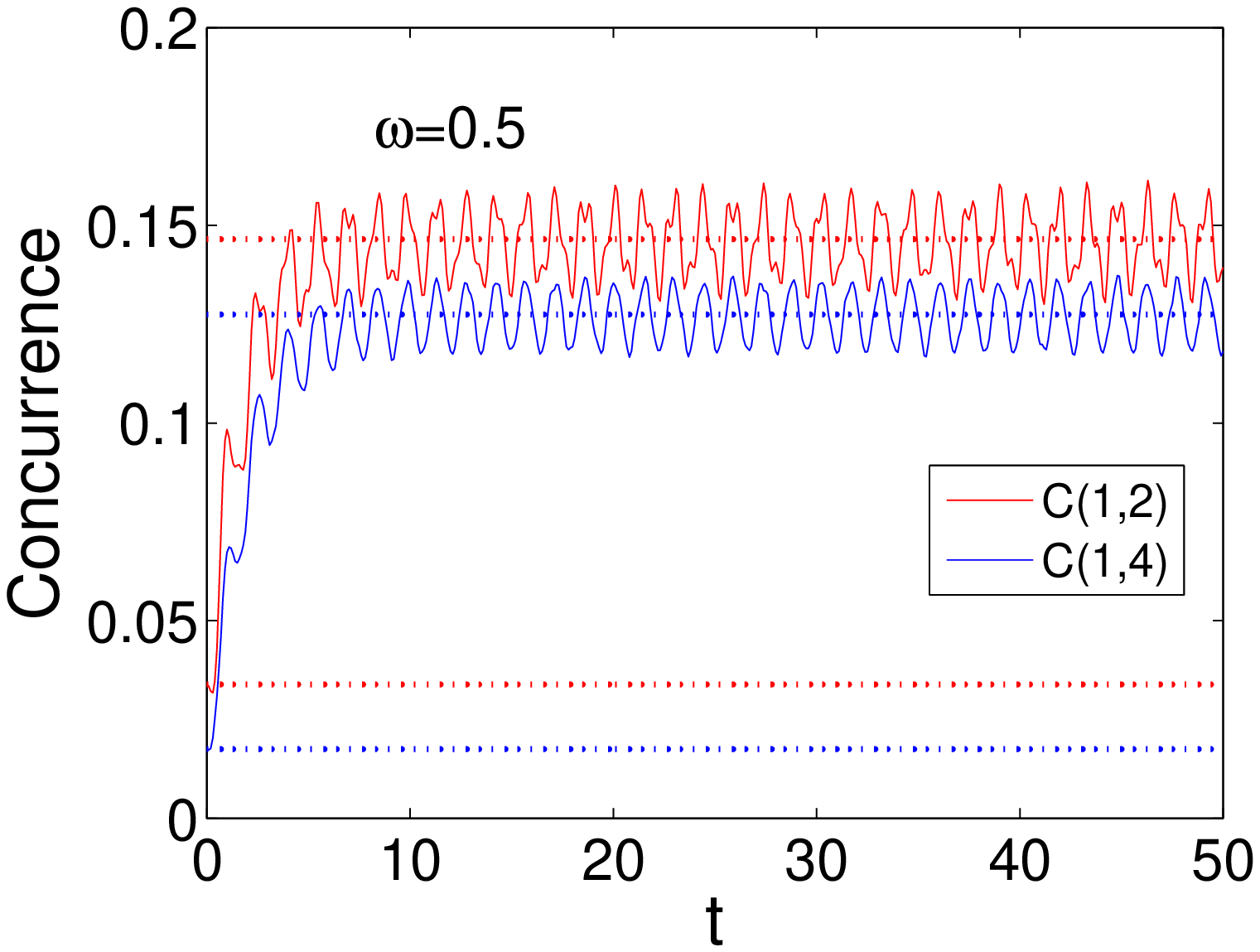}}\quad
   \subfigure[]{\label{fig:he_d}\includegraphics[width=6 cm]{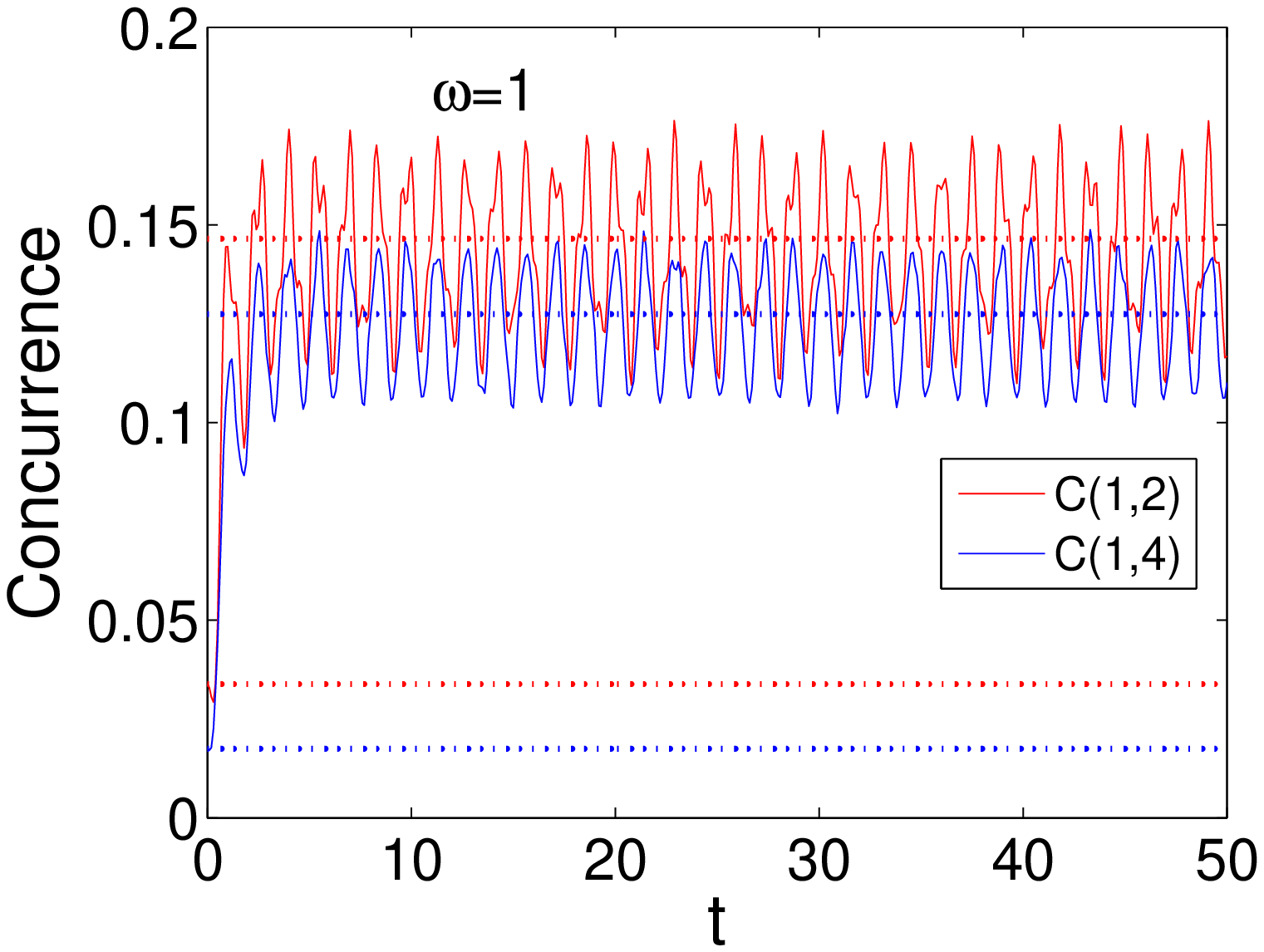}}
   \caption{{\protect\footnotesize (Color online) Dynamics of the concurrences C(1,2) and C(1,4) in applied exponential magnetic fields of various frequencies $\omega=0.1$, $0.5$ \& $1$, with $a=1$, $b=2$.}}
 \label{exp_h}
 \end{minipage}
\end{figure}

\begin{figure}[htbp]
\begin{minipage}[c]{\textwidth}
 \centering
   \subfigure[]{\label{fig:he_e}\includegraphics[width=6 cm]{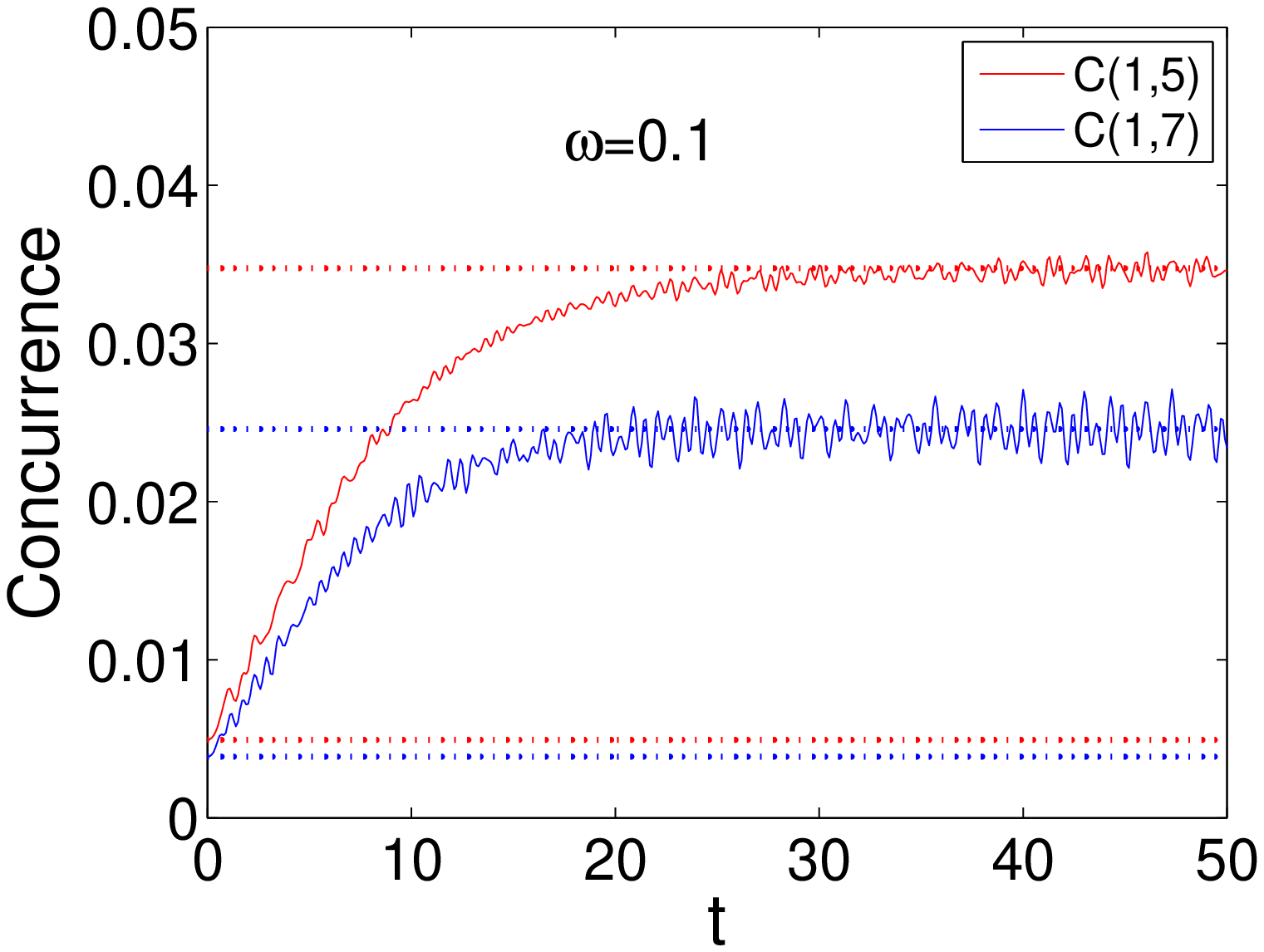}}\quad
   \subfigure[]{\label{fig:he_f}\includegraphics[width=6 cm]{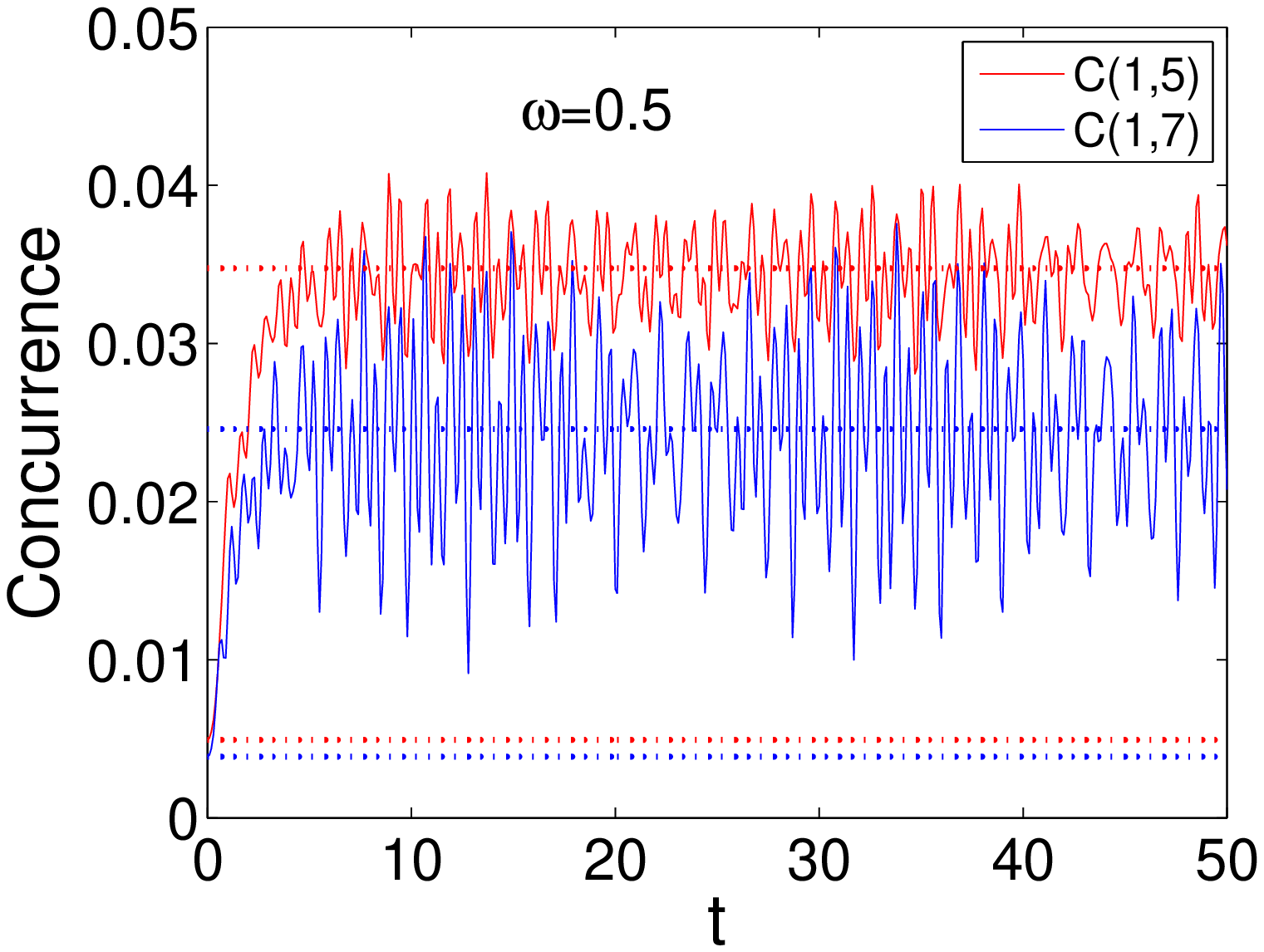}}
   \caption{{\protect\footnotesize (Color online) Dynamics of the concurrences C(1,5) and C(1,7) in applied exponential magnetic fields of various frequencies $\omega=0.1$, $0.5$ \& $1$, with $a=1$, $b=2$.}}
 \label{exp_h2}
 \end{minipage}
\end{figure}

The second kind of time-dependent magnetic field we will look at is the exponential one represented by
%%%%%%%%%%%%%%%%%%%%%%%%%%%%%%%%%%%%%%%%%%%%
\be\label{eq_magnetic fields_exp}
h(t)=\left\{
\begin{array}{lr}
a & \qquad t\leq t_0 \\
b+(a-b)e^{-\omega t} & \qquad t>t_0
\end{array}
\right.
\ee
%%%%%%%%%%%%%%%%%%%%%%%%%%%%%%%%%%%%%%%%%%
It is a more general form than the step function; when $\omega\rightarrow\infty$, the exponential function turns into a step function. Fig.~\ref{exp_h} highlights the effect of $\omega$ on the concurrences C(1,4) and C(1,2). As one can see the concurrence has the trend of following the shape of exponential function where it increases suddenly as the higher value is turned on and reaches certain asymptotic equilibrium value. As the transition parameter $\omega$ increases the concurrence shows an oscillation which increases rapidly as we further increase $\omega$ where in that case the behavior resembles the step function case. It is interesting to see that the concurrence asymptotic equilibrium value, under the exponential magnetic field, coincides with the concurrence corresponding to the higher constant magnetic field, $h=b$, (the blue (red) straight line corresponding to C(1,4) (C(1,2)), which means that the two-dimensional Ising spin system shows an ergodic behavior in contrary to the one-dimensional system. In all cases the edge concurrence C(1,2) is higher than the central one C(1,4) as expected as the latter shares the entanglement with higher number of other spin pairs. Various $a$ and $b$ combinations are tested just as what we did for the step function with similar results to that in Fig.~\ref{exp_h}. Of course, changing the value of the final magnetic field lead to a change of the equilibrium value of the concurrence following it up and down. The concurrences C(1,5) and C(1,7) show very similar behavior as C(1,2) and C(1,4) as shown in Fig. \ref{exp_h2} but with smaller magnitude.
%%%%%%%%%%%%%%%%%%%%%%%%%%%%%%%%%%%%%%%%%%%%%%%%%%%%%%%%%%%%%%%%%%%%%%%%%%%%%%%%%%%
%%%%%%%%%%%%%%%%%%%%%%%%%%%%%%%%%%%%%%%%%%%%%%%%%%%%%%%%%%%%%%%%%%%%%%%%%%%%%%%%%%%%%

In the sense of tuning entanglement, exponentially changed magnetic field can be used to vary the entanglement from one value to the other smoothly as long as long as its transition rate is slow enough. For instance $\omega=0.1$, that is ten percent of the interchange coupling $J$ in energy scale, is a good choice to accomplish this task as we will explain latter.

%%%%%%%%%%%%%%%%%%%%%%%%%%%%%%%%%%%%%%%%%%%%%%%%%%%%%%%%%%%%%%%%%%%%%%%%%%%%%%%%%%%
%%%%%%%%%%%%%%%%%%%%%%%%%%%%%%%%%%%%%%%%%%%%%%%%%%%%%%%%%%%%%%%%%%%%%%%%%%%%%%%%%%%%%

%%%%%%%%%%%%%%%%%%%%%%%%%%%%%%%%%%%%%%%%%%%%%%%%%%%%%%%%%%%%%%%%%%%%%%%%%
\subsection{Hyperbolic magnetic field}
%%%%%%%%%%%%%%%%%%%%%%%%%%%%%%%%%%%%%%%%%%%%%%%%%%%%%%%%%%%%%%%%%%%%%%%%%

As we have seen in the previous two subsections, applying a step function or a rapid exponential function may disturb the system and lead to a strongly oscillating concurrence. Now let us apply another form of external magnetic field, namely hyperbolic which provides a smoother interaction with the system and reduces disturbances. The hyperbolic field is represented by
\be\label{eq_magnetic fields_hyper}
h(t)=\left\{
\begin{array}{lr}
a & \qquad t\leq t_0 \\
\frac{(b-a)}{2}[\tanh(\omega t)+1]+a & \qquad t>t_0
\end{array}
\right.
\ee
%%%%%%%%%%%%%%%%%%%%%%%%%%%%%%%%%%%%%%%%%%%%%%%%%%%%%%%%%%%%%%%%%%%%%%%%%%%%%%%%%%%%%%%%%%%%%%%%%%%%%
\begin{figure}[htbp]
\begin{minipage}[c]{\textwidth}
 \centering
   \subfigure[]{\label{fig:ht_a}\includegraphics[width=6 cm]{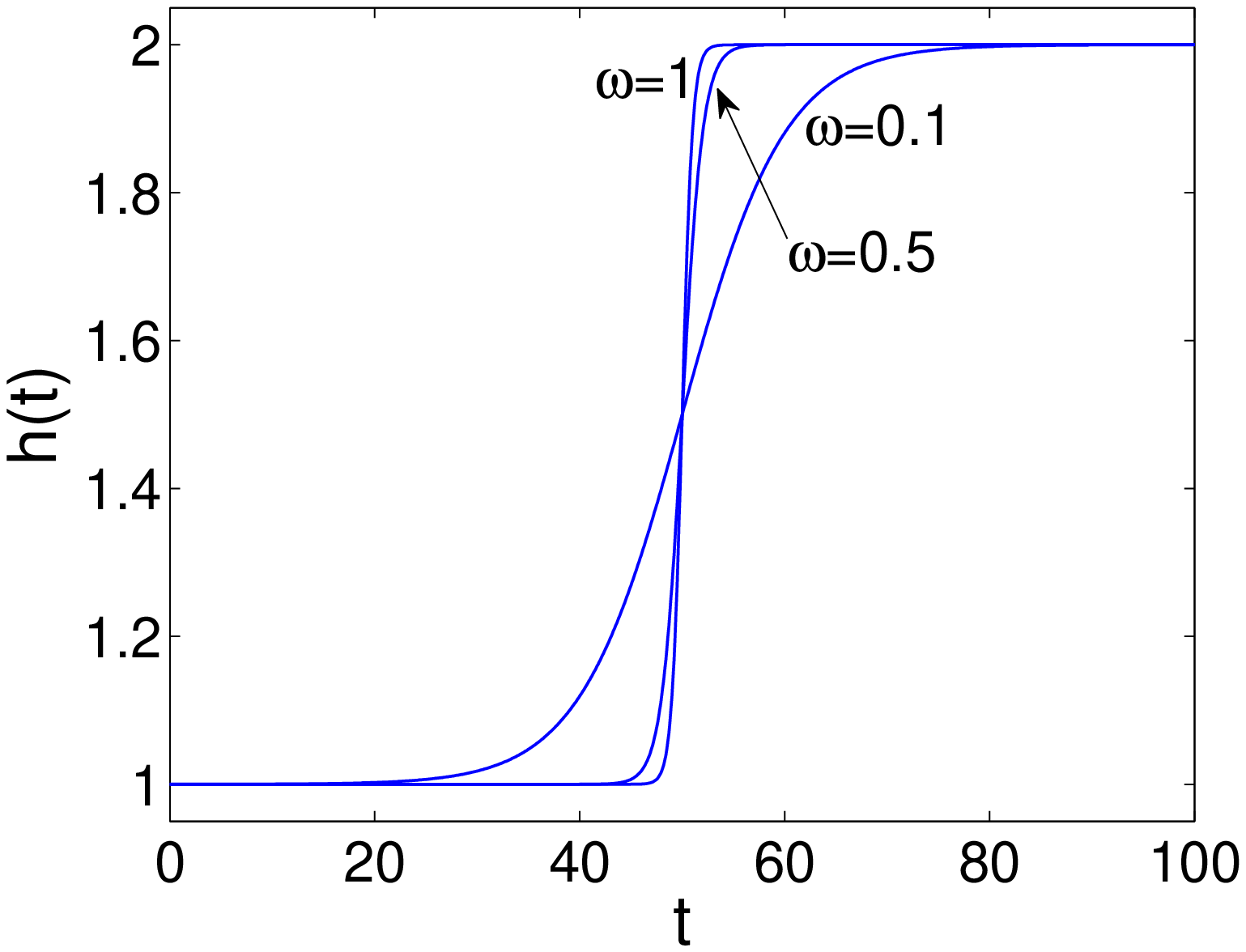}}\quad
   \subfigure[]{\label{fig:ht_b}\includegraphics[width=6 cm]{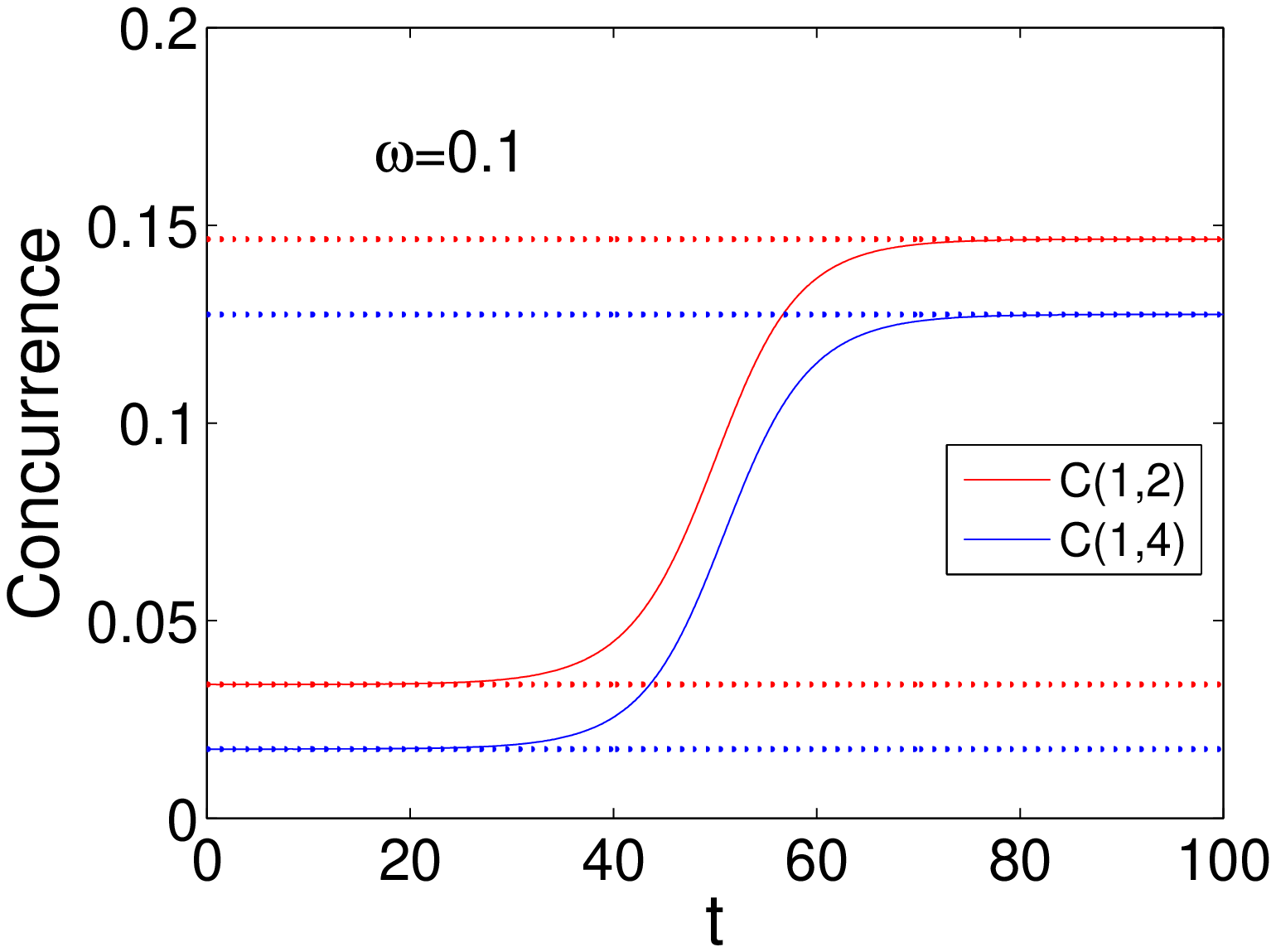}}\\
   \subfigure[]{\label{fig:ht_c}\includegraphics[width=6 cm]{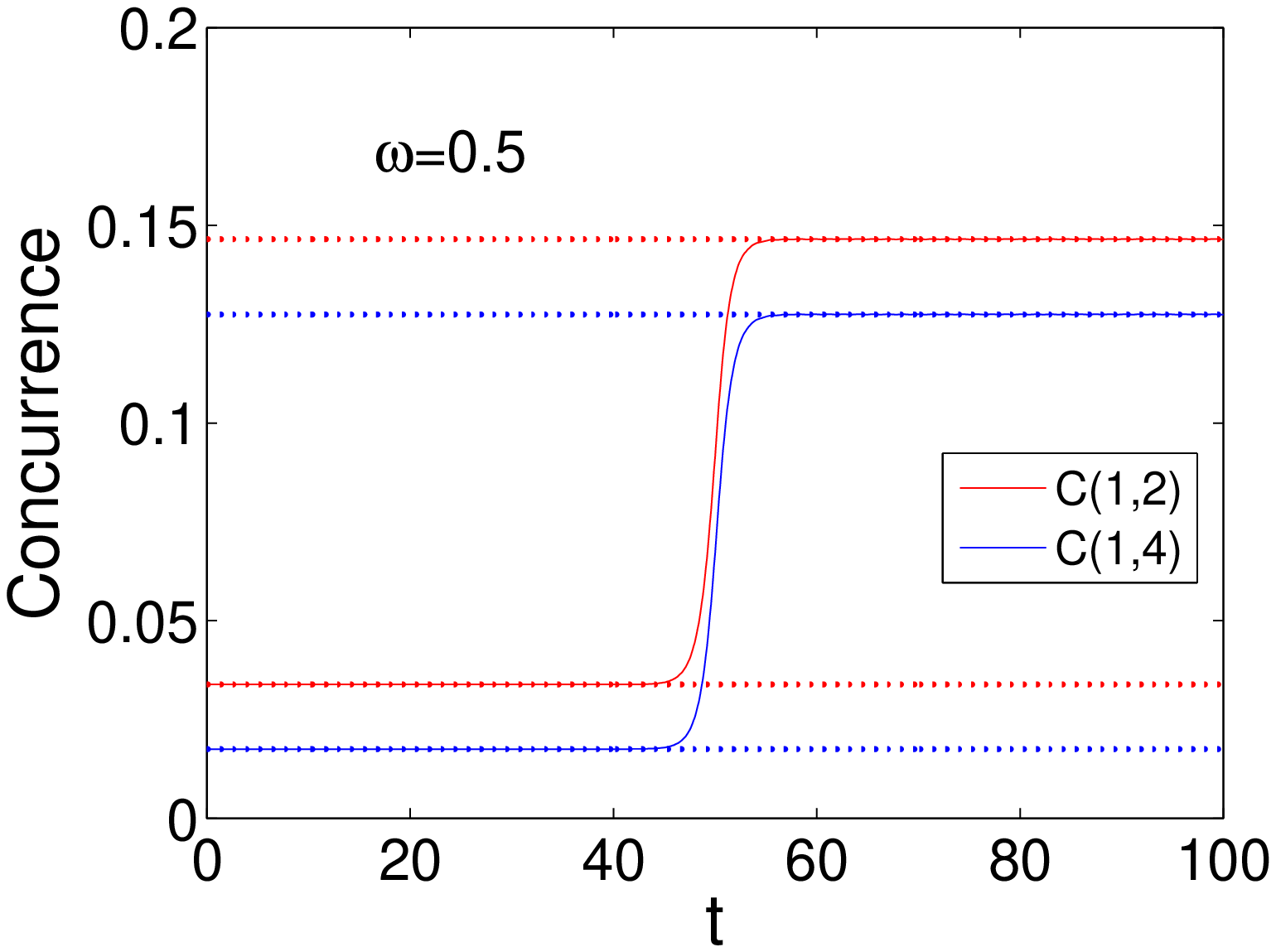}}\quad
   \subfigure[]{\label{fig:ht_d}\includegraphics[width=6 cm]{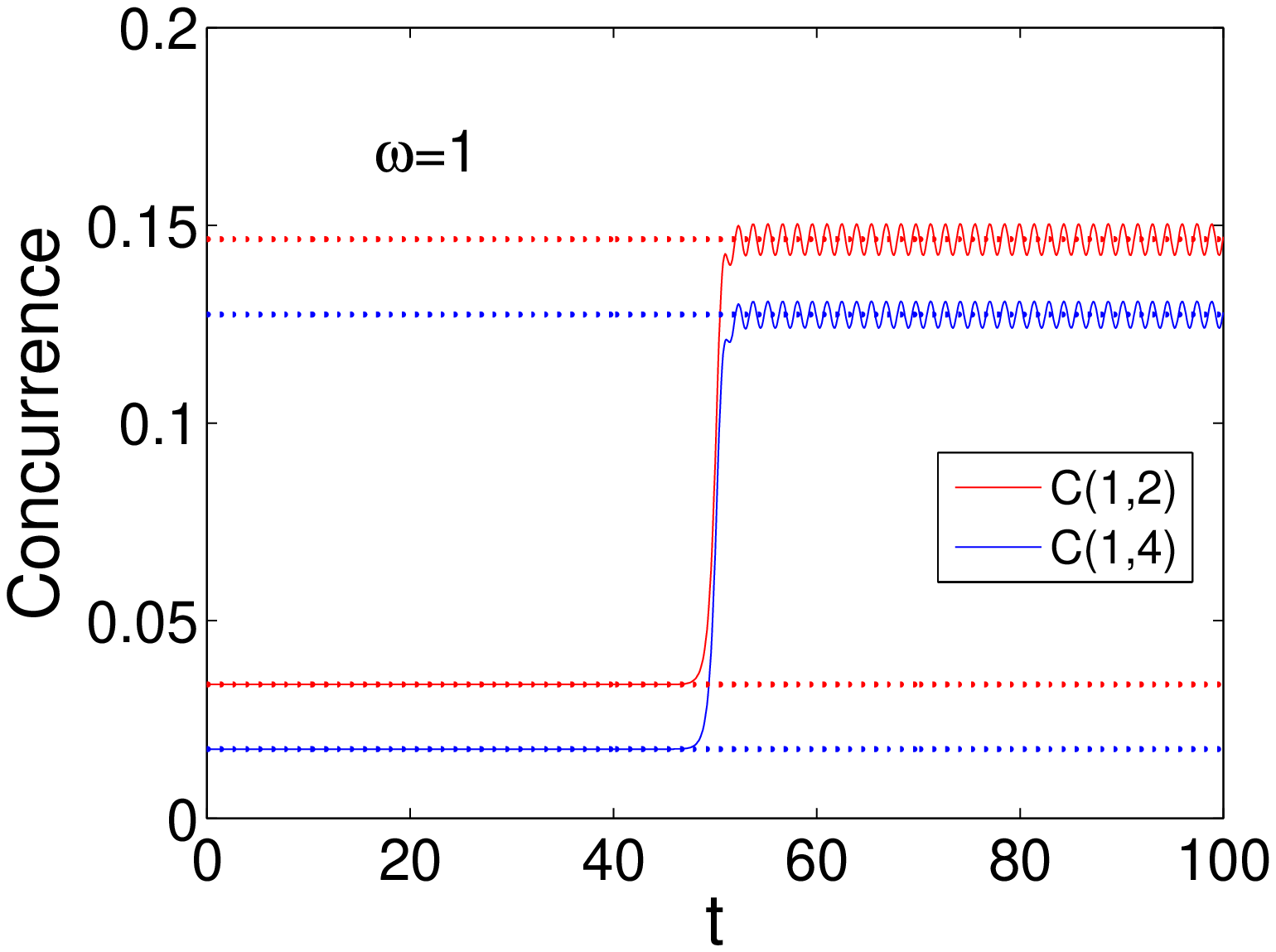}}
   \caption{{\protect\footnotesize D(Color online) Dynamics of the concurrences C(1,2) and C(1,4) in applied hyperbolic magnetic fields of various frequencies $\omega=0.1$, $0.5$ \& $1$, with $a=1$, $b=2$.}}
 \label{tanh_h}
 \end{minipage}
\end{figure}
%%%%%%%%%%%%%%%%%%%%%%%%%%%%%%%%%%%%%%%%%%%%%%%%%%%%%%%%%%%%%%%%%%%%%%%%%%%%%%%%%%%%%%%%%%%%%%%%%%%%%%
%%%%%%%%%%%%%%%%%%%%%%%%%%%%%%%%%%%%%%%%%%%%%%%%%%%%%%%%%%%%%%%%%%%%%%%%%%%%%%%%%%%%%%%%%%%%%%%%%%%%%
\begin{figure}[htbp]
\begin{minipage}[c]{\textwidth}
 \centering
   \subfigure[]{\label{fig:ht_e}\includegraphics[width=6 cm]{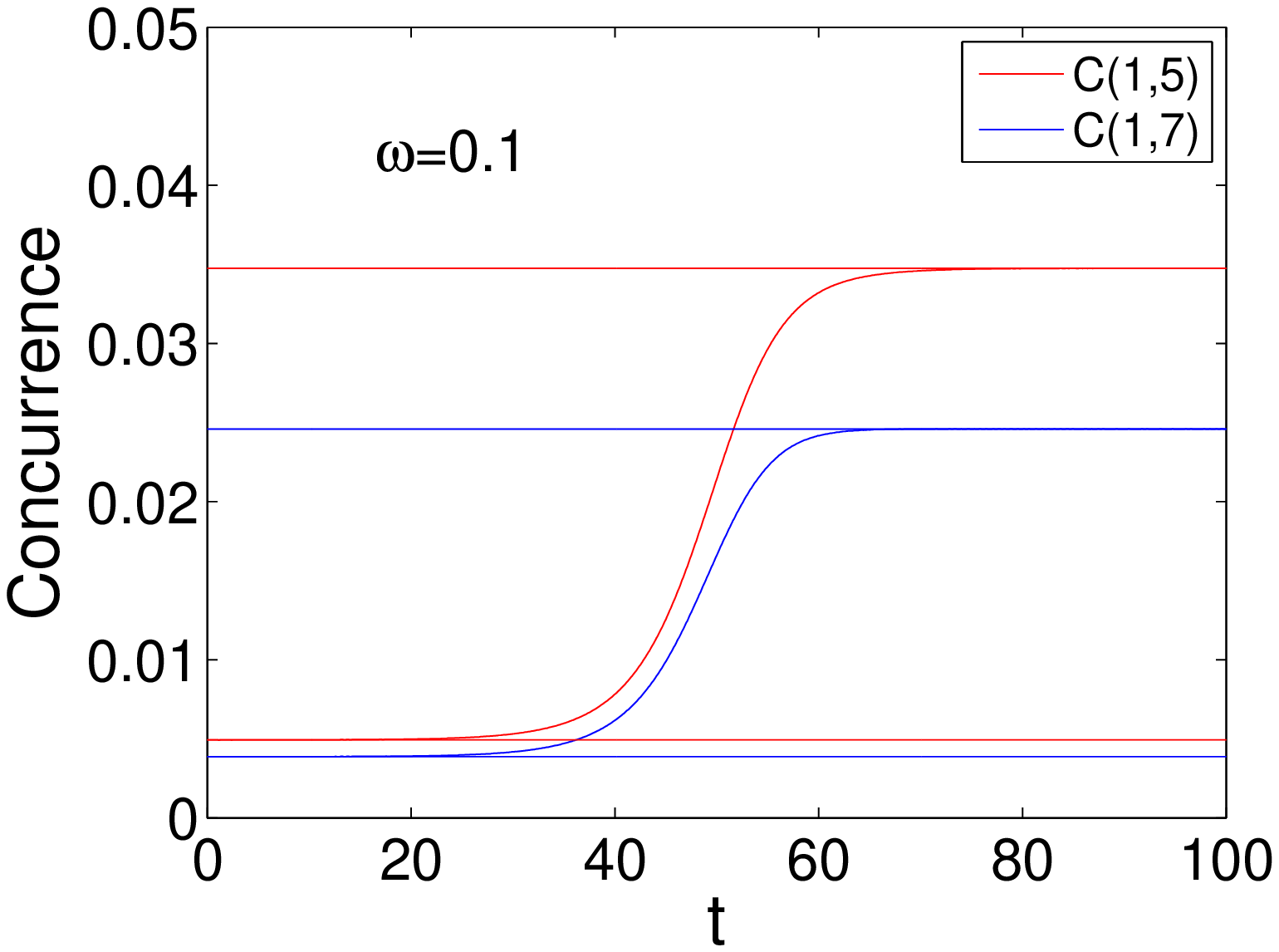}}\quad
   \subfigure[]{\label{fig:ht_f}\includegraphics[width=6 cm]{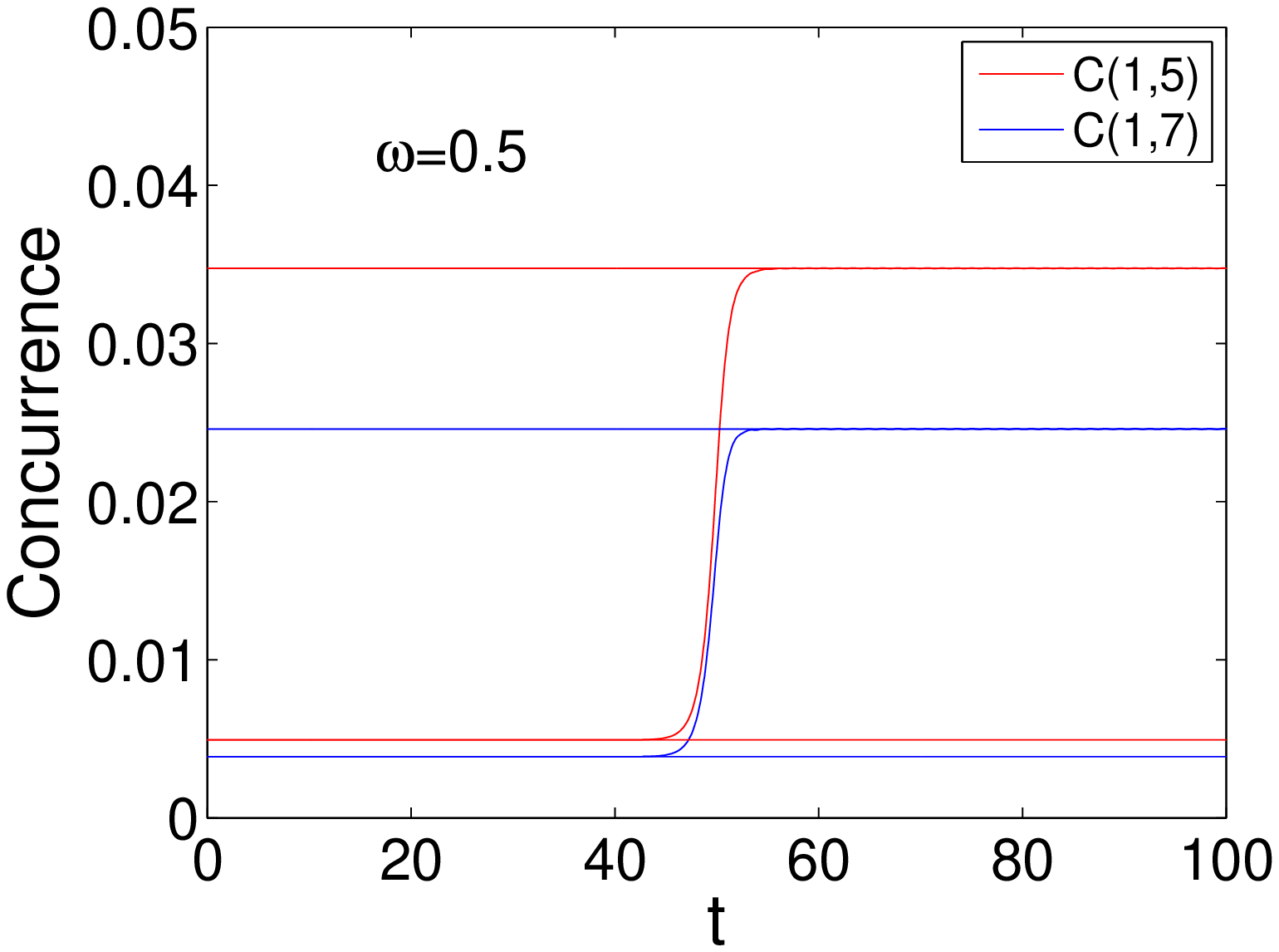}}
   \caption{{\protect\footnotesize (Color online) Dynamics of the concurrences C(1,5) and C(1,7) in applied hyperbolic magnetic fields of various frequencies $\omega=0.1$, $0.5$ \& $1$, with $a=1$, $b=2$.}}
 \label{tanh_h2}
 \end{minipage}
\end{figure}
%%%%%%%%%%%%%%%%%%%%%%%%%%%%%%%%%%%%%%%%%%%%%%%%%%%%%%%%%%%%%%%%%%%%%%%%%%%%%%%%%%%%%%%%%%%%%%%%%%%%%%

Figure.~\ref{tanh_h} explores the behavior of the concurrence C(1,4) and C(1,2) of the system under the hyperbolic field at different transition constant values $\omega$. Comparing the behavior of the concurrence in this case with the exponential case at the same frequencies, it is clear the similarity but with much higher controllability for the hyperbolic field with sharper asymptotic concurrence value. The system confirms its ergodic behavior as can be seen. Again the next nearest concurrences C(1,5) and C(1,7) show the same behavior as the nearest neighbor concurrences as shown in
Figure.~\ref{tanh_h2}.

%%%%%%%%%%%%%%%%%%%%%%%%%%%%%%%%%%%%%%%%%%%%%%%%%%%%%%%%%%%%%%%%%%%%%%%%%
\subsection{Periodic magnetic fields}
%%%%%%%%%%%%%%%%%%%%%%%%%%%%%%%%%%%%%%%%%%%%%%%%%%%%%%%%%%%%%%%%%%%%%%%%%

\begin{figure}[htbp]
\begin{minipage}[c]{\textwidth}
 \centering
   \subfigure[]{\label{fig:hs_a}\includegraphics[width=6 cm]{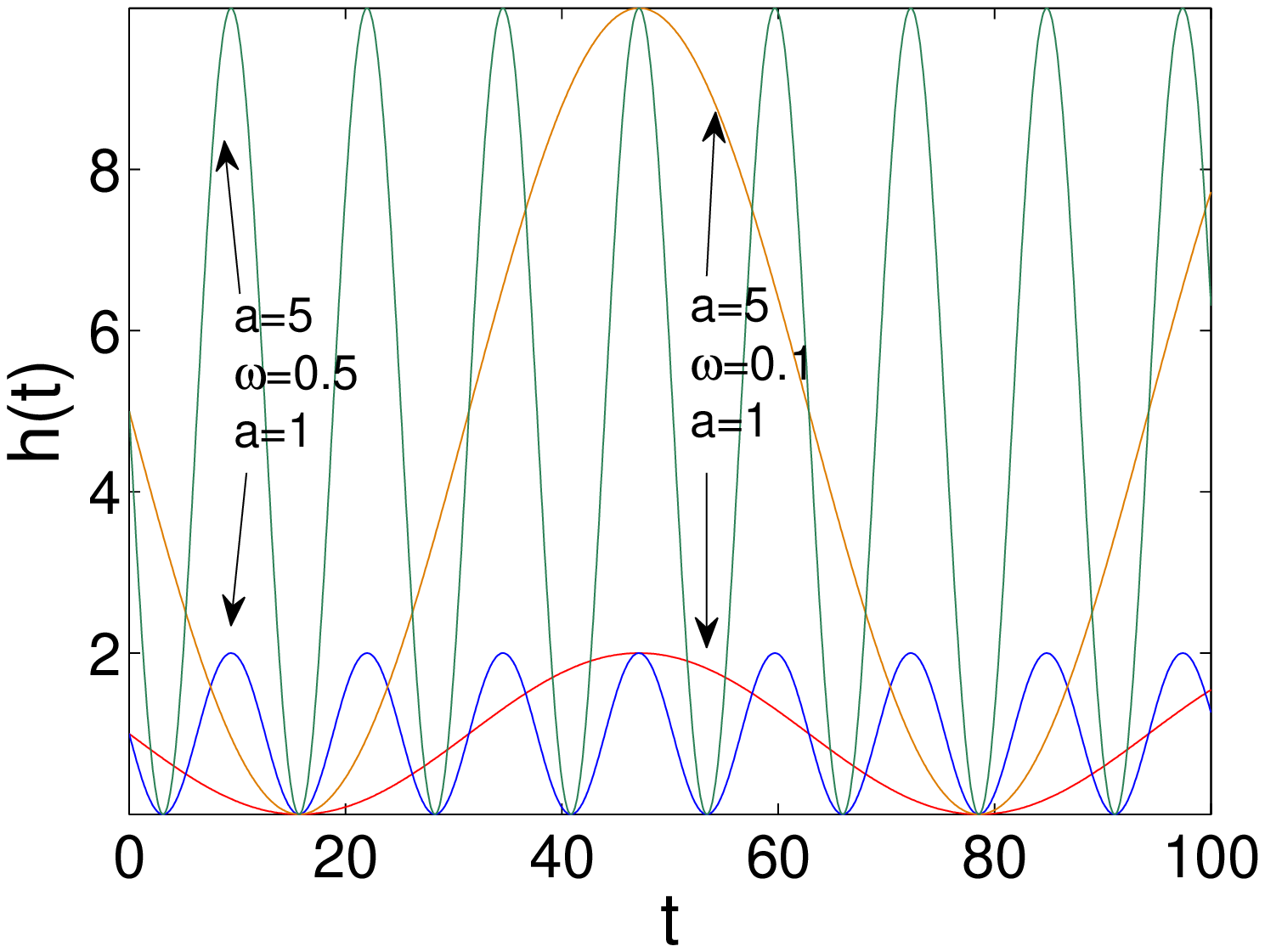}}\quad
   \subfigure[]{\label{fig:hs_b}\includegraphics[width=6 cm]{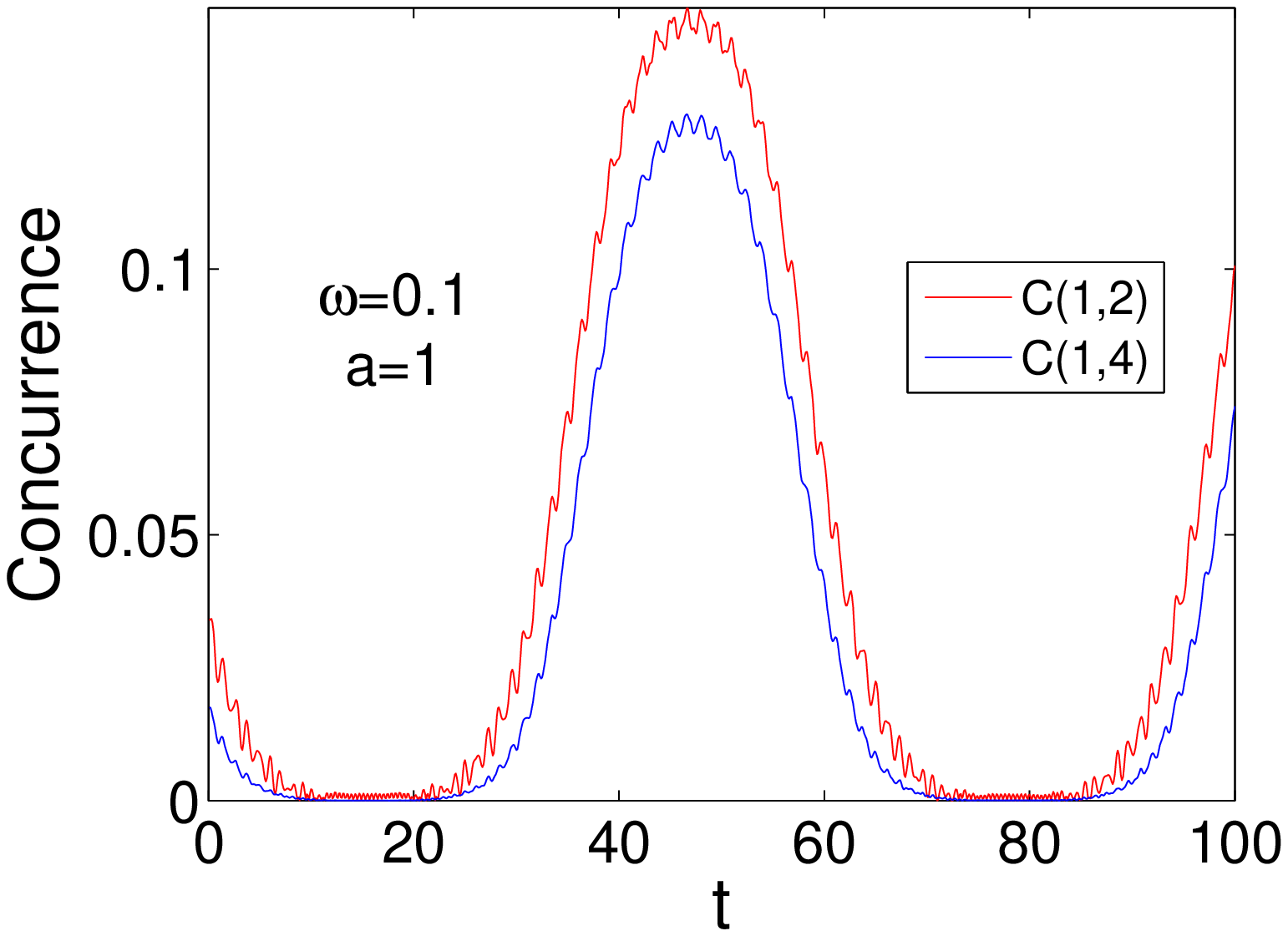}}\\
   \subfigure[]{\label{fig:hs_c}\includegraphics[width=6 cm]{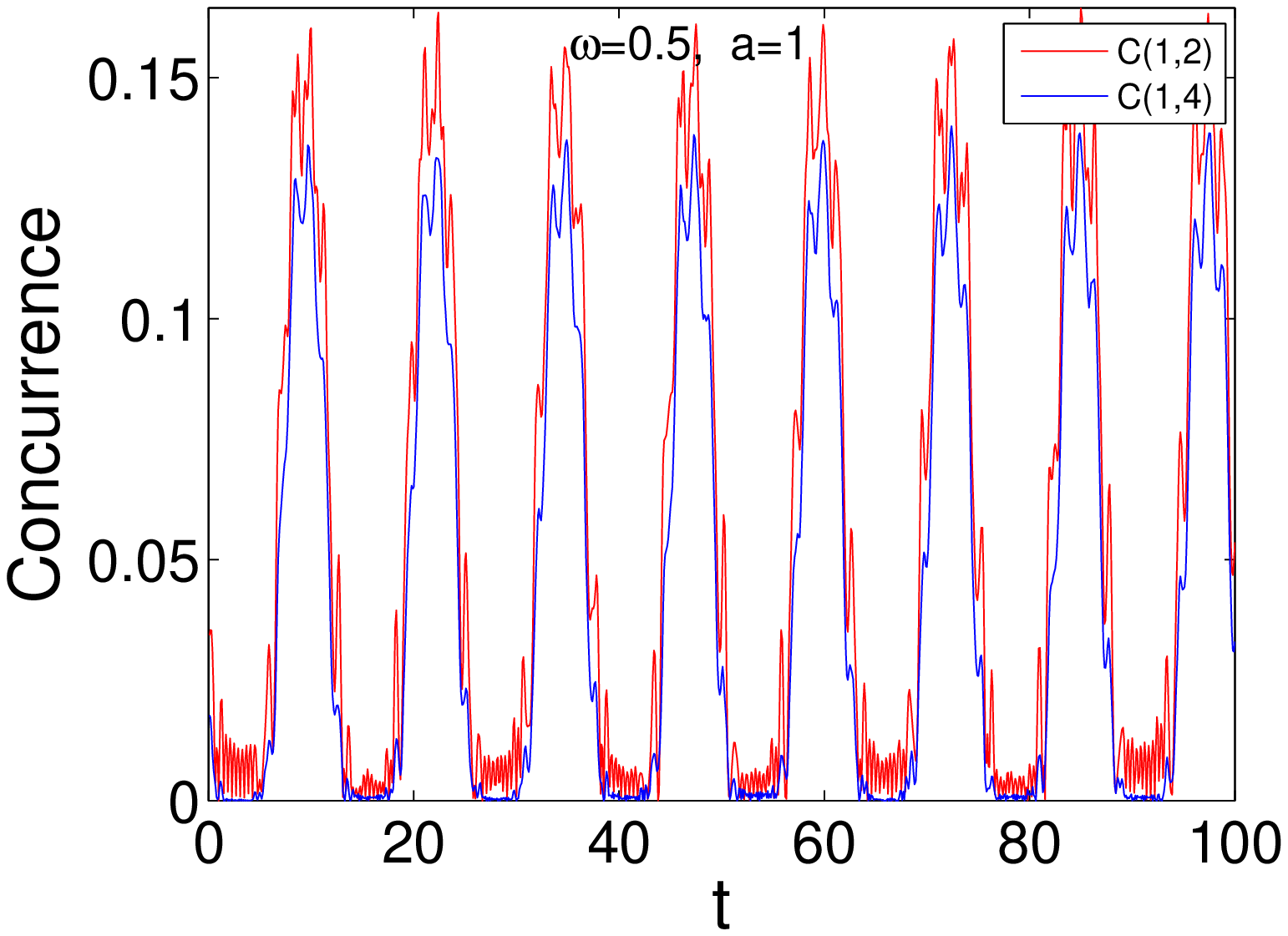}}\quad
   \subfigure[]{\label{fig:hs_d}\includegraphics[width=6 cm]{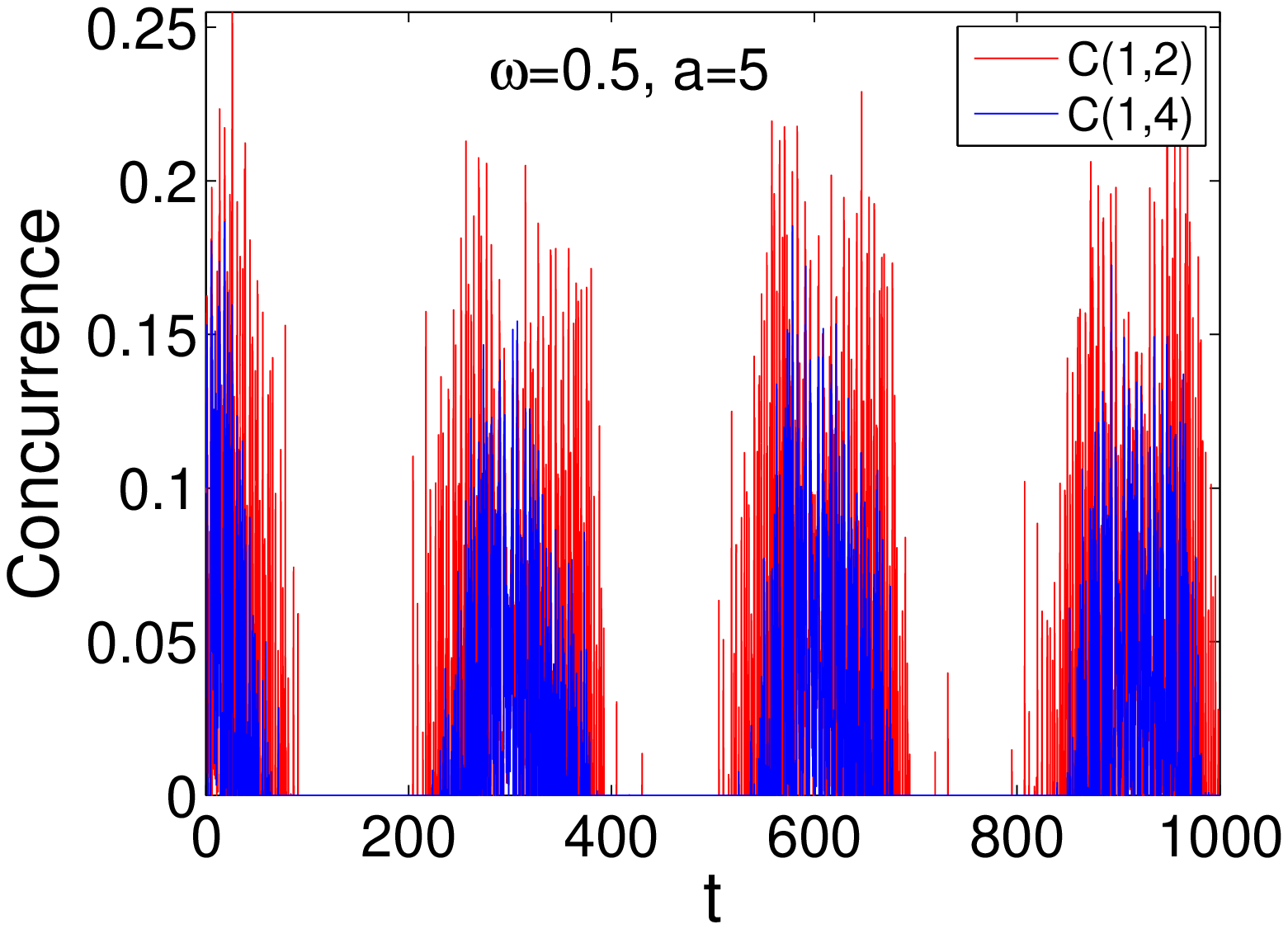}}
   \caption{{\protect\footnotesize (Color online) Dynamics of the concurrences C(1,2) and C(1,4) in applied sine magnetic fields of various frequencies and field strength $\omega=0.1$ \& $a=1$, $\omega=0.5$ \& $a=1$, $\omega=0.5$ \& $a=5$.}}
 \label{sin_h}
 \end{minipage}
\end{figure}

\begin{figure}[htbp]
\begin{minipage}[c]{\textwidth}
 \centering
   \subfigure[]{\label{fig:hs_e}\includegraphics[width=6 cm]{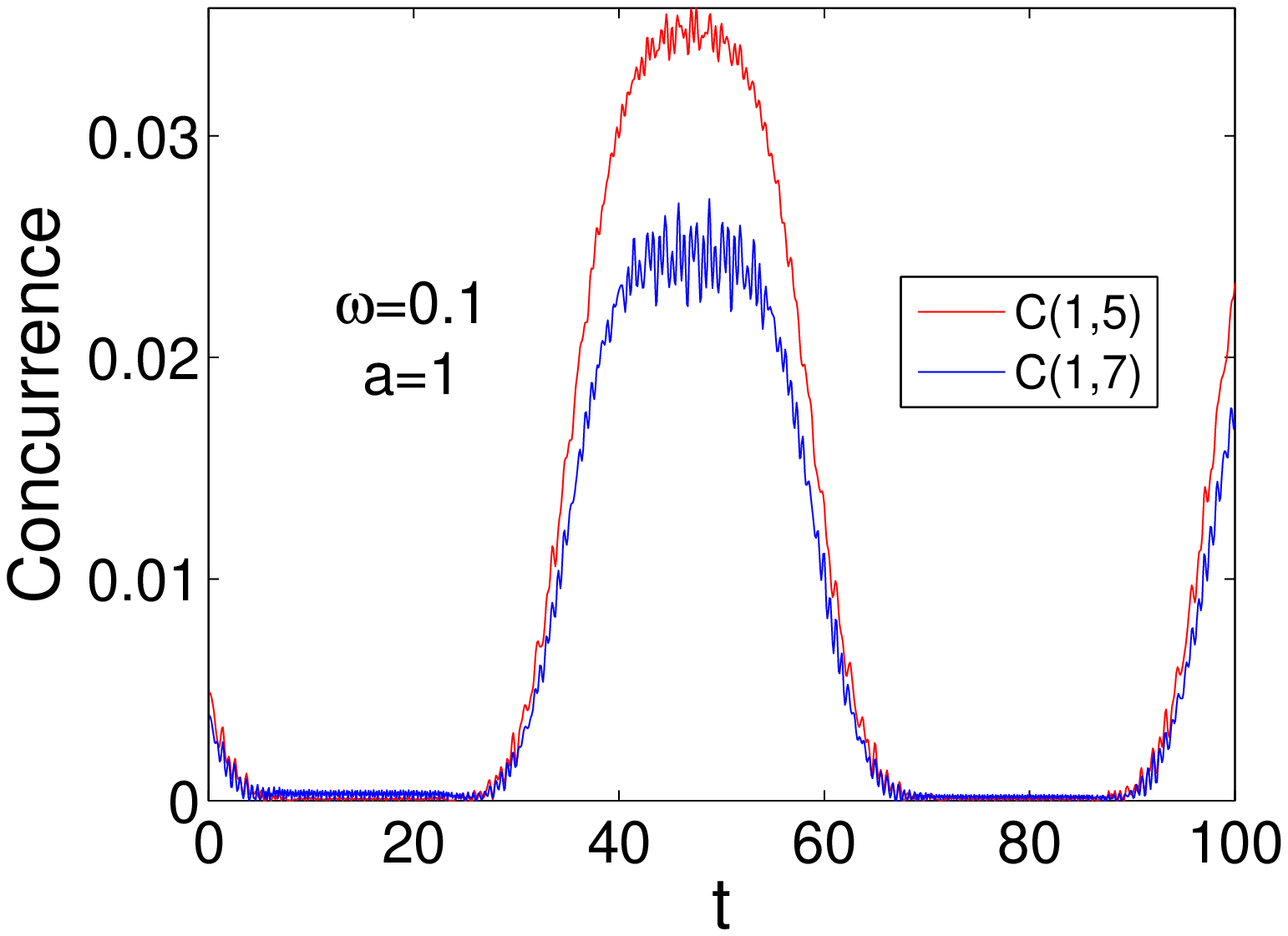}}\quad
   \subfigure[]{\label{fig:hs_f}\includegraphics[width=5.5 cm]{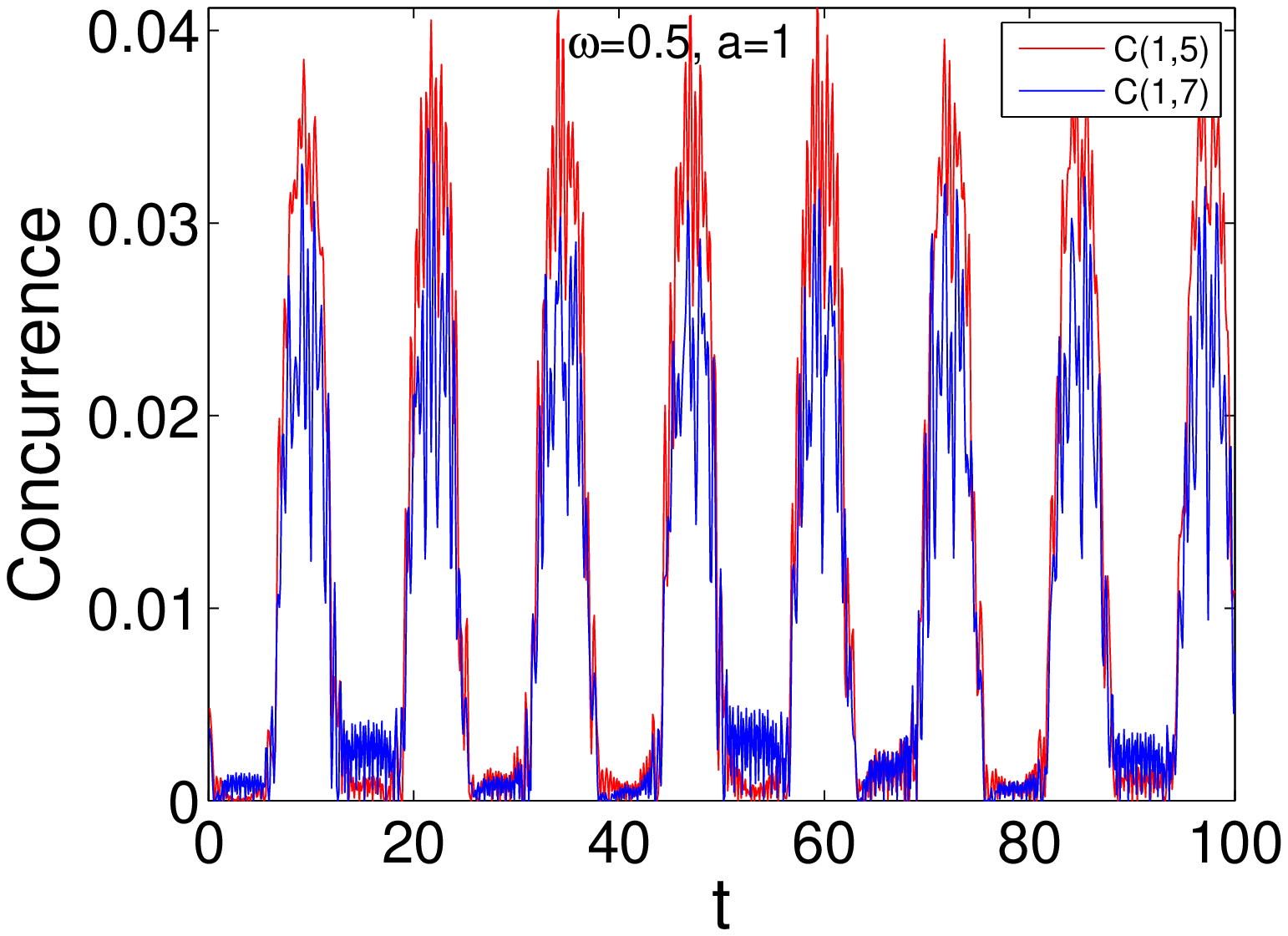}}
   \caption{{\protect\footnotesize (Color online) Dynamics of the concurrences C(1,5) and C(1,7) in applied sine magnetic fields of various frequencies and field strength $\omega=0.1$ \& $a=1$, $\omega=0.5$ \& $a=1$.}}
 \label{sin_h2}
 \end{minipage}
\end{figure}

\begin{figure}[htbp]
\begin{minipage}[c]{\textwidth}
 \centering
   \subfigure[]{\label{fig:hc_a}\includegraphics[width=6 cm]{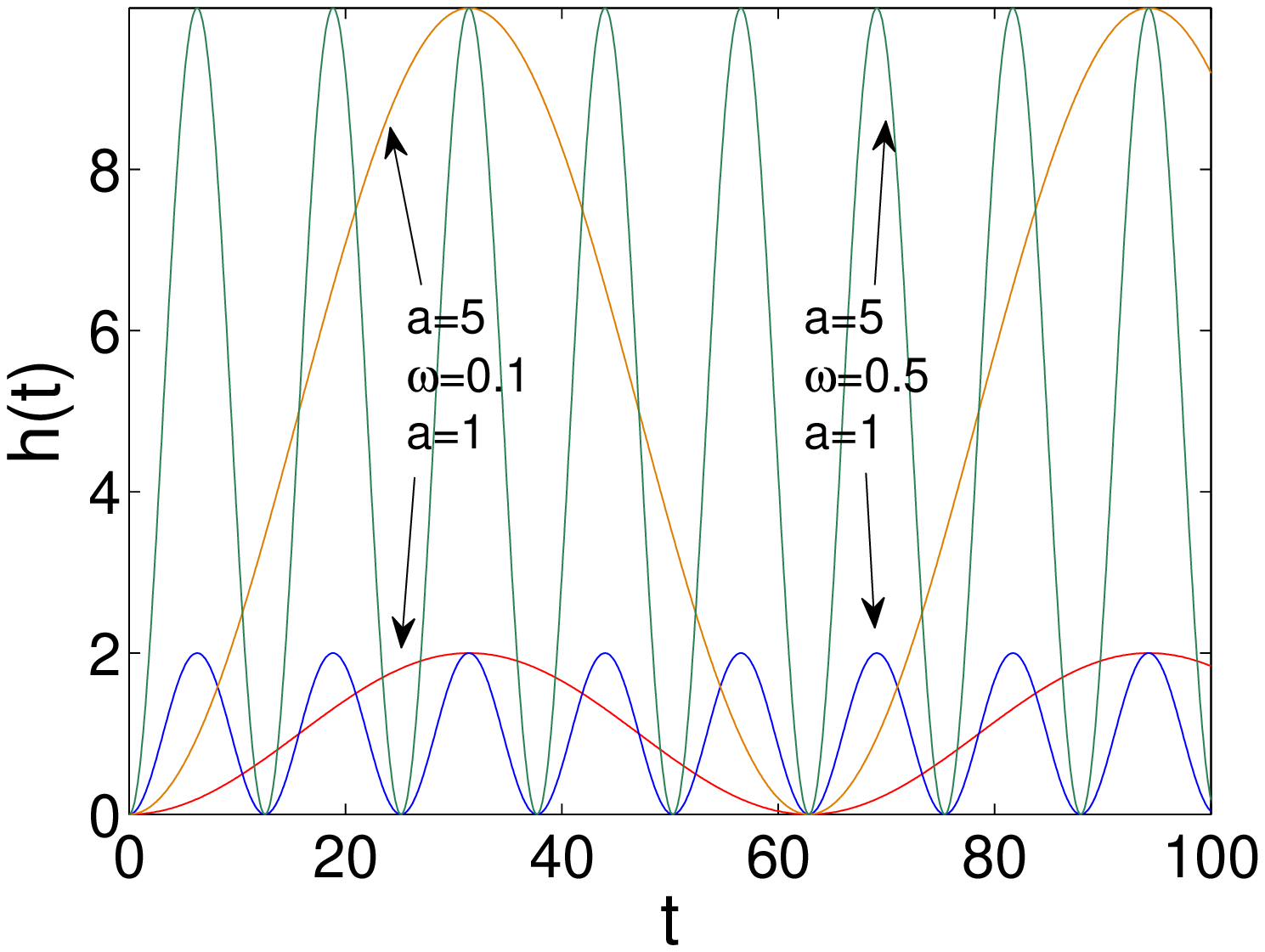}}\quad
   \subfigure[]{\label{fig:hc_b}\includegraphics[width=6 cm]{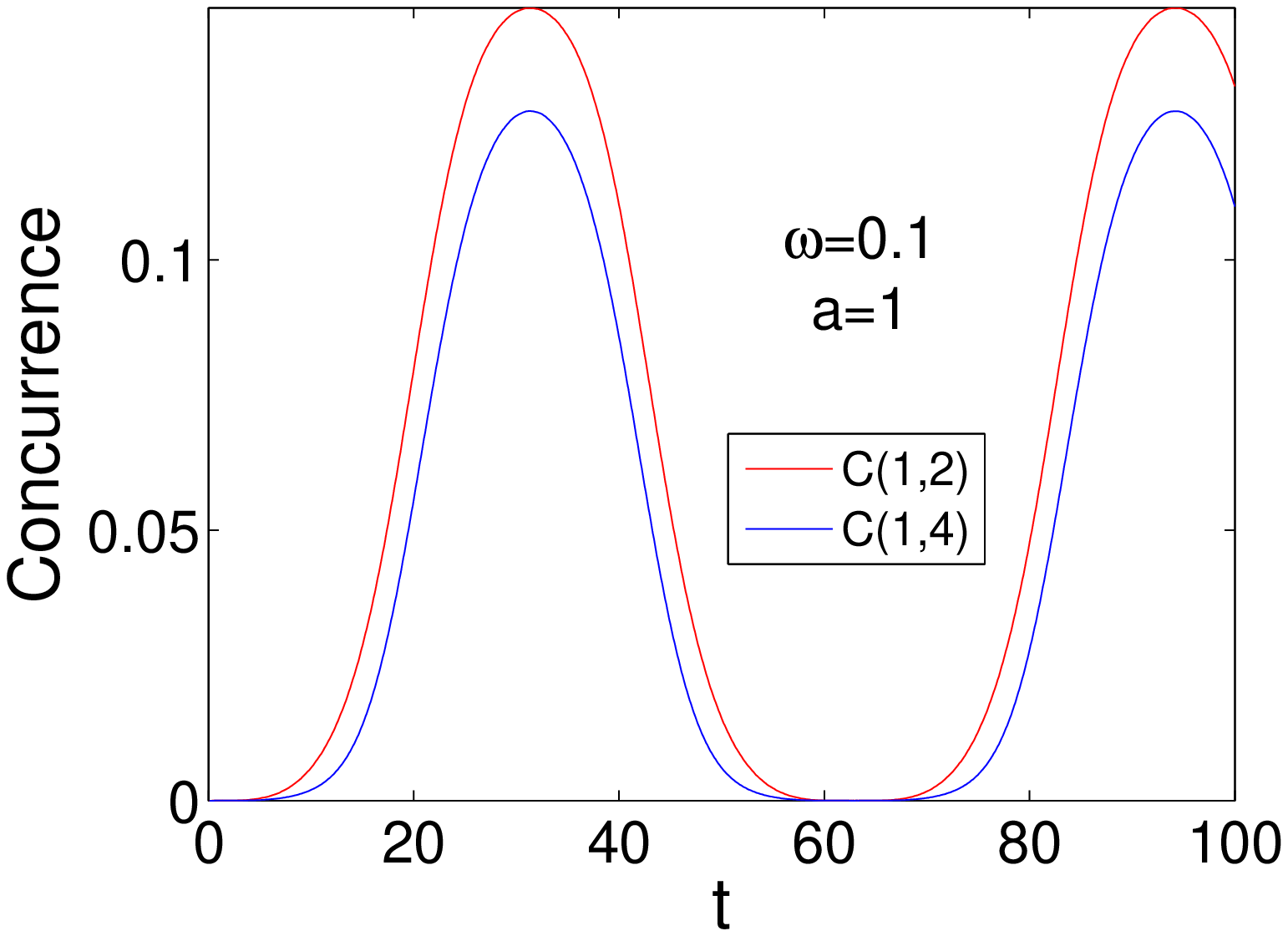}}\\
   \subfigure[]{\label{fig:hc_c}\includegraphics[width=6 cm]{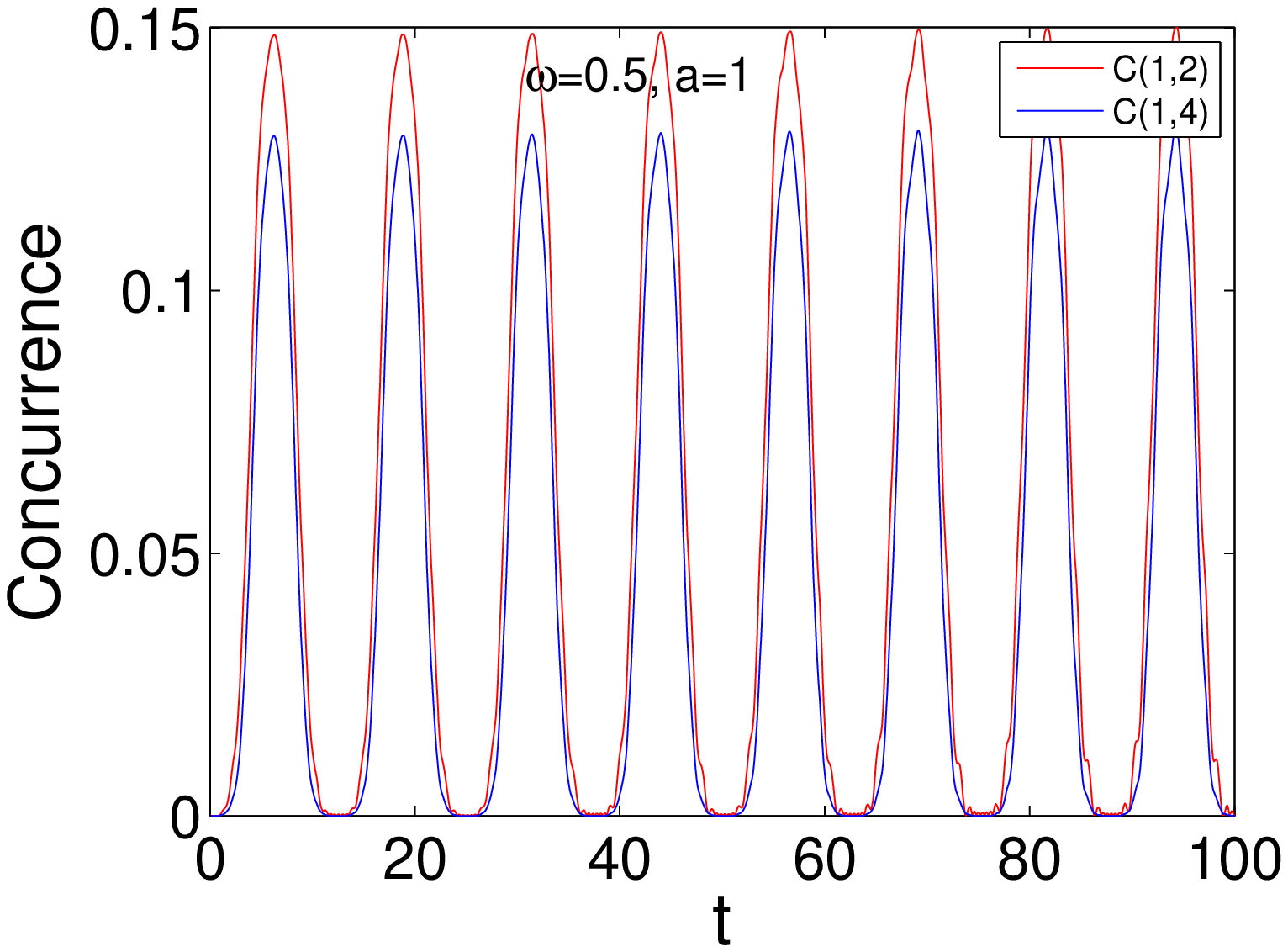}}\quad
   \subfigure[]{\label{fig:hc_d}\includegraphics[width=6 cm]{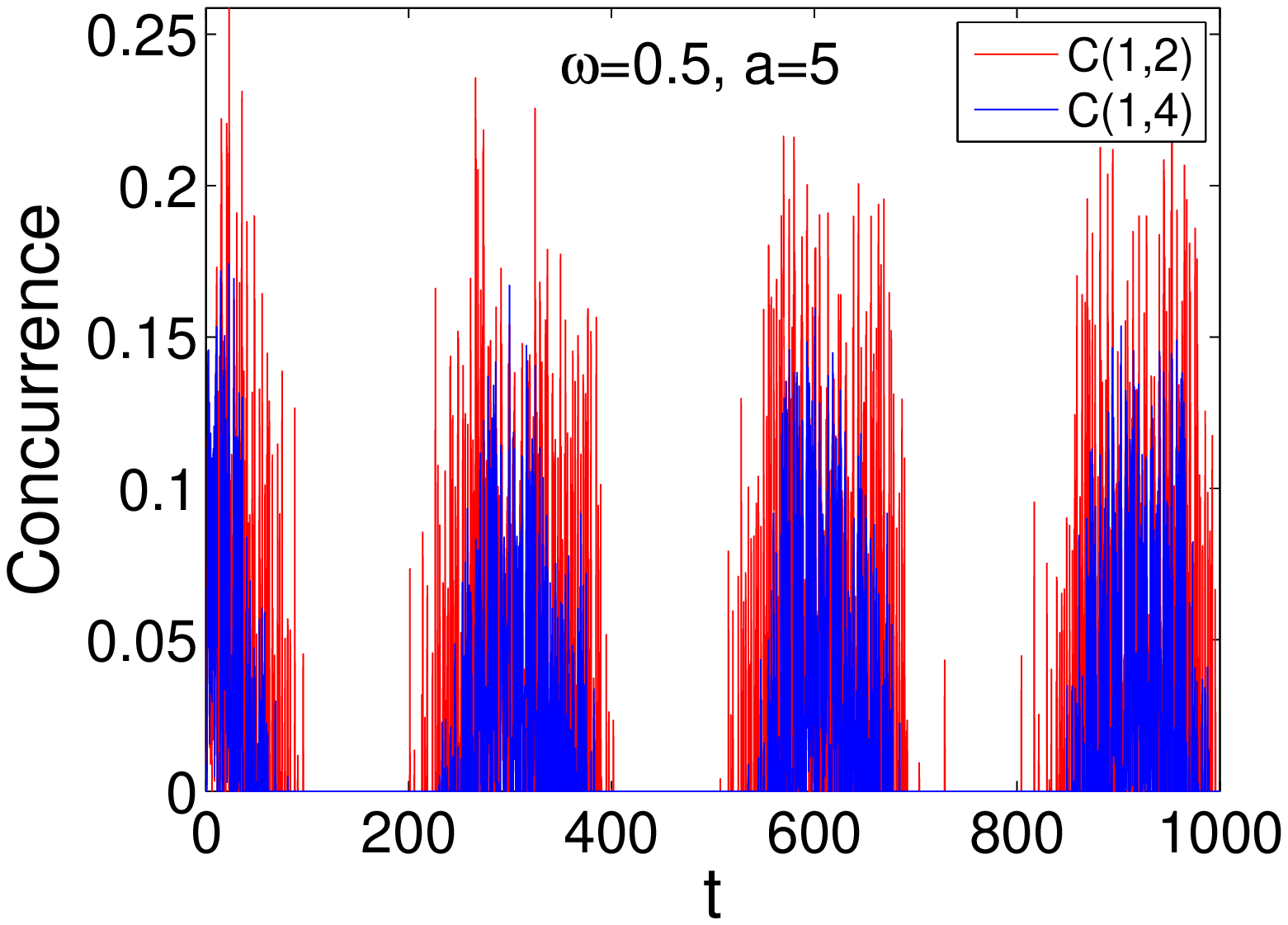}}
   \caption{{\protect\footnotesize (Color online) Dynamics of the concurrences C(1,2) and C(1,4) in applied cosine magnetic fields of various frequencies and field strength $\omega=0.1$ \& $a=1$, $\omega=0.5$ \& $a=1$, $\omega=0.5$ \& $a=5$.}}
 \label{cos_h}
 \end{minipage}
\end{figure}

\begin{figure}[htbp]
\begin{minipage}[c]{\textwidth}
 \centering
   \subfigure[]{\label{fig:hc_e}\includegraphics[width=6 cm]{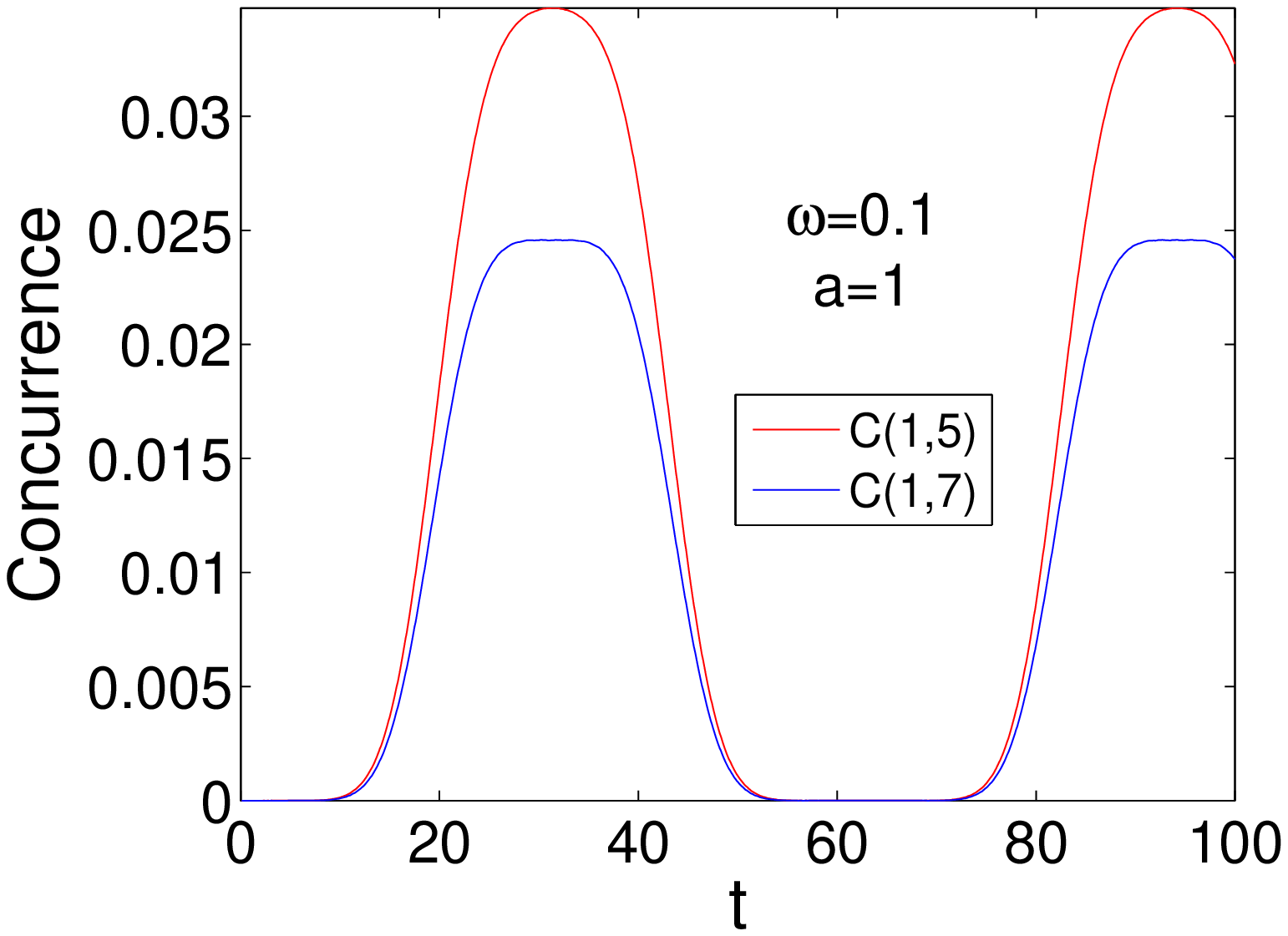}}\quad
   \subfigure[]{\label{fig:hc_f}\includegraphics[width=5.5 cm]{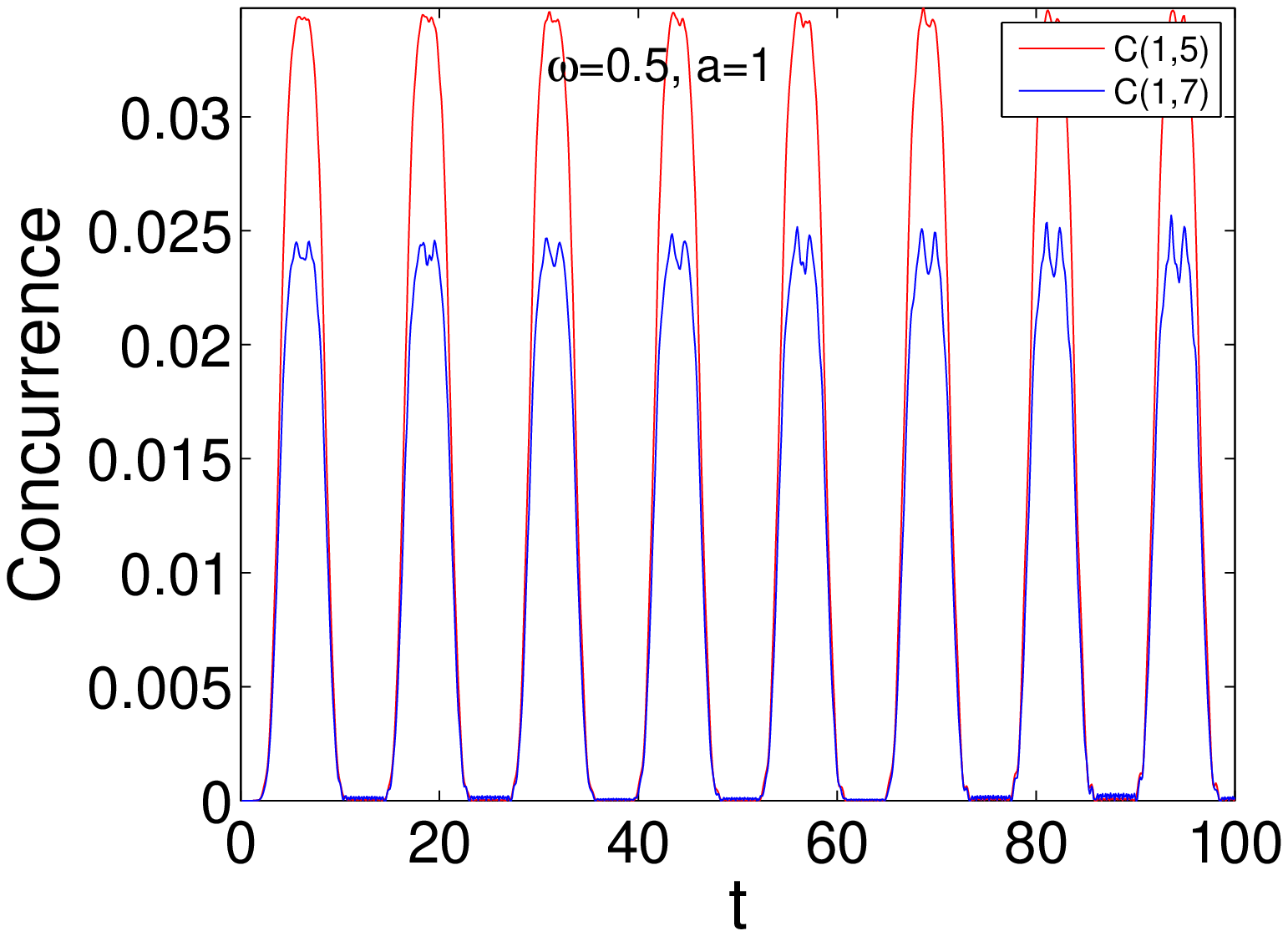}}
   \caption{{\protect\footnotesize (Color online) Dynamics of the concurrences C(1,5) and C(1,7) in applied cosine magnetic fields of various frequencies and field strength $\omega=0.1$ \& $a=1$, $\omega=0.5$ \& $a=1$.}}
 \label{cos_h2}
 \end{minipage}
\end{figure}

In this section we test the dynamics of the system under a different form of external magnetic field, namely periodic. It is represented by the sinusoidal function form
\be
h(t)=\left\{
\begin{array}{lr}
a & \qquad t\leq t_0 \\
a-a\sin(\omega t+\phi) & \qquad t>t_0
\end{array}
\right.
\ee
where $\phi$ represents an initial phase of the function, which determines the initial value of the applied magnetic field. For $\phi=0$ we obtain a $\sin (\omega t)$ function, which is studied in Figs.~\ref{sin_h} and \ref{sin_h2}, while for $\phi=\pi/2$ we get a $\cos (\omega t)$ which is explored in Figs.~\ref{cos_h} and \ref{cos_h2}.
%%%%%%%%%%%%%%%%%%%%%%%%%%%%%%%%%%%%%%%%%%%%%%%%%%%%%%%%%%%%%%%%%%%%%%%%%%%%%%%%%%%%%%%%%%%%%%%%%%%%%%%%%%%%%%%%%%%%%%%%%%%%%%%%%%%%
%%%%%%%%%%%%%%%%%%%%%%%%%%%%%%%%%%%%%%%%%%%%%%%%%%%%%%%%%%%%%%%%%%%%%%%%%%%%%%%%%%%%%%%%%%%%%%%%%%%%%%%%%%%%%%%%%%%%%%%%%%%%%%%%%%%%%%

The influential factors in the sinusoidal field are amplitude $a$ and angular frequency $\omega$. Figure \ref{fig:hs_b} is a good start to analyze them. When $\omega=0.1$ and $a=1$, both are small, concurrence varies up and down in the same frequency as the field, but little dents and big catches come out as both $\omega$ and $a$ increase as shown in Figs. \ref{fig:hs_c} and \ref{fig:hs_d}. As the frequency of the field increases it becomes too fast for the system to follow causing more imperfection in the concurrence oscillation. On the other hand, as we have seen in the previous magnetic field forms, larger magnetic fields don't necessarily bring larger concurrence, so forth dents appear as we increase $a$. Even larger magnetic fields totally break the pattern as can be seen in Fig. \ref{fig:hs_d}. It is interesting to see that the larger amplitudes are not so disturbing to the nearest concurrences as they are for the next nearest neighbors as can be concluded from Fig. \ref{sin_h2}.

\begin{figure}[htbp]
\begin{minipage}[c]{\textwidth}
 \centering
   \subfigure[]{\label{fig:hpi4_a}\includegraphics[width=6 cm]{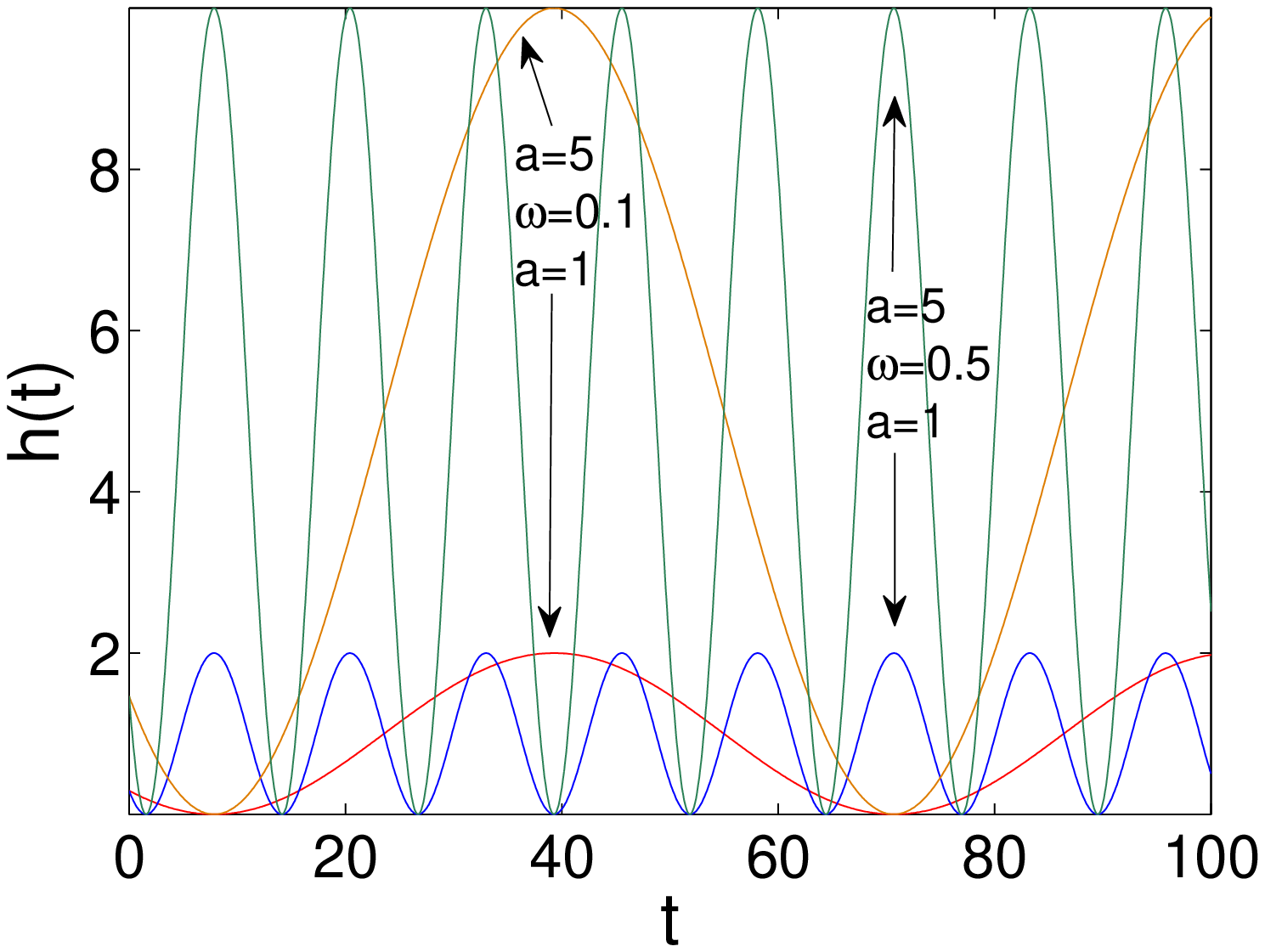}}\quad
   \subfigure[]{\label{fig:hpi4_b}\includegraphics[width=6 cm]{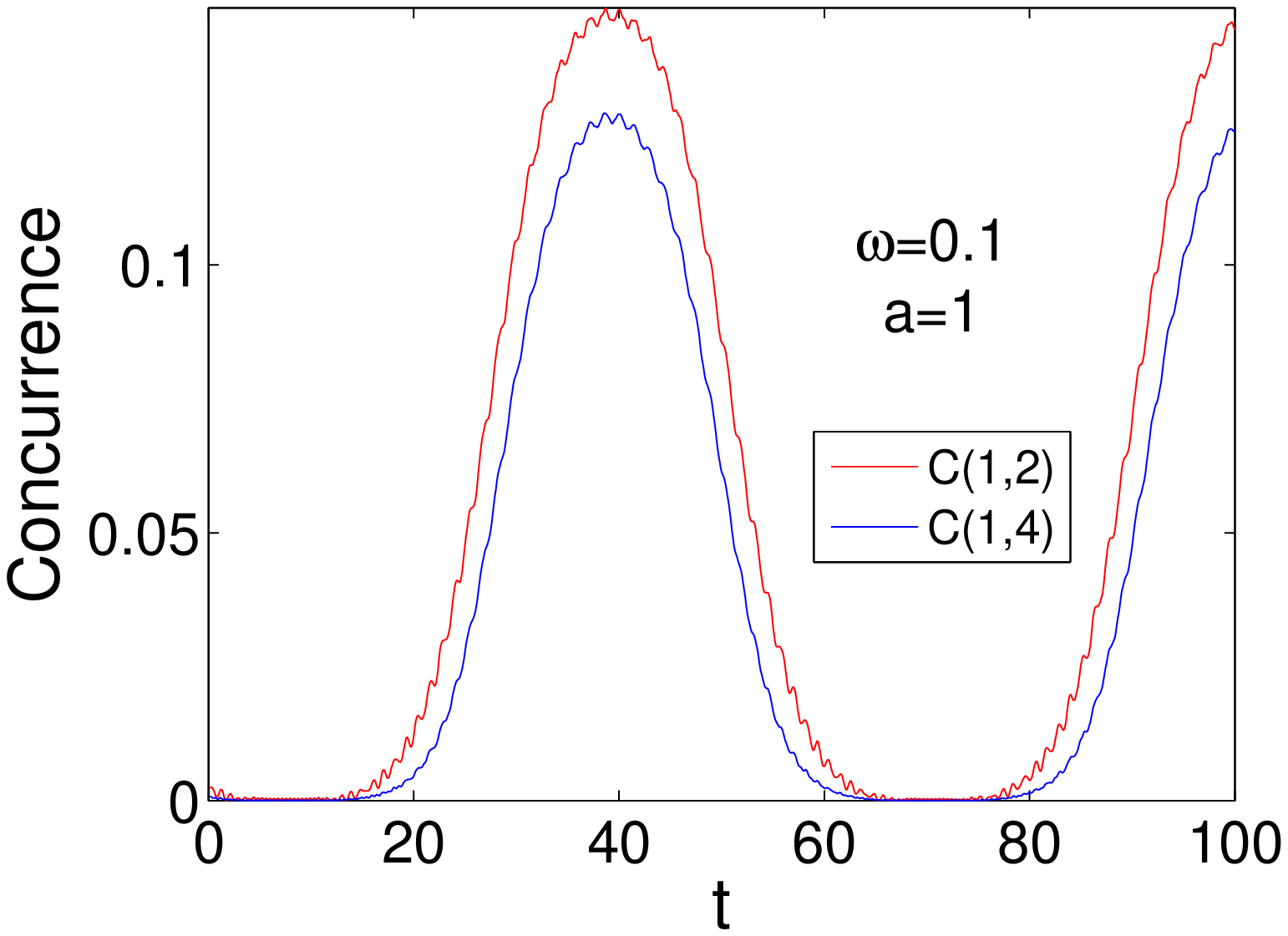}}\\
   \subfigure[]{\label{fig:hpi4_c}\includegraphics[width=6 cm]{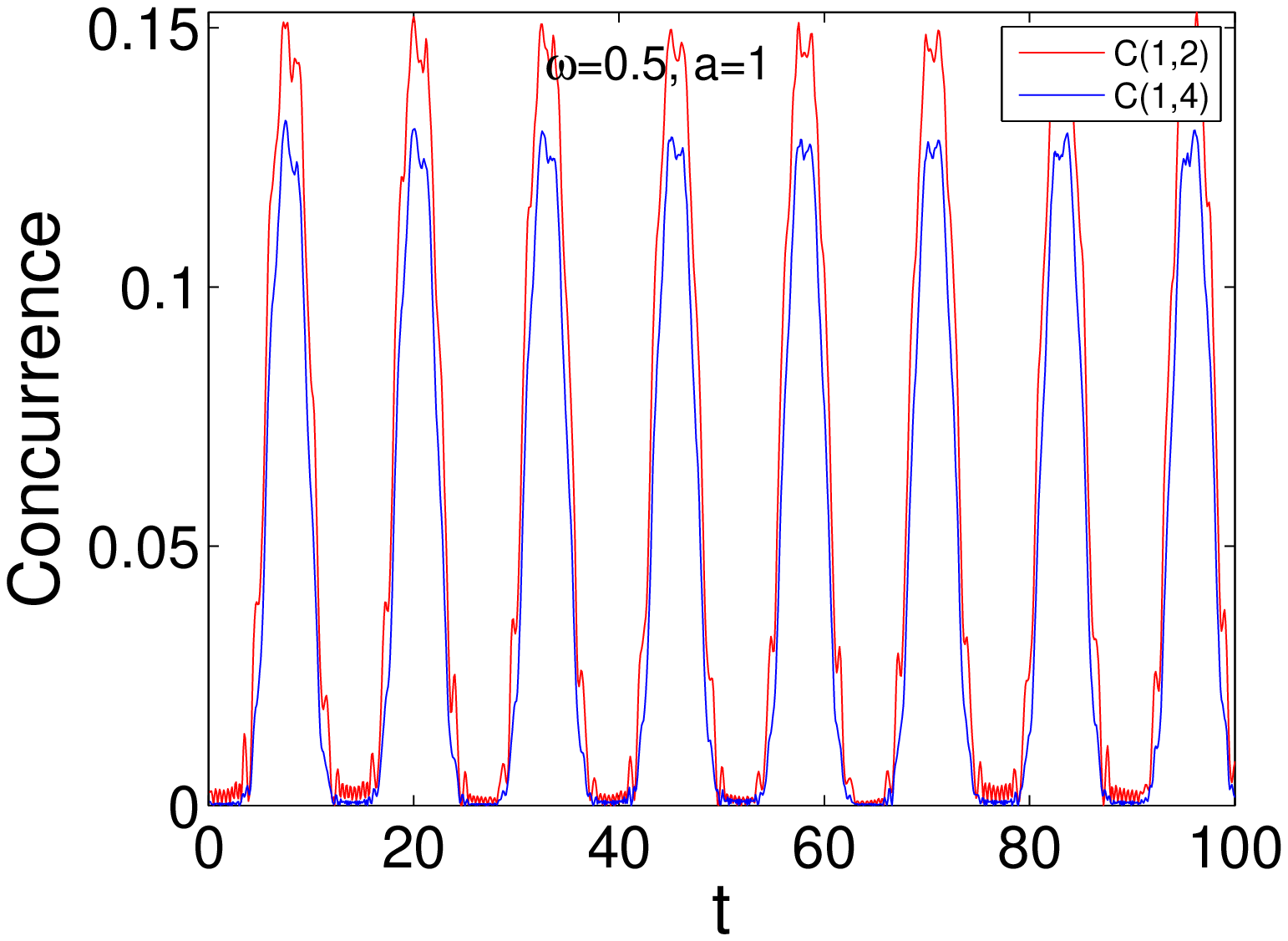}}\quad
   \subfigure[]{\label{fig:hpi4_d}\includegraphics[width=6 cm]{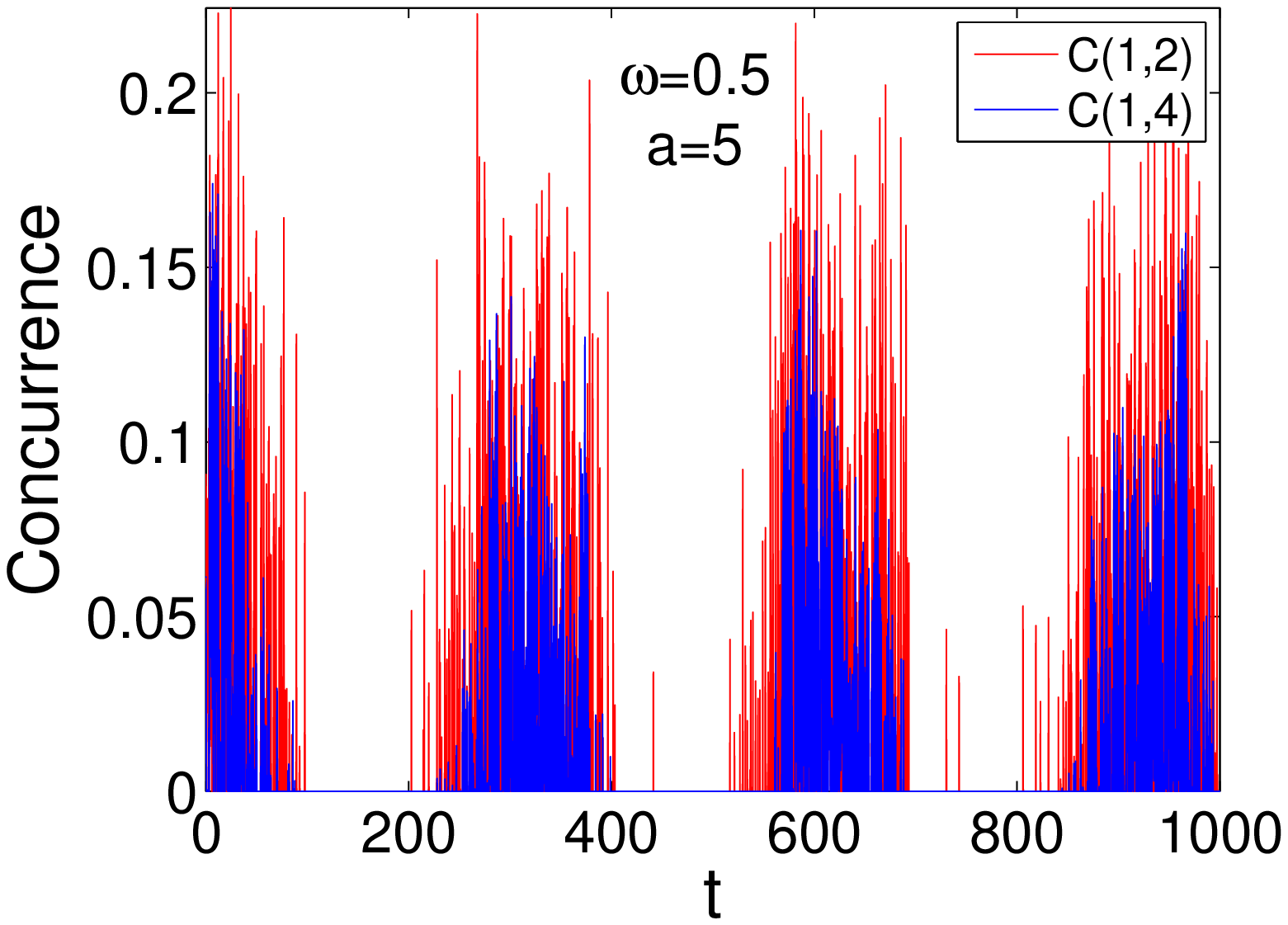}}
   \caption{{\protect\footnotesize (Color online) Dynamics of the concurrences C(1,2) and C(1,4) in applied sinusoidal magnetic fields ($\phi=\pi/4$) of various frequencies and field strength $\omega=0.1$ \& $a=1$, $\omega=0.5$ \& $a=1$, $\omega=0.5$ \& $a=5$.}}
 \label{pi4_h}
 \end{minipage}
\end{figure}

The critical effect of the initial phase, which determines the initial value of the field, can be seen in Figs. \ref{cos_h} and \ref{cos_h2} where the phase was chosen to be $\pi/2$ resulting a $\cos(\omega t)$ magnetic field. As one can notice the magnetic field is initially zero and the concurrence is closely following the magnetic filed with much sharper and much less distorted profile than the sin case. In fact, testing a middle value for the phase, between 0 and $\pi/2$, showed a concurrence with a middle profile between the sin and cos cases (Fig. \ref{pi4_h}). Therefore, the smaller the initial value of the external magnetic field, better if zero, the less distorted is the concurrence profile.

%%%%%%%%%%%%%%%%%%%%%%%%%%%%%%%%%%%%%%%%%%%%%%%%%%%%%%%%%%%%%%%%%%%%%%%%%%%%%%%%%%%%%%%%%%%%%%%%%%%%%%%%%%%%%%%%%%%%%%%%%%%%%%%%%%%%
%%%%%%%%%%%%%%%%%%%%%%%%%%%%%%%%%%%%%%%%%%%%%%%%%%%%%%%%%%%%%%%%%%%%%%%%%%%%%%%%%%%%%%%%%%%%%%%%%%%%%%%%%%%%%%%%%%%%%%%%%%%%%%%%%%%%%%

%%%%%%%%%%%%%%%%%%%%%%%%%%%%%%%%%%%%%%%%%%%%%%%%%%%%%%%%%%%%%%%%%%%%%%%%%%%%%%%%%%%%%%%%%%%%%%%%%%%%%%%%%%%%%%%%%%%%%%%%%%%%%%%%%%%%
%%%%%%%%%%%%%%%%%%%%%%%%%%%%%%%%%%%%%%%%%%%%%%%%%%%%%%%%%%%%%%%%%%%%%%%%%%%%%%%%%%%%%%%%%%%%%%%%%%%%%%%%%%%%%%%%%%%%%%%%%%%%%%%%%%%%%%

%%%%%%%%%%%%%%%%%%%%%%%%%%%%%%%%%%%%%%%%%%%%%%%%%%%%%%%%%%%%%%%%%%%%%%%%%%%%%%%%%%%%%%%%%%%%%%%%%%%%%%%%%%%%%%%%%%%%%%%%%%%%%%%%%%%%
%%%%%%%%%%%%%%%%%%%%%%%%%%%%%%%%%%%%%%%%%%%%%%%%%%%%%%%%%%%%%%%%%%%%%%%%%%%%%%%%%%%%%%%%%%%%%%%%%%%%%%%%%%%%%%%%%%%%%%%%%%%%%%%%%%%%%%

%%%%%%%%%%%%%%%%%%%%%%%%%%%%%%%%%%%%%%%%%%%%%%%%%%%%%%%%%%%%%%%%%%%%%%%%%%%%%%%%%%%%%%%%%%%%%%%%%%%%%%%%%%%%%%%%%%%%%%%%%%%%%%%%%%%%
%%%%%%%%%%%%%%%%%%%%%%%%%%%%%%%%%%%%%%%%%%%%%%%%%%%%%%%%%%%%%%%%%%%%%%%%%%%%%%%%%%%%%%%%%%%%%%%%%%%%%%%%%%%%%%%%%%%%%%%%%%%%%%%%%%%%%%

%%%%%%%%%%%%%%%%%%%%%%%%%%%%%%%%%%%%%%%%%%%%%%%%%%%%%%%%%%%%
\section{Dynamics of thermal entanglement}
%%%%%%%%%%%%%%%%%%%%%%%%%%%%%%%%%%%%%%%%%%%%%%%%%%%%%%%%%%%%

In this section we intend to take a glance at the properties of entanglement in many-body system at finite temperatures. The states describing a system in thermal equilibrium state at absolute temperature T, are determined by the Hamiltonian of the system and the inverse temperature $\beta=1/k T$, where $k$ is Boltzmann constant. The thermal density matrix of the system is $\rho=Z^{-1}e^{-\beta H}$, where $Z=tr(e^{-\beta H})$ is the partition function of the system. Consider our step system and micro-step system shown in Figs. \ref{Magnetic_fields_Step} and \ref{Step_by_step}, where any form of magnetic field can be represented by a sequence of step functions.

Assuming that the magnetic field has been divided into a sequence of steps, $a, b, c, \dots$ at times $t_0, t_1, t_2, \cdots$, then at $t=t_0$, the system is in an initial equilibrium state described by,
\be
H(a)=-\sum_{<i,j>}\sigma_{i}^{x}\sigma_{j}^{x}-a\Sigma_{i}\sigma_{i}^{z},
\ee
and
\be
\rho(t_0)=\frac{\sum_i e^{-\beta E_i(a)}|\phi_i\rangle\langle\phi_i|}{\sum_i e^{-\beta E_i(a)}}.
\ee
At $t_1 > t_0$, the system evolves under the new magnetic field $b$ such that
\be
H(b)=-\sum_{<i,j>}\sigma_{i}^{x}\sigma_{j}^{x}-b\Sigma_{i}\sigma_{i}^{z},
\ee
\be
\rho(t_1)= e^{-iH(b)(t_1-t_0)}\rho(0)e^{iH(b)(t_1-t_0)}
\ee
Similarly at $t_2 > t_1$, we have
\be
H(c) = -\sum_{<i,j>}\sigma_{i}^{x}\sigma_{j}^{x}-c\Sigma_{i}\sigma_{i}^{z},
\ee
and
\be
\rho(t_2) = e^{-iH(c)(t_2-t_1)}\rho(t_1)e^{iH(c)(t_2-t_1)}.
\ee
Continuing in the same way along this sequence we can obtain the density matrix at any time $t$, which can be used to evaluate the concurrence as explained earlier.

%%%%%%%%%%%%%%%%%%%%%%%%%%%%%%%%%%%%%%%%%%%%%%%%%%%%%%%%%%%%%%%%%%%%%%%%%%%%%%%%%%%%%%%%%%%%%%%%%%%%%%%%%%%%%%%%%%%%%%%%%%%%%%%%%%%%
\begin{figure}[htbp]
\begin{minipage}[c]{\textwidth}
 \centering
   \subfigure[]{\label{fig:Step_KT1}\includegraphics[width=7 cm]{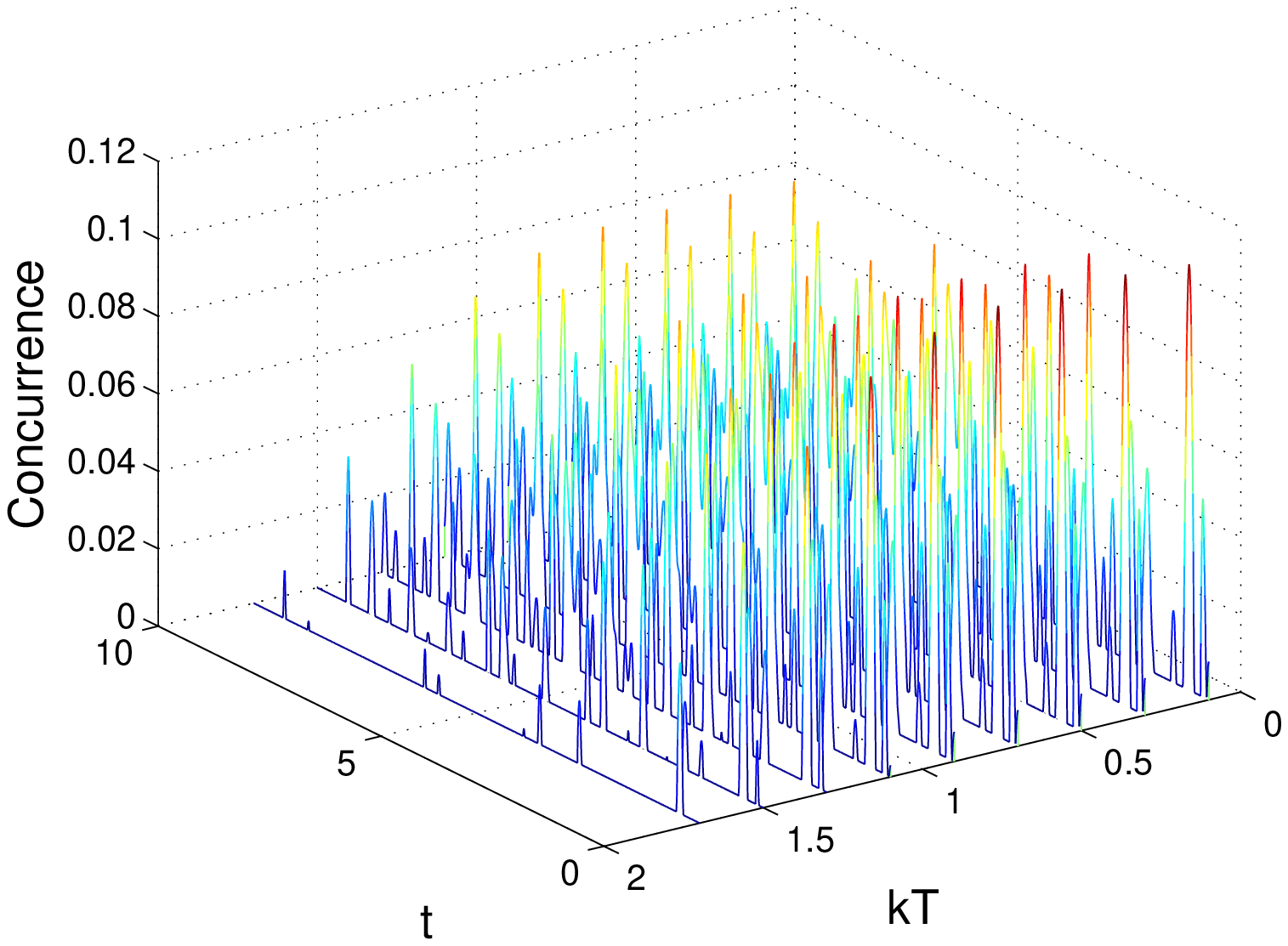}}\quad
   \subfigure[]{\label{fig:Step_KT2}\includegraphics[width=7 cm]{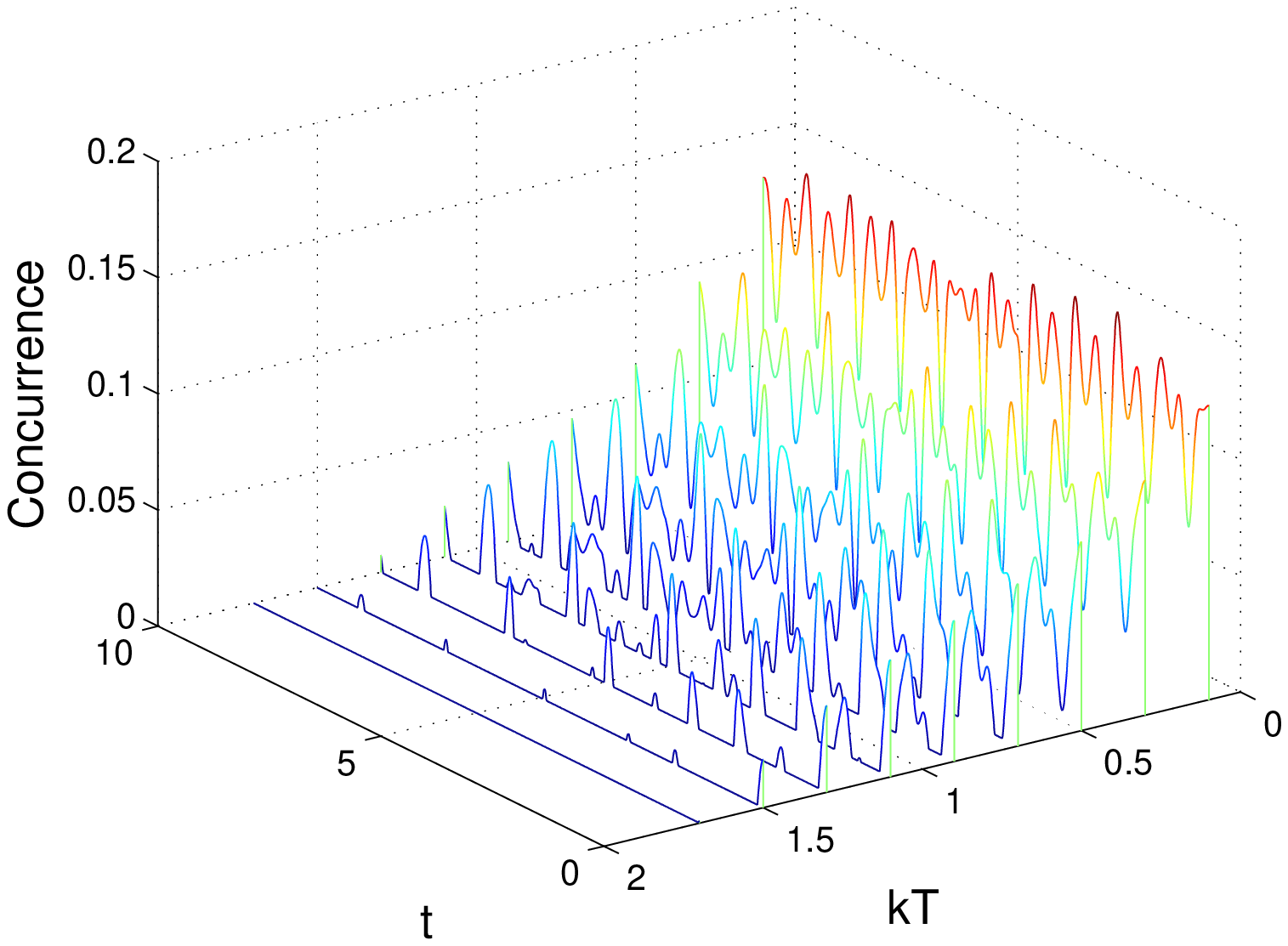}}
   \caption{{\protect\footnotesize The time evolution of the concurrence as a function of the temperature $kT$ under an applied step magnetic field, where time $t$ is in the unit of $J^{-1}$ and $kT$ in the unit of $J$.}}
 \label{Step_KT}
 \end{minipage}
\end{figure}
%%%%%%%%%%%%%%%%%%%%%%%%%%%%%%%%%%%%%%%%%%%%%%%%%%%%%%%%%%%%%%%%%%%%%%%%%%%%%%%%%%%%%%%%%%%%%%%%%%%%%%%%%%%%%%%%%%%%%%%%%%%%%%%%%%%%%%

%%%%%%%%%%%%%%%%%%%%%%%%%%%%%%%%%%%%%%%%%%%%%%%%%%%%%%%%%%%%%%%%%%%%%%%%%%%%%%%%%%%%%%%%%%%%%%%%%%%%%%%%%%%%%%%%%%%%%%%%%%%%%%%%%%%%
\begin{figure}[htbp]
\begin{minipage}[c]{\textwidth}
 \centering
   \subfigure[]{\label{fig:Exp_KT1}\includegraphics[width=7 cm]{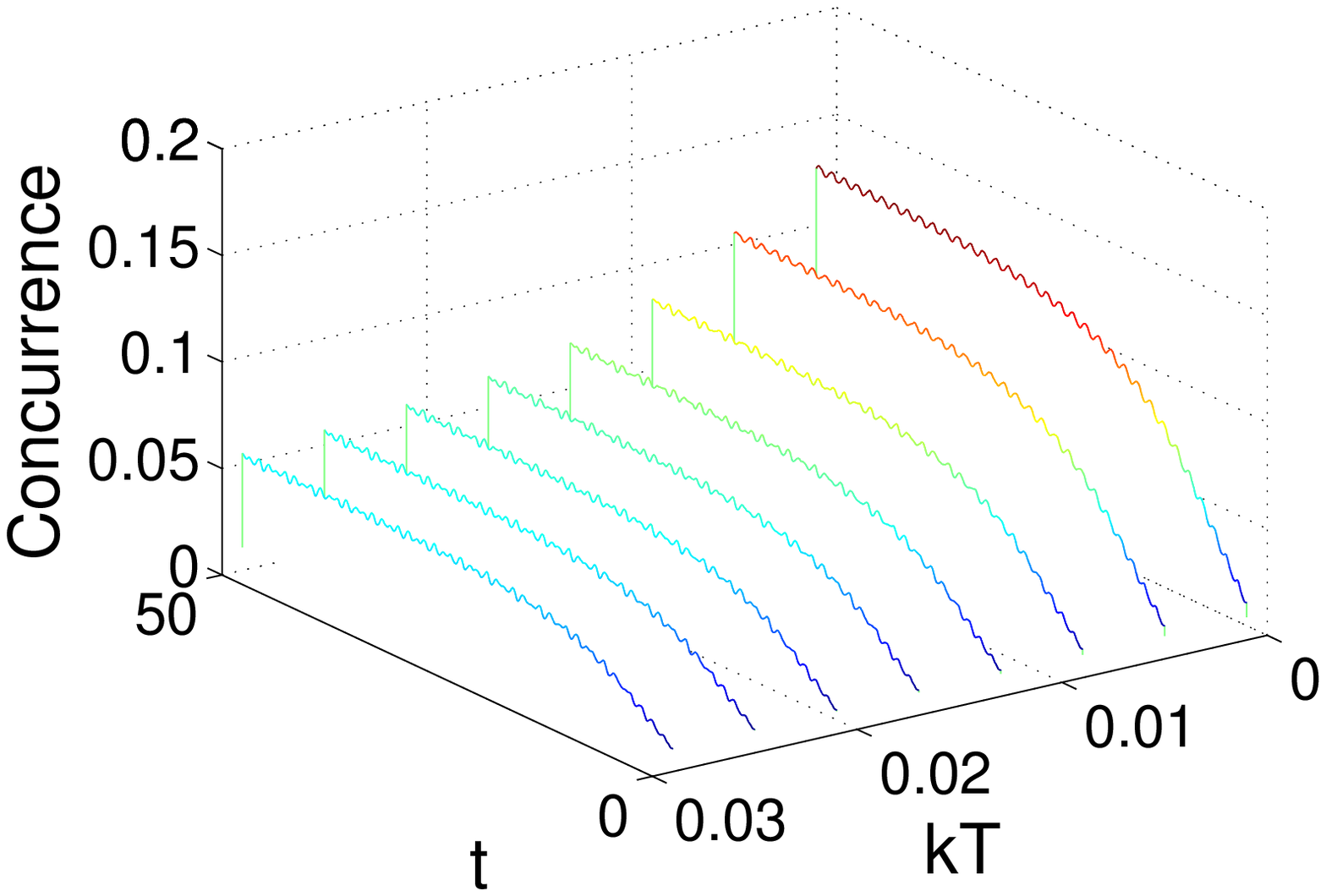}}\quad
   \subfigure[]{\label{fig:Exp_KT2}\includegraphics[width=7 cm]{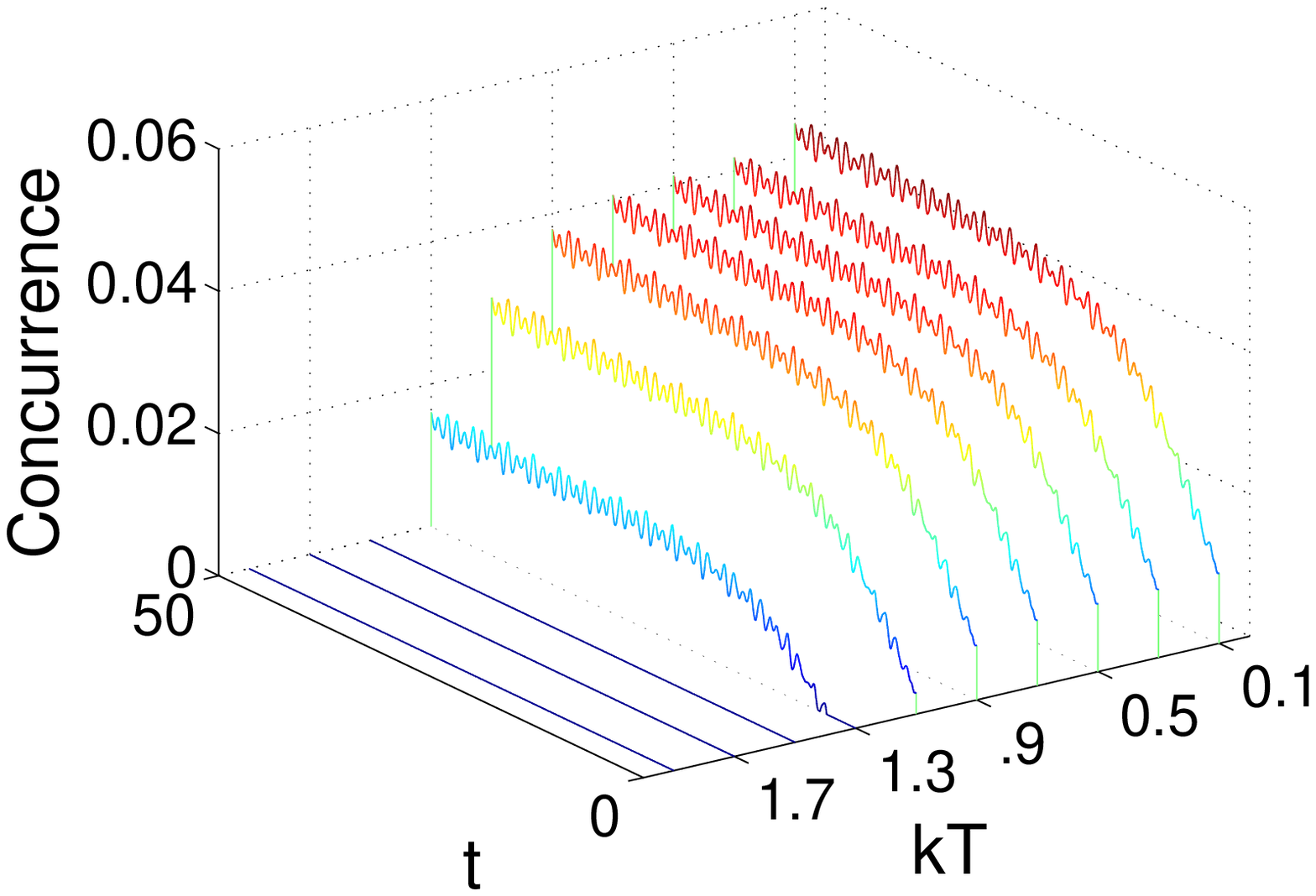}}
   \caption{{\protect\footnotesize The time evolution of the concurrence as a function of the temperature $kT$ under an applied exponential magnetic field, where time $t$ is in the unit of $J^{-1}$ and $kT$ in the unit of $J$.}}
 \label{Exp_KT}
 \end{minipage}
\end{figure}
%%%%%%%%%%%%%%%%%%%%%%%%%%%%%%%%%%%%%%%%%%%%%%%%%%%%%%%%%%%%%%%%%%%%%%%%%%%%%%%%%%%%%%%%%%%%%%%%%%%%%%%%%%%%%%%%%%%%%%%%%%%%%%%%%%%%%%

%%%%%%%%%%%%%%%%%%%%%%%%%%%%%%%%%%%%%%%%%%%%%%%%%%%%%%%%%%%%%%%%%%%%%%%%%%%%%%%%%%%%%%%%%%%%%%%%%%%%%%%%%%%%%%%%%%%%%%%%%%%%%%%%%%%%
\begin{figure}[htbp]
\begin{minipage}[c]{\textwidth}
 \centering
   \subfigure[]{\label{fig:tanh_KT1}\includegraphics[width=7 cm]{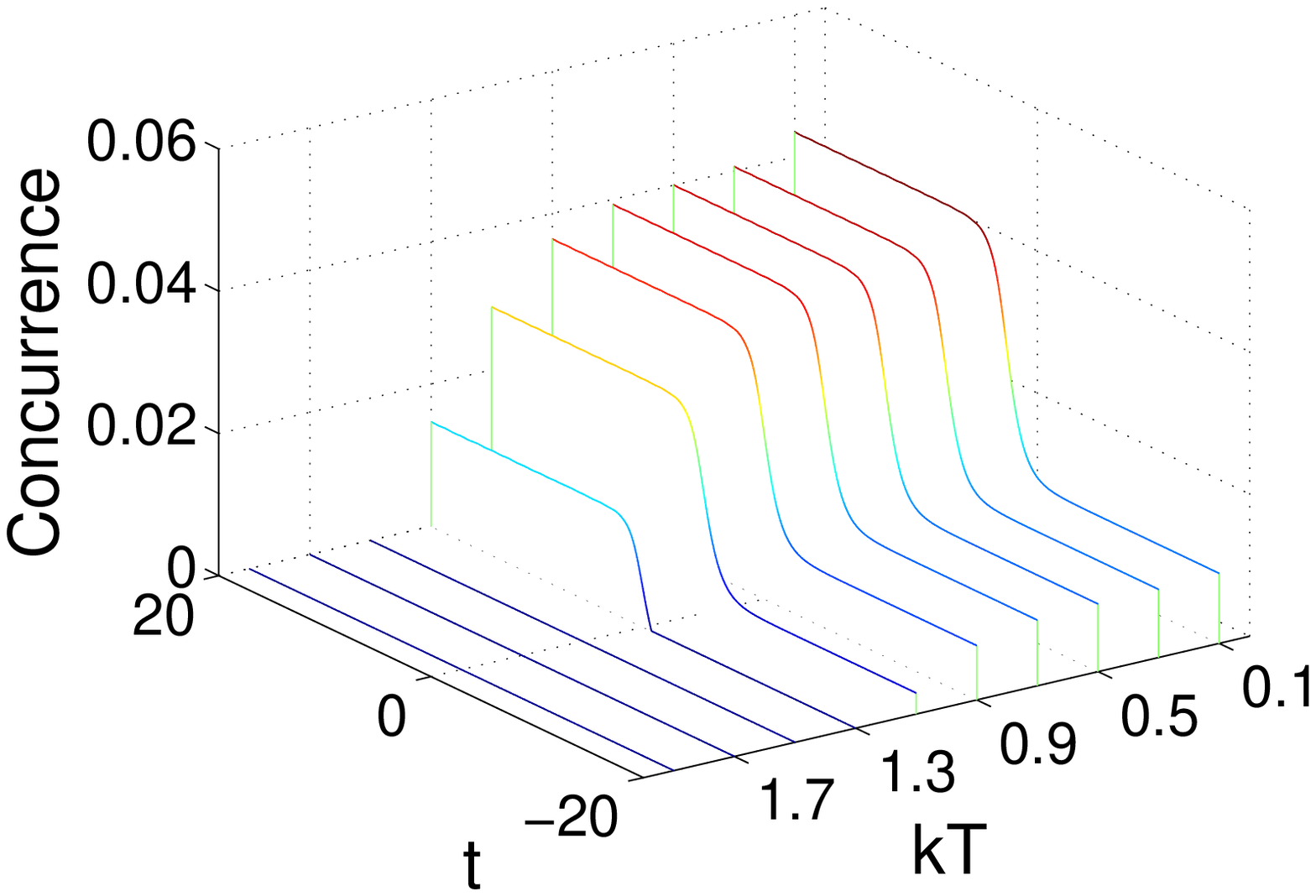}}\quad
   \subfigure[]{\label{fig:tanh_KT2}\includegraphics[width=7 cm]{tem_tanh_a=1_b=2_w=0.5.eps}}
   \caption{{\protect\footnotesize The time evolution of the concurrence as a function of the temperature $kT$ under an applied hyperbolic magnetic field, where time $t$ is in the unit of $J^{-1}$ and $kT$ in the unit of $J$. }}
 \label{tanh_KT}
 \end{minipage}
\end{figure}
%%%%%%%%%%%%%%%%%%%%%%%%%%%%%%%%%%%%%%%%%%%%%%%%%%%%%%%%%%%%%%%%%%%%%%%%%%%%%%%%%%%%%%%%%%%%%%%%%%%%%%%%%%%%%%%%%%%%%%%%%%%%%%%%%%%%%%

One can see from the formulation that the thermal equilibrium (relaxation) only enters at $t=0$, where we assume that the time scale of the dynamics under study is much smaller than the thermal relaxation time.

Figure \ref{Step_KT} shows how concurrence evolves versus time and temperature under step magnetic field $h(t)=a$  at $t\leq t_0$, and $h(t)=b$ at $t>t_0$. Adjusting the step values, lead to similar behaviors. They all oscillate through time which is consistent with the results at zero temperature. When the temperature increases, oscillations keep the shape but are weakened obviously. As can be seen for either the step $a=1$, $b=2$, or the step $a=2$, $b=3$, the  concurrence disappears around $kT=1.75$. $kT=1.75$ means when the energy associated with temperature is 1.75 times of the exchange interaction $J$ of the system, concurrence will be ``killed'', which shows how fragile the entanglement is in this spin system. Figures \ref{Exp_KT} and \ref{tanh_KT} show a very similar behavior of the concurrence of the system under other forms of the external magnetic field namely exponential and hyperbolic respectively. The general profile of the concurrence resembles the zero-temperature case specially at low temperatures. Remarkably, as the temperature increases the concurrence decays very rabidly close to the zero temperature as shown in Figs. \ref{fig:Exp_KT1} and \ref{fig:tanh_KT1} before it completely vanishes at about the same temperature as the exponential case (Figs. \ref{fig:Exp_KT2} and \ref{fig:tanh_KT2}).

\begin{figure}\centering\label{tem_t0_decay_a=0.1-15}
\includegraphics[scale=0.6]{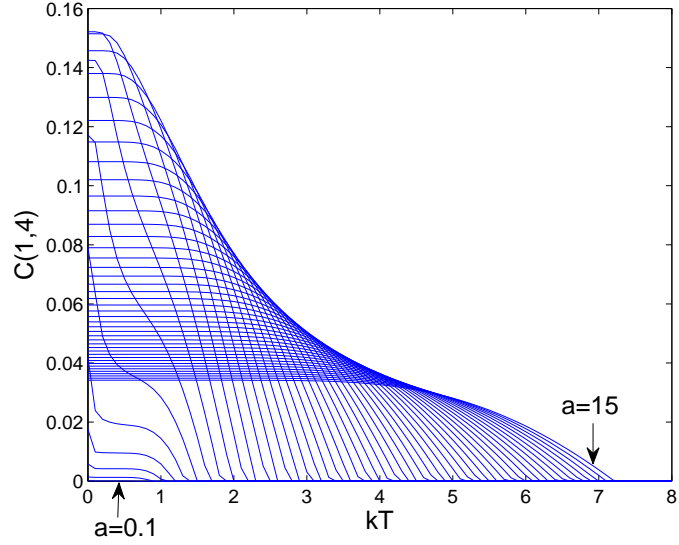}
\caption{Concurrence vs temperature at certain constant magnetic field strength $a$, where $a$ varies from $0.1$ to $15$ in the unit of $J$.}
\end{figure}

To understand the critical temperature, at which the concurrence vanishes, better we draw the graph ``concurrence $C$ v.s. temperature $kT$ under certain constant magnetic field strength $a$'' (Fig. \ref{tem_t0_decay_a=0.1-15}). As the magnetic field gets big, the critical temperature gets big too. When $a$ is small, zero temperature concurrence is small, and stays stable as temperature increases a bit, but for not long it drops dramatically. When $a$ is close to the ``maximum point'' (Fig. \ref{before_after_cross_mp_14}), the zero temperature concurrence is relatively large, and drops all the way down as temperature increases. When $a$ passes that point, though the concurrence starts smaller and smaller, it owns the ability to be more and more stable over temperature change, and finally become zero at relatively larger temperature. This is telling us that sacrificing certain amount of concurrence (choosing a large $a$) may make the concurrence more robust against thermal fluctuations.

In the attempt to understand the vanishing of concurrence, we found that if only the ground state doublets are included (others excluded by force), which are not exactly degenerate due to finite size effect, the entanglement never vanishes at finite temperature. However, after the one-particle excitation continuum states are included, with higher energy states projected out by force, the entanglement disappears at certain temperatures. This tells us that the mixing of the ground state and the low-energy excited states will cause the system lose its entanglement. Physically, the population of the excited state $n$ is determined by the ratio $(E_n-E_0)/T$, and remains almost zero for $T<E_n-E_0$ and becomes significantly nonzero as $T>E_n-E_0$. Therefore in the well gapped regions (very small fields and large fields) the entanglement sustains the $T=0$ values up to a certain $T^*$ and then drops to zero; while in the critical region the entanglement starts dropping immediately after $T>0$ without a plateau region. Numerically, however, $T^*$ is, thought related, not simply equal to the excitation gap. This might be so because the size of the system prevents it from revealing what only statistically holds true. Another question to be studied is why the concurrence should come to zero in a `sudden' way like undergoing a phase transition, in contrast to an exponential decay as one expects with a crossover.

%%%%%%%%%%%%%%%%%%%%%%%%%%%%%%%%%%%%%%%%%%%%%%%%%%%%%%%%%%%%%%%%%%%%%%%%%
\section{Fermi's golden rule and adiabatic approximation}
%%%%%%%%%%%%%%%%%%%%%%%%%%%%%%%%%%%%%%%%%%%%%%%%%%%%%%%%%%%%%%%%%%%%%%%%%

For all exponential, hyperbolic and periodic, when the transition constant (frequency) $\omega$ is small, entanglement tends to follow the change of external magnetic field; when $\omega$ gets larger, entanglement gradually loses pace with the field. These phenomena can be explained by Fermi's golden rule and the adiabatic approximation.
In order for the entanglement to follow the change in the external field, one requires that the system not to deviate far from the ground state, which is the adiabatic approximation.

Physically, it is equivalent to the requirement that the characteristic frequency in the external field is much smaller than the gap, $E_1(t)-E_0(t)$. This may be demonstrated in the following way. From time-dependent perturbation theory, given a system at its ground state at $t=0$, the probability amplitude of the $n$th state at $t>0$ is given by
\bea c_n(t)\approx\frac{-i}{\hbar}\int_0^t{dt}\langle n|H'(t)|0\rangle\exp(i(E_n-E_o)t).\eea
The transition probability from ground state to the $n$th state is
\bea P_{0n}(t)=|c_n(t)|^2\approx\frac{|S^z_{n0}|^2}{\hbar^2}|h(E_n-E_0)|^2,\eea
where we have
\bea H'(t)=h(t)\sum_i\sigma_z^i=h(t)S^z,\eea
and
\bea h(\omega')=\int_{-\infty}^{\infty}h(t)\exp(i\omega' t)dt.\eea
This is the Fermi's golden rule, that the system only absorbs perturbations at frequencies that match the excitations energies. Both data shown for the exponential form and the harmonic form of external field can be explained using this principle. For example, if one has
\be
h(t)=\left\{
\begin{array}{lr}
1 & \qquad t\leq 0 \\
1-e^{-\omega t} & \qquad t>0
\end{array}
\right.,
\ee
Fourier transform gives
\bea |h(\omega')|^2=\frac{1}{(\omega'^2+\omega^2)}.\eea
The transition probability is
\bea P_{0n}\approx\frac{|S^z_{n0}|^2}{\hbar^2}\frac{1}{(\omega'^2+\omega^2)}.\eea
From this formula we can see that if one has $\omega\ll E_1-E_0$, the transition rate is very small and the system is able to follow the change of the magnetic field.

The validity of the adiabatic approximation may fail in two cases. First, when the external field is changing too fast for the system to follow, say, $\omega\gg E_n-E_0$ for certain $n$, and the system has a significant probability of being excited to the $n$th state thus losing the ground state entanglement. One can see this type of breaking for exponential (hyperbolic) form of external field with large $\omega$ and for step function (which corresponds to $\omega\to\infty$ of the exponential form) and for the case of harmonic field with high frequency. The second type of adiabatic approximation breaking takes place if the strength of the external field is too strong that during its change, the system has to cross its phase boundary. Suppose the system is able to follow the field and stays on the ground state up to the critical field, then right at the transition the gap between the ground state and a one-particle excitation continuum is closing therefore an arbitrarily slow field can significantly send the system to various excited states. Although this picture is in principle more relevant in the thermodynamic limit than to the small sized system in the current study, this type of adiabatic breaking is still observed in Fig. \ref{fig:hs_d} \& \ref{fig:hc_d}.

%%%%%%%%%%%%%%%%%%%%%%%%%%%%%%%%%%%%%%%%%%%%%%%%%%%%%%%%%%%%
\section{Extension of projection method to larger systems}
%%%%%%%%%%%%%%%%%%%%%%%%%%%%%%%%%%%%%%%%%%%%%%%%%%%%%%%%%%%%

At the end of section ``Step by step projection'', we mentioned the summation of eq. (\ref{projection_sum}) covers all the eigenstates. For large systems, such as 19-site system, this step is not realistic. Now if we consider a system at zero temperature and at $t=0$ one starts to change the field. One can prove that not all excited states need to be included in the projection, because most excited states have zero overlap with the ground state by symmetry. The global symmetries of the system are: spatial six-fold rotation $C_6$, spin-flip operation $Z_2$, spatial reflection about the x-axis $m_x$ and the y-axis $m_y$. Group theory tells us that the Hamiltonian can be block diagonalized and each block corresponds to an irreducible representation of the symmetry group. If the ground state lies in the $i$-th block, then since the time evolution operator $U(t)=T\exp(-i\int_0^tH(t)dt)$ commutes with all symmetry operators, the end state $|\psi(t)\rangle$ still lies in the same irreducible representation. Let us illustrate this idea by considering a simpler symmetry group constituted by only $C_6$ and $Z_2$, then the group has twelve one-dimensional irreducible representations. Each representation corresponds to the eigenvalues of the two operators: $((-1)^m,\exp(i\frac{n\pi}{3}))$, where $m=1,2$ and $n=1,2...6$. Suppose $|\psi_0\rangle$ belongs the sector denoted by $(m,n)$, and $|\phi_{m'n'}(t)\rangle$ is an eigenstate of $H(t)$ in the $(m',n')$ sector. Then we have
\be
[C_6,U(t)]=0,
\ee
\be
\langle\phi_{m'n'}(t)|[C_6,U(t)]|\psi(0)\rangle=(e^{im'\pi/3}-e^{im\pi/3})\langle\phi_{m'n'}|\psi(t)\rangle=0.
\ee
Therefore the overlap is nonzero only if $m'=m$. Similarly we can prove $n'=n$. The simplified symmetry group can divide the original Hilbert space into (unequally) twelve parts, and the dynamics is fully captured in only one of them; the additional $m_{x,y}$ symmetries can help further reduce the number of states one need to consider for the zero temperature dynamics.

Although symmetry has helped largely reduce the calculation cost, the remaining problem may still be too extensive to solve. It is reasonable to apply an approximation: only states that have energies not much higher than the ground state are included in the projection if the external field does not vary too fast. Consider the case when the field changes very slowly, hence the adiabatic approximation is valid and one needs to include only the ground state in the calculation. Further, if the field changes faster than the excitation gap but not as fast as ranging the whole excitation spectrum, it would be enough if one only includes the first few excited states (possibly the first band of excitations) to cover the most important time scales.

\section{Conclusion and future directions}
We have investigated the dynamics of entanglement in a two-dimensional triangular Ising spin lattice in an external time-dependent magnetic field. The spins are coupled to each other through nearest neighbor exchange interaction. We studied nearest neighbor and next nearest neighbor concurrences of the system under different time-dependent forms of the external magnetic field; step, exponential, hyperbolic and periodic. In contrary to the one dimensional Ising spin system, the two-dimensional system shows an ergodic behavior under the effect of the time dependent magnetic fields. The step magnetic field causes great disturbance to the system and leads to rapidly oscillating concurrence with amplitude and frequency that depend on the magnetic field step values. The System shows more controllability under the effect of the other forms of magnetic fields where the concurrence profile follows the shape of the applied magnetic field very closely particularly for small magnetic field strength and small transition constant for the exponential and hyperbolic fields and frequency for the periodic field. As the values of these parameters increase the concurrence breaks the pattern and rapid oscillations take place. The initial value of the applied periodic field, independent of the amplitude, is very critical to the oscillating concurrence profile, smaller initial value yields less distorted oscillation. Studying the entanglement at zero and finite temperature revealed that it sustains the same profile under the different magnetic fields as the temperature increases but with reduced magnitude. The effect of the temperature is very devastating to the entanglement of the system which decays rapidly as the temperature increases. Interestingly, though small value of the magnetic field strength leads to small concurrence, it is found to be more robust against thermal fluctuations than the larger field strength. In future, we would like to study a larger size two-dimensional spin lattice to examine the effect of the size on the different properties of the system \cite{Kais2003}. Also we would like to study a more generalized spin system where the coupling among the spins in the other directions, rather than the z direction, is taken into account as well as the effect of a time dependent exchange coupling.

Furthermore the dynamics of a real system is determined not only by its internal Hamiltonian but also by its environment. The rich and varied physics of spin systems make spin baths fundamentally interesting \cite{Sadiek2008}. We plan to study the reduced dynamics of the center spin with the rest as environment, also entanglements and decoherence for a reduced system with two interacting spins and the rest as the environment, in the 7 sites and 19 sites spin systems.

\section*{Acknowledgments}
We are grateful to the Army Research Office (ARO) for support of this work at Purdue and to the Deanship of Scientific Research for support at King Saud University.


\begin{thebibliography}{37}
\expandafter\ifx\csname natexlab\endcsname\relax\def\natexlab#1{#1}\fi
\expandafter\ifx\csname bibnamefont\endcsname\relax
  \def\bibnamefont#1{#1}\fi
\expandafter\ifx\csname bibfnamefont\endcsname\relax
  \def\bibfnamefont#1{#1}\fi
\expandafter\ifx\csname citenamefont\endcsname\relax
  \def\citenamefont#1{#1}\fi
\expandafter\ifx\csname url\endcsname\relax
  \def\url#1{\texttt{#1}}\fi
\expandafter\ifx\csname urlprefix\endcsname\relax\def\urlprefix{URL }\fi
\providecommand{\bibinfo}[2]{#2}
\providecommand{\eprint}[2][]{\url{#2}}

\bibitem[{\citenamefont{Peres}(1993)}]{Peres1993}
\bibinfo{author}{\bibfnamefont{A.}~\bibnamefont{Peres}},
  \emph{\bibinfo{title}{Quantum Theory: Concepts and Methods}}
  (\bibinfo{publisher}{Kluwer, Dordrecht, The Netherlands},
  \bibinfo{year}{1993}).

\bibitem[{\citenamefont{Kais}(2007)}]{Kais2007}
\bibinfo{author}{\bibfnamefont{S.}~\bibnamefont{Kais}}, in
  \emph{\bibinfo{booktitle}{Reduced-Density-Matrix Mechanics - with Application
  to Many-Electron Atoms and Molecules}} (\bibinfo{publisher}{John Wiley \&
  Sons Inc}, \bibinfo{address}{New York}, \bibinfo{year}{2007}), vol.
  \bibinfo{volume}{134} of \emph{\bibinfo{series}{Advances in Chemical
  Physics}}, pp. \bibinfo{pages}{493--535}.

\bibitem[{\citenamefont{Sondhi et~al.}(1997)\citenamefont{Sondhi, Girvin,
  Carini, and Shahar}}]{Sondhi1997}
\bibinfo{author}{\bibfnamefont{S.~L.} \bibnamefont{Sondhi}},
  \bibinfo{author}{\bibfnamefont{S.~M.} \bibnamefont{Girvin}},
  \bibinfo{author}{\bibfnamefont{J.~P.} \bibnamefont{Carini}},
  \bibnamefont{and} \bibinfo{author}{\bibfnamefont{D.}~\bibnamefont{Shahar}},
  \bibinfo{journal}{Rev. Mod. Phys.} \textbf{\bibinfo{volume}{69}},
  \bibinfo{pages}{315} (\bibinfo{year}{1997}).

\bibitem[{\citenamefont{Osborne and Nielsen}(2002)}]{Osborne2002}
\bibinfo{author}{\bibfnamefont{T.~J.} \bibnamefont{Osborne}} \bibnamefont{and}
  \bibinfo{author}{\bibfnamefont{M.~A.} \bibnamefont{Nielsen}},
  \bibinfo{journal}{Phys. Rev. A} \textbf{\bibinfo{volume}{66}},
  \bibinfo{pages}{032110} (\bibinfo{year}{2002}).

\bibitem[{\citenamefont{Zhang et~al.}(2009)\citenamefont{Zhang, Cucchietti,
  Chandrashekar, Laforest, Ryan, Ditty, Hubbard, Gamble, and
  Laflamme}}]{ZhangJF2009}
\bibinfo{author}{\bibfnamefont{J.}~\bibnamefont{Zhang}},
  \bibinfo{author}{\bibfnamefont{F.~M.} \bibnamefont{Cucchietti}},
  \bibinfo{author}{\bibfnamefont{C.~M.} \bibnamefont{Chandrashekar}},
  \bibinfo{author}{\bibfnamefont{M.}~\bibnamefont{Laforest}},
  \bibinfo{author}{\bibfnamefont{C.~A.} \bibnamefont{Ryan}},
  \bibinfo{author}{\bibfnamefont{M.}~\bibnamefont{Ditty}},
  \bibinfo{author}{\bibfnamefont{A.}~\bibnamefont{Hubbard}},
  \bibinfo{author}{\bibfnamefont{J.~K.} \bibnamefont{Gamble}},
  \bibnamefont{and} \bibinfo{author}{\bibfnamefont{R.}~\bibnamefont{Laflamme}},
  \bibinfo{journal}{Phys. Rev. A} \textbf{\bibinfo{volume}{79}},
  \bibinfo{pages}{012305} (\bibinfo{year}{2009}).

\bibitem[{\citenamefont{Osenda et~al.}(2003)\citenamefont{Osenda, Huang, and
  Kais}}]{Osenda2003}
\bibinfo{author}{\bibfnamefont{O.}~\bibnamefont{Osenda}},
  \bibinfo{author}{\bibfnamefont{Z.}~\bibnamefont{Huang}}, \bibnamefont{and}
  \bibinfo{author}{\bibfnamefont{S.}~\bibnamefont{Kais}},
  \bibinfo{journal}{Phys. Rev. A} \textbf{\bibinfo{volume}{67}},
  \bibinfo{pages}{062321} (\bibinfo{year}{2003}).

\bibitem[{\citenamefont{Huang et~al.}(2004)\citenamefont{Huang, Osenda, and
  Kais}}]{HuangZ2004}
\bibinfo{author}{\bibfnamefont{Z.}~\bibnamefont{Huang}},
  \bibinfo{author}{\bibfnamefont{O.}~\bibnamefont{Osenda}}, \bibnamefont{and}
  \bibinfo{author}{\bibfnamefont{S.}~\bibnamefont{Kais}},
  \bibinfo{journal}{Phys. Lett. A} \textbf{\bibinfo{volume}{322}},
  \bibinfo{pages}{137 } (\bibinfo{year}{2004}).

\bibitem[{\citenamefont{Nielsen and Chuang}(2000)}]{Nielsen2000}
\bibinfo{author}{\bibfnamefont{M.}~\bibnamefont{Nielsen}} \bibnamefont{and}
  \bibinfo{author}{\bibfnamefont{I.}~\bibnamefont{Chuang}},
  \emph{\bibinfo{title}{Quantum Computation and Quantum Communication}}
  (\bibinfo{publisher}{Cambridge Univ. Press, Cambridge},
  \bibinfo{year}{2000}).

\bibitem[{\citenamefont{Boumeester et~al.}(2000)\citenamefont{Boumeester,
  Ekert, and Zeilinger}}]{Boumeester2000}
\bibinfo{editor}{\bibfnamefont{D.}~\bibnamefont{Boumeester}},
  \bibinfo{editor}{\bibfnamefont{A.}~\bibnamefont{Ekert}}, \bibnamefont{and}
  \bibinfo{editor}{\bibfnamefont{A.}~\bibnamefont{Zeilinger}}, eds.,
  \emph{\bibinfo{title}{The Physics of Quantum Information: Quantum
  Cryptography, Quantum Teleportation, Quantum Computing}}
  (\bibinfo{publisher}{Springer, Berlin}, \bibinfo{year}{2000}).

\bibitem[{\citenamefont{Barenco et~al.}(1995)\citenamefont{Barenco, Deutsch,
  Ekert, and Jozsa}}]{Barenco1995}
\bibinfo{author}{\bibfnamefont{A.}~\bibnamefont{Barenco}},
  \bibinfo{author}{\bibfnamefont{D.}~\bibnamefont{Deutsch}},
  \bibinfo{author}{\bibfnamefont{A.}~\bibnamefont{Ekert}}, \bibnamefont{and}
  \bibinfo{author}{\bibfnamefont{R.}~\bibnamefont{Jozsa}},
  \bibinfo{journal}{Phys. Rev. Lett.} \textbf{\bibinfo{volume}{74}},
  \bibinfo{pages}{4083} (\bibinfo{year}{1995}).

\bibitem[{\citenamefont{Vandersypen et~al.}(2001)\citenamefont{Vandersypen,
  Steffen, Breyta, Yannoni, Sherwood, and Chuang}}]{Vandersypen2001}
\bibinfo{author}{\bibfnamefont{L.}~\bibnamefont{Vandersypen}},
  \bibinfo{author}{\bibfnamefont{M.}~\bibnamefont{Steffen}},
  \bibinfo{author}{\bibfnamefont{G.}~\bibnamefont{Breyta}},
  \bibinfo{author}{\bibfnamefont{C.}~\bibnamefont{Yannoni}},
  \bibinfo{author}{\bibfnamefont{M.}~\bibnamefont{Sherwood}}, \bibnamefont{and}
  \bibinfo{author}{\bibfnamefont{I.}~\bibnamefont{Chuang}},
  \bibinfo{journal}{Nature} \textbf{\bibinfo{volume}{414}},
  \bibinfo{pages}{883} (\bibinfo{year}{2001}).

\bibitem[{\citenamefont{Chuang et~al.}(1998)\citenamefont{Chuang, Gershenfeld,
  and Kubinec}}]{Chuang1998}
\bibinfo{author}{\bibfnamefont{I.~L.} \bibnamefont{Chuang}},
  \bibinfo{author}{\bibfnamefont{N.}~\bibnamefont{Gershenfeld}},
  \bibnamefont{and} \bibinfo{author}{\bibfnamefont{M.}~\bibnamefont{Kubinec}},
  \bibinfo{journal}{Phys. Rev. Lett.} \textbf{\bibinfo{volume}{80}},
  \bibinfo{pages}{3408} (\bibinfo{year}{1998}).

\bibitem[{\citenamefont{Jones et~al.}(1998)\citenamefont{Jones, Mosca, and
  Hansen}}]{Jones1998}
\bibinfo{author}{\bibfnamefont{J.}~\bibnamefont{Jones}},
  \bibinfo{author}{\bibfnamefont{M.}~\bibnamefont{Mosca}}, \bibnamefont{and}
  \bibinfo{author}{\bibfnamefont{R.}~\bibnamefont{Hansen}},
  \bibinfo{journal}{Nature} \textbf{\bibinfo{volume}{393}},
  \bibinfo{pages}{344} (\bibinfo{year}{1998}).

\bibitem[{\citenamefont{Cirac and Zoller}(1995)}]{Cirac1995}
\bibinfo{author}{\bibfnamefont{J.~I.} \bibnamefont{Cirac}} \bibnamefont{and}
  \bibinfo{author}{\bibfnamefont{P.}~\bibnamefont{Zoller}},
  \bibinfo{journal}{Phys. Rev. Lett.} \textbf{\bibinfo{volume}{74}},
  \bibinfo{pages}{4091} (\bibinfo{year}{1995}).

\bibitem[{\citenamefont{Turchette et~al.}(1995)\citenamefont{Turchette, Hood,
  Lange, Mabuchi, and Kimble}}]{Turchette1995}
\bibinfo{author}{\bibfnamefont{Q.~A.} \bibnamefont{Turchette}},
  \bibinfo{author}{\bibfnamefont{C.~J.} \bibnamefont{Hood}},
  \bibinfo{author}{\bibfnamefont{W.}~\bibnamefont{Lange}},
  \bibinfo{author}{\bibfnamefont{H.}~\bibnamefont{Mabuchi}}, \bibnamefont{and}
  \bibinfo{author}{\bibfnamefont{H.~J.} \bibnamefont{Kimble}},
  \bibinfo{journal}{Phys. Rev. Lett.} \textbf{\bibinfo{volume}{75}},
  \bibinfo{pages}{4710} (\bibinfo{year}{1995}).

\bibitem[{\citenamefont{Averin et~al.}(1997)\citenamefont{Averin, Korotkov,
  Manninen, and Pekola}}]{Averin1997}
\bibinfo{author}{\bibfnamefont{D.~V.} \bibnamefont{Averin}},
  \bibinfo{author}{\bibfnamefont{A.~N.} \bibnamefont{Korotkov}},
  \bibinfo{author}{\bibfnamefont{A.~J.} \bibnamefont{Manninen}},
  \bibnamefont{and} \bibinfo{author}{\bibfnamefont{J.~P.}
  \bibnamefont{Pekola}}, \bibinfo{journal}{Phys. Rev. Lett.}
  \textbf{\bibinfo{volume}{78}}, \bibinfo{pages}{4821} (\bibinfo{year}{1997}).

\bibitem[{\citenamefont{Wei et~al.}(2010)\citenamefont{Wei, Kais, and
  Chen}}]{WeiQ2010}
\bibinfo{author}{\bibfnamefont{Q.}~\bibnamefont{Wei}},
  \bibinfo{author}{\bibfnamefont{S.}~\bibnamefont{Kais}}, \bibnamefont{and}
  \bibinfo{author}{\bibfnamefont{Y.~P.} \bibnamefont{Chen}},
  \bibinfo{journal}{J. Chem. Phys.} \textbf{\bibinfo{volume}{132}},
  \bibinfo{pages}{121104} (\bibinfo{year}{2010}).

\bibitem[{\citenamefont{Zurek}(1991)}]{Zurek1991}
\bibinfo{author}{\bibfnamefont{W.}~\bibnamefont{Zurek}},
  \bibinfo{journal}{Phys. Today} \textbf{\bibinfo{volume}{44}},
  \bibinfo{pages}{36} (\bibinfo{year}{1991}).

\bibitem[{\citenamefont{Shor}(1995)}]{Shor1995}
\bibinfo{author}{\bibfnamefont{P.~W.} \bibnamefont{Shor}},
  \bibinfo{journal}{Phys. Rev. A} \textbf{\bibinfo{volume}{52}},
  \bibinfo{pages}{R2493} (\bibinfo{year}{1995}).

\bibitem[{\citenamefont{Bacon et~al.}(2000)\citenamefont{Bacon, Kempe, Lidar,
  and Whaley}}]{Bacon2000}
\bibinfo{author}{\bibfnamefont{D.}~\bibnamefont{Bacon}},
  \bibinfo{author}{\bibfnamefont{J.}~\bibnamefont{Kempe}},
  \bibinfo{author}{\bibfnamefont{D.~A.} \bibnamefont{Lidar}}, \bibnamefont{and}
  \bibinfo{author}{\bibfnamefont{K.~B.} \bibnamefont{Whaley}},
  \bibinfo{journal}{Phys. Rev. Lett.} \textbf{\bibinfo{volume}{85}},
  \bibinfo{pages}{1758} (\bibinfo{year}{2000}).

\bibitem[{\citenamefont{DiVincenzo et~al.}(2000)\citenamefont{DiVincenzo,
  Bacon, Kempe, Burkard, and Whaley}}]{Divincenzo2000}
\bibinfo{author}{\bibfnamefont{D.~P.} \bibnamefont{DiVincenzo}},
  \bibinfo{author}{\bibfnamefont{D.}~\bibnamefont{Bacon}},
  \bibinfo{author}{\bibfnamefont{J.}~\bibnamefont{Kempe}},
  \bibinfo{author}{\bibfnamefont{G.}~\bibnamefont{Burkard}}, \bibnamefont{and}
  \bibinfo{author}{\bibfnamefont{K.~B.} \bibnamefont{Whaley}},
  \bibinfo{journal}{Nature} \textbf{\bibinfo{volume}{408}},
  \bibinfo{pages}{339} (\bibinfo{year}{2000}).

\bibitem[{\citenamefont{Doronin et~al.}(2002)\citenamefont{Doronin, Fel'dman,
  and Lacelle}}]{Doronin2002}
\bibinfo{author}{\bibfnamefont{S.}~\bibnamefont{Doronin}},
  \bibinfo{author}{\bibfnamefont{E.}~\bibnamefont{Fel'dman}}, \bibnamefont{and}
  \bibinfo{author}{\bibfnamefont{S.}~\bibnamefont{Lacelle}},
  \bibinfo{journal}{Chem. Phys. Lett.} \textbf{\bibinfo{volume}{353}},
  \bibinfo{pages}{226} (\bibinfo{year}{2002}).

\bibitem[{\citenamefont{Lages et~al.}(2005)\citenamefont{Lages, Dobrovitski,
  Katsnelson, De~Raedt, and Harmon}}]{Lages2005}
\bibinfo{author}{\bibfnamefont{J.}~\bibnamefont{Lages}},
  \bibinfo{author}{\bibfnamefont{V.~V.} \bibnamefont{Dobrovitski}},
  \bibinfo{author}{\bibfnamefont{M.~I.} \bibnamefont{Katsnelson}},
  \bibinfo{author}{\bibfnamefont{H.~A.} \bibnamefont{De~Raedt}},
  \bibnamefont{and} \bibinfo{author}{\bibfnamefont{B.~N.}
  \bibnamefont{Harmon}}, \bibinfo{journal}{Phys. Rev. E}
  \textbf{\bibinfo{volume}{72}}, \bibinfo{pages}{026225}
  (\bibinfo{year}{2005}).

\bibitem[{\citenamefont{Huang and Kais}(2005)}]{HuangZ2005}
\bibinfo{author}{\bibfnamefont{Z.}~\bibnamefont{Huang}} \bibnamefont{and}
  \bibinfo{author}{\bibfnamefont{S.}~\bibnamefont{Kais}},
  \bibinfo{journal}{Int. J. Quantum Information} \textbf{\bibinfo{volume}{3}},
  \bibinfo{pages}{483} (\bibinfo{year}{2005}).

\bibitem[{\citenamefont{Huang and Kais}(2006)}]{HuangZ2006}
\bibinfo{author}{\bibfnamefont{Z.}~\bibnamefont{Huang}} \bibnamefont{and}
  \bibinfo{author}{\bibfnamefont{S.}~\bibnamefont{Kais}},
  \bibinfo{journal}{Phys. Rev. A} \textbf{\bibinfo{volume}{73}},
  \bibinfo{pages}{022339} (\bibinfo{year}{2006}).

\bibitem[{\citenamefont{Sadiek et~al.}(2010)\citenamefont{Sadiek, Alkurtass,
  and Aldossary}}]{Sadiek2010}
\bibinfo{author}{\bibfnamefont{G.}~\bibnamefont{Sadiek}},
  \bibinfo{author}{\bibfnamefont{B.}~\bibnamefont{Alkurtass}},
  \bibnamefont{and}
  \bibinfo{author}{\bibfnamefont{O.}~\bibnamefont{Aldossary}},
  \bibinfo{journal}{Phys. Rev. A} \textbf{\bibinfo{volume}{82}},
  \bibinfo{pages}{052337} (\bibinfo{year}{2010}).

\bibitem[{\citenamefont{Lieb et~al.}(1961)\citenamefont{Lieb, Schultz, and
  Mattis}}]{Lieb1961}
\bibinfo{author}{\bibfnamefont{E.}~\bibnamefont{Lieb}},
  \bibinfo{author}{\bibfnamefont{T.}~\bibnamefont{Schultz}}, \bibnamefont{and}
  \bibinfo{author}{\bibfnamefont{D.}~\bibnamefont{Mattis}},
  \bibinfo{journal}{Ann. Phys.} \textbf{\bibinfo{volume}{16}},
  \bibinfo{pages}{407} (\bibinfo{year}{1961}).

\bibitem[{\citenamefont{Sachdev}(2001)}]{Sachdev2001}
\bibinfo{author}{\bibfnamefont{S.}~\bibnamefont{Sachdev}},
  \emph{\bibinfo{title}{Quantum Phase Transitions}}
  (\bibinfo{publisher}{Cambridge Univ. Press, Cambridge},
  \bibinfo{year}{2001}).

\bibitem[{\citenamefont{Xavier}(2010)}]{Xavier2010}
\bibinfo{author}{\bibfnamefont{J.~C.} \bibnamefont{Xavier}},
  \bibinfo{journal}{Phys. Lett. B} \textbf{\bibinfo{volume}{81}},
  \bibinfo{pages}{224404} (\bibinfo{year}{2010}).

\bibitem[{\citenamefont{Silva-Valencia
  et~al.}(2005)\citenamefont{Silva-Valencia, Xavier, and
  Miranda}}]{Silva-Valencia2005}
\bibinfo{author}{\bibfnamefont{J.}~\bibnamefont{Silva-Valencia}},
  \bibinfo{author}{\bibfnamefont{J.}~\bibnamefont{Xavier}}, \bibnamefont{and}
  \bibinfo{author}{\bibfnamefont{E.}~\bibnamefont{Miranda}},
  \bibinfo{journal}{Phys. Rev. B} \textbf{\bibinfo{volume}{71}},
  \bibinfo{pages}{024405} (\bibinfo{year}{2005}).

\bibitem[{\citenamefont{Capraro and Gros}(2002)}]{Capraro2002}
\bibinfo{author}{\bibfnamefont{F.}~\bibnamefont{Capraro}} \bibnamefont{and}
  \bibinfo{author}{\bibfnamefont{C.}~\bibnamefont{Gros}},
  \bibinfo{journal}{Eur. Phys. J. B} \textbf{\bibinfo{volume}{29}},
  \bibinfo{pages}{35} (\bibinfo{year}{2002}).

\bibitem[{\citenamefont{Xu et~al.}(2010)\citenamefont{Xu, Kais, Naumov, and
  Sameh}}]{XuQ2010}
\bibinfo{author}{\bibfnamefont{Q.}~\bibnamefont{Xu}},
  \bibinfo{author}{\bibfnamefont{S.}~\bibnamefont{Kais}},
  \bibinfo{author}{\bibfnamefont{M.}~\bibnamefont{Naumov}}, \bibnamefont{and}
  \bibinfo{author}{\bibfnamefont{A.}~\bibnamefont{Sameh}},
  \bibinfo{journal}{Phys. Rev. A} \textbf{\bibinfo{volume}{81}},
  \bibinfo{pages}{022324} (\bibinfo{year}{2010}).

\bibitem[{\citenamefont{Cohen-Tannoudji
  et~al.}(2005)\citenamefont{Cohen-Tannoudji, Diu, and
  Lalo\"{e}}}]{Cohen-Tannoudji2005}
\bibinfo{author}{\bibfnamefont{C.}~\bibnamefont{Cohen-Tannoudji}},
  \bibinfo{author}{\bibfnamefont{B.}~\bibnamefont{Diu}}, \bibnamefont{and}
  \bibinfo{author}{\bibfnamefont{F.}~\bibnamefont{Lalo\"{e}}},
  \emph{\bibinfo{title}{Quantum Mechanics}} (\bibinfo{publisher}{John Wiley \&
  Sons Inc}, \bibinfo{year}{2005}).

\bibitem[{\citenamefont{Osterloh et~al.}(2002)\citenamefont{Osterloh, Amico,
  Falci, and Fazio}}]{Osterloh2002}
\bibinfo{author}{\bibfnamefont{A.}~\bibnamefont{Osterloh}},
  \bibinfo{author}{\bibfnamefont{L.}~\bibnamefont{Amico}},
  \bibinfo{author}{\bibfnamefont{G.}~\bibnamefont{Falci}}, \bibnamefont{and}
  \bibinfo{author}{\bibfnamefont{R.}~\bibnamefont{Fazio}},
  \bibinfo{journal}{Nature} \textbf{\bibinfo{volume}{416}},
  \bibinfo{pages}{608} (\bibinfo{year}{2002}).

\bibitem[{\citenamefont{Wootters}(1998)}]{Wooters1998}
\bibinfo{author}{\bibfnamefont{W.~K.} \bibnamefont{Wootters}},
  \bibinfo{journal}{Phys. Rev. Lett.} \textbf{\bibinfo{volume}{80}},
  \bibinfo{pages}{2245} (\bibinfo{year}{1998}).

\bibitem[{\citenamefont{Kais and Serra}(2003)}]{Kais2003}
\bibinfo{author}{\bibfnamefont{S.}~\bibnamefont{Kais}} \bibnamefont{and}
  \bibinfo{author}{\bibfnamefont{P.}~\bibnamefont{Serra}}
  (\bibinfo{publisher}{John Wiley \& Sons Inc}, \bibinfo{address}{New York},
  \bibinfo{year}{2003}), vol. \bibinfo{volume}{{125}} of
  \emph{\bibinfo{series}{Advances in Chemical Physics}}, pp.
  \bibinfo{pages}{{1--99}}.

\bibitem[{\citenamefont{Sadiek et~al.}(2008)\citenamefont{Sadiek, Huang,
  Aldossary, and Kais}}]{Sadiek2008}
\bibinfo{author}{\bibfnamefont{G.}~\bibnamefont{Sadiek}},
  \bibinfo{author}{\bibfnamefont{Z.}~\bibnamefont{Huang}},
  \bibinfo{author}{\bibfnamefont{O.}~\bibnamefont{Aldossary}},
  \bibnamefont{and} \bibinfo{author}{\bibfnamefont{S.}~\bibnamefont{Kais}},
  \bibinfo{journal}{Mol. Phys.} \textbf{\bibinfo{volume}{106}},
  \bibinfo{pages}{1777} (\bibinfo{year}{2008}).

\end{thebibliography}
\end{document}